\documentclass[graybox]{svmult}
\usepackage{makeidx}         
\usepackage{graphicx} 
\usepackage{multicol}
\usepackage[bottom]{footmisc}

\usepackage{newtxtext}       %
\usepackage[varvw]{newtxmath} 
\usepackage{xcolor}
\usepackage{tabularx}
\usepackage{bm}   

\makeindex             

\begin{document}
\renewcommand{\labelenumii}{\arabic{enumi}.\arabic{enumii}}

\title*{A Unified Semiclassical Framework for Ultrafast Competitive Electron Transfer in Multiredox Molecular Systems}
\titlerunning{Ultrafast Competitive ET in Multiredox Molecular Systems}
\author{Serguei V. Feskov\orcidID{0000-0001-5020-8211} and\\ Anatoly I. Ivanov\orcidID{0000-0002-4420-5863}}
\institute{S. V. Feskov \at Volgograd State University, Universitetsky prosp., 100, Volgograd, Russia \email{serguei.feskov@volsu.ru}
\and A. I. Ivanov \at Volgograd State University, Universitetsky prosp., 100, Volgograd, Russia \email{anatoly.ivanov@volsu.ru}}

\maketitle

\abstract{
  Ultrafast multistage electron transfer (ET) in molecular systems with multiple redox centers is fundamental to photochemical energy conversion, including processes in natural photosynthesis, molecular optoelectronics, and organic photovoltaics. These systems often operate under nonequilibrium conditions, where solvent relaxation, intramolecular vibrations, and competing ET pathways jointly determine reaction kinetics and product yields. In this chapter, we present a unified semiclassical framework for modeling ultrafast competitive ET in multiredox compounds embedded in polar environments with complex relaxation dynamics. The approach constructs diabatic free energy surfaces (FESs) in a multidimensional coordinate space that integrates both polarization and relaxation components of the environment within a unified representation. Electron dynamics are described using a stochastic point-transition method that captures the coupling between nonadiabatic quantum transitions and classical nuclear motion. The formalism generalizes and unifies several established semiclassical models — including the Najbar-Tachiya, Zusman-Beratan, and Sumi-Marcus approaches — and supports efficient simulation of multistage ET cascades. As an application, we investigate ultrafast charge separation in donor–acceptor–acceptor (DA$_1$A$_2$) triads, showing how hot charge shift to a secondary acceptor can suppress nonequilibrium charge recombination. Numerical simulations reveal how reorganization energies, vibrational coupling, molecular geometry, and bending angle collectively influence ET efficiency. The proposed framework offers a general and scalable tool for the rational design of photofunctional molecular systems. 
}


\section{Introduction}
\label{sec:intro}

Electron transfer (ET) is a fundamental process in chemical physics that underpins a wide range of technologies, including solar energy conversion, molecular electronics, and electrocatalysis \cite{ponseca_cr_17, lee_cr_23, lawrence_nrb_23, scattergood_dt_14, fukuzumi_acr_14, bottari_ccr_21, ballabio_acs_22, machin2023, wang_as_24, vauthey_jppa_06, souza_cc_09, ostroverkhova_cr_16, santos_cr_22, xin_nrp_19, fisher_jacs_24}. In many applications, achieving fast and directional ET is important, as it competes with energy-dissipative or chemically unproductive pathways that limit overall efficiency and selectivity. This is particularly true for photoinduced ET, where charge-transfer reactions occur on femtosecond to picosecond timescales and are strongly influenced by solvent polarization and intramolecular vibrational relaxation.

\subsection{Nonequilibrium Nuclear Configurations}

In multistage ET processes, nonequilibrium nuclear configurations typically result from electronic transitions, such as optical excitation or preceding charge transfer steps, that shift the system away from thermal equilibrium. These nonequilibrium states can substantially influence the kinetics and pathways of subsequent ET events by altering the energy landscape and the coupling between electronic and nuclear degrees of freedom \cite{cho_jcp_95, ivanov_cp_99, feskov_jpca_08, zimmermann_jpcb_01, wan_cpl_05, feskov_cpl_07, feskov_jpca_09, kumpulainen_cr_17}. As a result, nonequilibrium ET often deviates significantly from classical thermally activated behavior, necessitating theoretical approaches that account for the dynamic evolution of the nuclear environment. These effects have been extensively investigated through ultrafast spectroscopic experiments and advanced theoretical modeling \cite{torieda_jpca_04, barykov_jpcc_15, feskov_jppc_16, kundu_jpcl_18, matyushov_jml_18, lu_nc_20, siplivy_jcp_20}.

\subsection{Multiredox systems}

The challenge of controlling photoinduced ET on ultrafast timescales is particularly important in organic photovoltaic and photocatalytic systems, where efficient charge separation (CS) must outcompete recombination pathways. In multiredox macromolecular systems, such as molecular triads and extended donor–acceptor cascades, multiple ET steps can occur either sequentially or in parallel \cite{santoro_mol_22, douhal_jppc_16}. In such architectures, accurately predicting and optimizing ET efficiency requires models that incorporate both nuclear relaxation dynamics and electronic couplings across multiple intermediate states \cite{feskov_jppc_16, lebard_jpcb_09}.

\subsection{Photosynthetic Reaction Centers}

Natural photosynthetic reaction centers (RCs) exemplify the remarkable efficiency of multistage ET, where photoinduced charge separation proceeds through a series of ET events with near-unity quantum yields \cite{blankenship_book_21, cherepanov2022}. These biological systems motivate the design of artificial compounds that replicate multistage directional ET. However, achieving this in synthetic systems requires theoretical models that go beyond single-step, equilibrium-based descriptions — models that can capture nonequilibrium effects, multistate competition, and environmental complexity.

\subsection{Symmetry-breaking charge transfer}

In recent years, considerable attention has been directed toward symmetric donor–\-acceptor architectures, including quadrupolar A–D–A and D–A–D compounds, octupolar star-shaped systems such as D(–A)$_3$ and A(–D)$_3$, as well as symmetric dimers. These molecular systems exhibit a diverse range of excited-state phenomena, most notably symmetry-breaking charge transfer (SBCT) and excimer formation, which has become the subject of extensive experimental and theoretical investigation \cite{Wasielewski2020, Vauthey_PCCP23, Terenziani23, dereka_jpcl_24, Hariharan25, Ivanov18, Antipov22, Siplivy24, les24, MikhMikh24, IVANOVRev24, IvanovQubit}. A key feature of these systems is their ability to undergo ultrafast charge separation without requiring thermal activation \cite{Nazarov_SB20}, as evidenced by characteristic charge transfer timescales on the order of a few picoseconds — comparable to solvent relaxation times in polar environments \cite{Vauthey17}. While highly relevant to the broader study of photoinduced ET, these topics lie outside the scope of this chapter and will not be discussed further.

\subsection{Modeling Ultrafast Competitive Electron Transfer}

Here, we present a unified semiclassical framework for modeling ultrafast multistage ET in molecular systems containing multiple redox centers embedded in polar environments. Building on the stochastic point-transition approach \cite{zusman_cp_80}, the theory extends classical models, including those of Najbar \& Tachiya \cite{najbar_jpc_94}, Zusman \& Beratan \cite{zusman_jcp_99}, and Sumi \& Marcus \cite{sumi_jcp_86}, by explicitly incorporating:
\begin{itemize}
   \item multiple coupled ET steps across diabatic free energy surfaces;
   \item hot electron transfer proceeding from nonequilibrium nuclear configurations;
   \item complex multi-component dynamics of environmental relaxation.
\end{itemize}

We demonstrate how this framework can be used for simulation of ultrafast charge separation, quantification of ET quantum yields, and identification of molecular design parameters — including reorganization energies, vibrational coupling strengths, donor–acceptor distances, and overall geometry — that govern ET efficiency. Applications are illustrated for donor–acceptor–acceptor (DA$_1$A$_2$) triads inspired by Zn-porphyrin–imide systems, with relevance to light-activated molecular switches and artificial photosynthetic assemblies.

\section{General Theoretical Framework for Multistage ET in Nonequilibrium Environments}
\label{sec:theory}

Photoinduced ET between donor (D) and acceptor (A) molecules in solution is significantly influenced by the dielectric polarization of the surrounding medium. This interaction is typically characterized by a single parameter — the medium reorganization energy, $\lambda$. In the classical Marcus theory of outer-sphere ET \cite{marcus_jcp_56}, the activation free energy is given by $G^\sharp = \left( \lambda + \Delta G_{\mathrm{ET}} \right)^2 / 4\lambda$, where $\Delta G_{\mathrm{ET}}$ is the ET driving force. Together, $\lambda$ and $\Delta G_{\mathrm{ET}}$ define the barrier height and therefore the ET rate. In weakly coupled systems, the reaction is treated as a nonadiabatic transition between two parabolic diabatic free energy surfaces associated with the donor and acceptor states.

The Marcus framework assumes that the nuclear degrees of freedom remain in thermal equilibrium within the initial donor state before the ET event. Under such conditions, electron transfer is activated by thermally driven fluctuations of polarization. However, this equilibrium assumption is often violated in ultrafast photoreactions, where the initial Franck-Condon state is created by a short laser pulse, placing the system on an upper electronic free energy surface. In such cases, photoinduced ET can proceed on timescales comparable to or even shorter than the characteristic solvent and intramolecular relaxation times.

Electron transfer in polar solvents is often modeled using a single reaction coordinate, $z$, which represents the vertical free energy gap between the donor (reactant) and acceptor (product) electronic states: $z = \Delta G$. This coordinate effectively captures the influence of solvent fluctuations and enables the reduction of the many-body solvent environment to a one-dimensional representation. The resulting two-state, single-coordinate framework has proven useful in many foundational theoretical models \cite{marcus_jcp_56, zusman_cp_80, bixon_acp_99}.

However, in macromolecular systems where photoexcitation triggers a cascade of ultrafast ET steps among multiple redox centers, each individual ET event is associated with its own energy-gap coordinate, $z_k$. These reaction coordinates typically exhibit mutual dependence \cite{marchi_jacs_93, najbar_jpc_94, tang_jcp_94, cho_jcp_95, ando_jpcb_98, newton_ijc_04}, resulting in correlations and nonorthogonality across the multidimensional space of ET processes. Such coupling arises in both discrete and continuum treatments of the environment and must be accounted for to accurately model multistage ET dynamics \cite{newton_jpcb_15}.

\begin{figure}[ht]
\sidecaption[t]
   \includegraphics[scale=0.65]{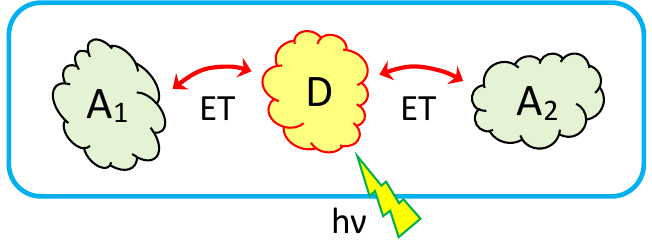}
   \caption{Schematic representation of competitive photoinduced electron transfer in A$_1$DA$_2$ compounds. Optical excitation promotes the electron donor (D) to an excited state, followed by subsequent ET pathways to the acceptors A$_1$ and A$_2$.}
   \label{fig:DAA_scheme}
\end{figure}

For instance, in a A$_1$DA$_2$ triad composed of a donor directly linked to two acceptor units (see Fig.~\ref{fig:DAA_scheme}), competitive ET from D to A$_1$ and A$_2$ proceeds along distinct energy-gap coordinates, $z_1$ and $z_2$. These coordinates are often significantly correlated, with the degree of correlation strongly dependent on the spatial geometry of the system and the polarity of the surrounding medium \cite{najbar_jpc_94}. A similar situation arises in DA$_1$A$_2$ triads, where charge separation toward the $|\text{D}^+ \text{A}_1 \text{A}_2^-\rangle$ state is governed not only by individual ET rates but also by the correlation between the reaction coordinates associated with the D → A$_1$ and A$_1$ → A$_2$ steps \cite{cho_jcp_95, ando_jpcb_98, newton_ijc_04, newton_jpcb_15, feskov_jpca_13, feskov_rjpca_16}. These examples highlight the role of solvent polarization, established during the earliest stages of ultrafast ET, in controlling the dynamics of subsequent electron transfer events.

Naturally, the correlation between the energy-gap coordinates $z_1$ and $z_2$ of two sequential ET steps manifests itself only when the second ET occurs on an ultrafast timescale, comparable to or shorter than the characteristic solvent relaxation time. In such cases, the second ET step proceeds from the nonequilibrium nuclear configuration established by the preceding charge transfer event. If $z_1$ and $z_2$ are orthogonal, deviations from equilibrium along $z_1$ do not influence the distribution along $z_2$, resulting in independent kinetics for the two steps. However, when $z_1$ and $z_2$ are non-orthogonal, the rate of the secondary ET depends on the degree of nuclear nonequilibrium along the $z_2$ coordinate, which, in turn, is determined by the angle between $z_1$ and $z_2$. This mechanism explains the influence of the reaction coordinate correlation on the kinetics of multistage reactions. A similar effect is observed in systems where multiple ET reactions occur in parallel \cite{feskov_rjpca_17}.

It is important to emphasize that this effect represents a notable exception to the Ostwald principle of the independence of elementary reactions. This principle, foundational to classical chemical kinetics, holds that the rate of an elementary step depends solely on the species directly involved and is unaffected by other concurrent reactions within the same system. However, in the context of ultrafast ET, this assumption may break down. When ET occurs on timescales comparable to or faster than solvent and intramolecular relaxation, the system deviates from equilibrium conditions, and the polarization established during earlier ET steps can significantly influence subsequent reaction pathways \cite{sumi_jcp_86, feskov_jppc_16}. By contrast, in slower multistage reactions where the nuclear environment equilibrates between steps, the independence of elementary reactions remains a valid approximation.

The mutual influence of consecutive ultrafast ET steps arising from correlations between their respective energy-gap reaction coordinates has been recognized and analyzed in previous studies \cite{marchi_jacs_93, najbar_jpc_94, tang_jcp_94, cho_jcp_95, ando_jpcb_98, newton_ijc_04, newton_jpcb_15}. However, these investigations have primarily focused on three-center systems involving only two ET steps. Extending such models to more complex, multi-center architectures is essential for advancing the understanding of functional molecular systems. 

In this section, we consider a general macromolecular system composed of $N$ fixed electron donor and acceptor subunits (redox centers). The rigid spatial arrangement of these sites allows us to neglect slow, large-amplitude intramolecular motions and focus instead on the dominant role of electrostatic interactions between the transferred charge and the polar environment in governing reaction energetics. Nonetheless, it is well established that high-frequency, small-amplitude intramolecular vibrations can significantly influence ultrafast ET by modulating activation barriers and affecting the quantum yield of nonequilibrium transitions \cite{jortner_jcp_88, akesson_jcp_92, barbara_sci_92, bagchi_acp_99}. Although the present formulation emphasizes solvent dynamics, the theoretical framework admits a natural extension to include ET-active quantum vibrational modes, which will be developed in later sections of this chapter.

The construction of diabatic free energy surfaces (FESs) corresponding to different electronic states of the system is an important component of any ET model. In multistage systems, the full set of FESs defines not only the activation barriers for individual ET steps but also governs the pathways of solvent relaxation between successive electronic transitions. As a foundation, we adopt the linear response approximation for the dielectric response of the medium. Within this framework, the FESs for electronic states are represented as quadratic functions of the energy-gap reaction coordinates associated with individual ET events \cite{najbar_jpc_94, cho_jcp_95}. These coordinates can be linearly transformed to a new basis in which the quadratic form becomes diagonal, yielding a set of independent generalized solvent coordinates. This diagonalization procedure has been explicitly demonstrated for two-center ET models \cite{najbar_jpc_94} and is generalized here for multistage systems.

We show that the dimensionality of the generalized configuration space is determined not by the number of possible ET transitions between redox sites, but rather by the total number of redox-active centers in the molecular system. To facilitate modeling, we introduce a systematic algorithm for constructing diabatic FESs by incrementally expanding the configuration space. The algorithm enables efficient and scalable simulation of multistage ET dynamics within the stochastic point-transition framework, which extends beyond the applicability of the standard Golden Rule formalism and its perturbative variants \cite{barzykin_acp_02}. The stochastic approach has previously been applied to describe ET in systems with multiple solvent relaxation modes as well as those involving ET-active high-frequency intramolecular vibrations \cite{ivanov_rcr_10, feskov_jppc_16}, and is well suited for addressing nonequilibrium effects in complex environments.

\subsection{Electronic States and Polarization Coordinates of Environment}
\label{subsec:el_states}

\begin{figure}[t]
   \includegraphics[scale=0.48]{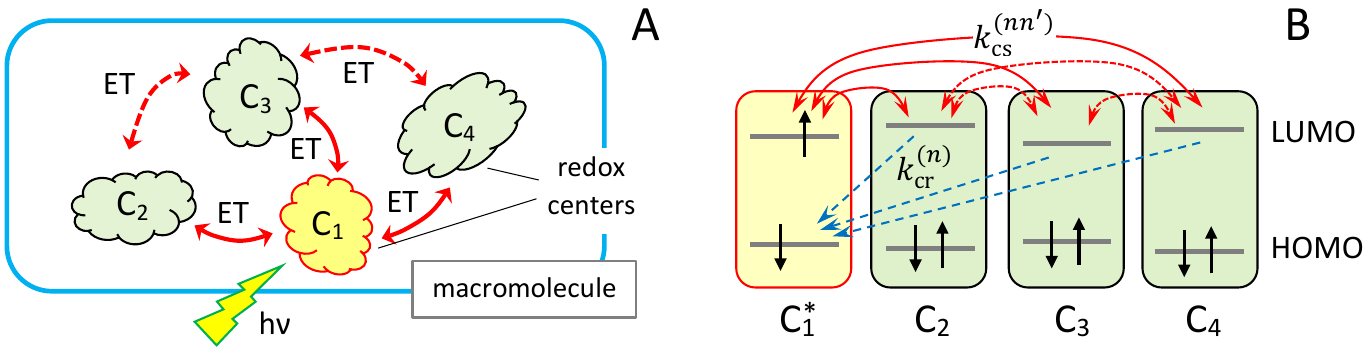}
   \caption{(A) Schematic representation of a macromolecular compound containing multiple redox centers C$_n$, where C$_1$ serves as the photoactive unit initiating a cascade of competitive electron transfer steps. (B) Energy-level diagram illustrating charge separation (CS) and recombination (CR) transitions between HOMO and LUMO orbitals of individual redox units. The quantities $k_{\mathrm{cs}}^{(nn^\prime)}$ and $k_{\mathrm{cr}}^{(n)}$ denote the rate constants of CS and CR transitions in the quasi-equilibrium regime.}
   \label{fig:scheme1}
\end{figure}

Consider $N$ redox centers C$_n$ ($n=\overline{1,N}$) within a macromolecule in a polar environment, as illustrated in Fig.~\ref{fig:scheme1}A. At $t = 0$, the chromophoric center C$_1$ absorbs a photon, initiating a sequence of nonadiabatic electronic transitions, including forward ET (charge separation) and backward ET (charge recombination). The specific pathway of the photoreaction is primarily determined by the characteristics of individual ET steps, particularly the rate constants for charge separation (CS) and recombination (CR) processes. However, during the initial nonequilibrium stage of the photoreaction the standard concept of the ET rate constant becomes inadequate. This transient regime, typically spanning several to tens of picoseconds, is marked by incomplete solvent and intramolecular vibrational relaxation.

Fig.~\ref{fig:scheme1}B illustrates the electronic energy levels associated with the HOMO and LUMO orbitals of the redox centers C$_n$. Immediately after photoexcitation, the system is promoted to an electronically excited state denoted by $\vert \psi_1 \rangle$. Subsequent charge separation results in configurations in which the electron is localized on one of the redox centers C$_n$ ($n \geq 2$), yielding charge-separated states $\vert \psi_n \rangle$. The relevant electronic configurations can be defined as
\begin{equation} \label{states}
  \begin{aligned}
     \vert \psi_0 \rangle =& \vert \text{C}_1 \text{C}_2 \dots \text{C}_N\rangle, \quad \text{ground state} \\
     \vert \psi_1 \rangle =& \vert \text{C}_1^* \text{C}_2 \dots \text{C}_N\rangle, \quad \text{excited state} \\
     \vert \psi_n \rangle =& \vert \text{C}_1^+  \dots \text{C}_n^- \dots \text{C}_N\rangle, \quad (n = \overline{2,N}),  \quad \text{CS states}. 
  \end{aligned}
\end{equation}
Charge separation, charge recombination and internal conversion processes are defined in terms of the $\vert \psi_{n} \rangle$ states as
\begin{equation} \label{ET_types}
  \begin{aligned}
     \text{CS}:& \vert \psi_{n} \rangle \to \vert \psi_{n^\prime} \rangle, \quad (n,n^\prime = 1\dots N, n \neq n^\prime, n^\prime \neq 1) \\
     \text{CR}:& \vert \psi_{n} \rangle \to \vert \psi_0 \rangle, \quad (n = 2\dots N) \\
     \text{IC}:& \vert \psi_1 \rangle \to \vert \psi_0 \rangle.
  \end{aligned}
\end{equation}

It is convenient to express the system Hamiltonian $\hat{H}$ as the sum of two components
\begin{equation}\label{sm_H}
    \hat{H} = \hat{H}_\mathrm{eq} + \hat{H}_\mathrm{ne},
\end{equation}
where $\hat{H}_\mathrm{eq}$ describes the macromolecule in equilibrium with the environment, while $\hat{H}_\mathrm{ne}$ accounts for nonequilibrium configurations of a polar medium around the redox centers. The equilibrium part of the Hamiltonian can be expressed as
\begin{equation}\label{H_eq}
  \begin{aligned}
    \hat{H}_\mathrm{eq} = \sum\limits_{n=0}^N \check{G}^{(n)} \vert \psi_n \rangle \langle \psi_n \vert &+ \sum\limits_{n=1}^N V^{(n)}_\mathrm{cr} \left(\vert \psi_n \rangle \langle \psi_{0} \vert + \text{h.c.} \right) +  \\ &+ \sum\limits_{n=2}^N \sum\limits_{n^\prime>n} V^{(nn^\prime)}_\mathrm{cs} \left(\vert \psi_n \rangle \langle \psi_{n^\prime} \vert + \text{h.c.} \right).
  \end{aligned}
\end{equation}
Here, $\check{G}^{(n)}$ denotes the free energy of the system in the electronic state $\vert \psi_n \rangle$ when it is in equilibrium with the polarization of the medium. The quantities $V^{(n)}_\mathrm{cr}$ and $V^{(nn^\prime)}_\mathrm{cs}$ are the electronic coupling energies governing charge recombination (CR) and charge separation (CS) transitions, respectively. Within the one-electron approximation, these couplings are expressed through orbital overlaps as
\begin{equation}\label{V_defs}
    \begin{aligned}
        V^{(n)}_\mathrm{cr} &= \langle \text{LUMO}_n \vert \hat{\mathcal{H}} \vert \text{HOMO}_1 \rangle, \\[4pt]
        V^{(nn^\prime)}_\mathrm{cs} &= \langle \text{LUMO}_n \vert \hat{\mathcal{H}} \vert \text{LUMO}_{n^\prime} \rangle,
    \end{aligned}
\end{equation}
where $\hat{\mathcal{H}}$ is the one-electron Hamiltonian. The LUMO orbitals appear in these matrix elements because they host the photoexcited electron and therefore directly participate in recombination with the donor HOMO as well as in electron transfer between different acceptor sites, as illustrated in Fig.~\ref{fig:scheme1}B.

The nonequilibrium part of the system Hamiltonian, $\hat{H}_\mathrm{ne}$, accounts for unbalanced dielectric polarization and can be expressed in terms of independent polarization coordinates $q_k$, following the approach developed in Ref.~\cite{feskov_jcp_18}. Assuming a linear response of the medium to charge redistribution within the macromolecule, $\hat{H}_\mathrm{ne}$ takes the form of a parabolic function in the $q_k$ coordinates
\begin{equation}\label{H_ne}
    \hat{H}_\mathrm{ne} = \sum\limits_{n=0}^N \vert \psi_n \rangle \langle \psi_n \vert \sum\limits_{k=1}^{K} \left(q_k - \check{q}_k^{(n)}\right)^2 = \sum\limits_{n=0}^N \vert \psi_n \rangle \langle \psi_n \vert \left\vert\, \bm{q} - \check{\bm{q}}^{(n)}\right\vert^2.
\end{equation}
Here, $\bm{q} = (q_1, q_2, \dots q_K)^\mathrm{T}$ and $\check{\bm{q}}^{(n)} = (\check{q}_1^{(n)}, \check{q}_2^{(n)}, \dots \check{q}_K^{(n)})^\mathrm{T}$ are vectors in the $K$-dimensional configuration space, and $\check{\bm{q}}^{(n)}$ represents the equilibrium polarization associated with the electronic state $\vert \psi_n \rangle$. The operator $\vert \psi_n \rangle \langle \psi_n \vert$ acts as a projector onto the state $\psi_n$, ensuring that the corresponding polarization contribution is applied only when the system occupies this state. Finally, $\left\vert \bm{q} \right\vert$ denotes the Euclidean norm of the vector $\bm{q}$, $\vert \bm{q} \vert = (q_1^2 + q_2^2 + \dots + q_K^2)^{1/2}$.

The $\bm{q}$-space is complete in the sense that it fully characterizes the state of the environment with respect to the CS/CR transitions. The number of the $q_k$ coordinates is in general determined by the number of redox centers $N$ and follows the relation \cite{feskov_jcp_18}
\begin{equation} \label{K}
    K = N - 1.
\end{equation}

\begin{figure}[t]
   \includegraphics[scale=0.48]{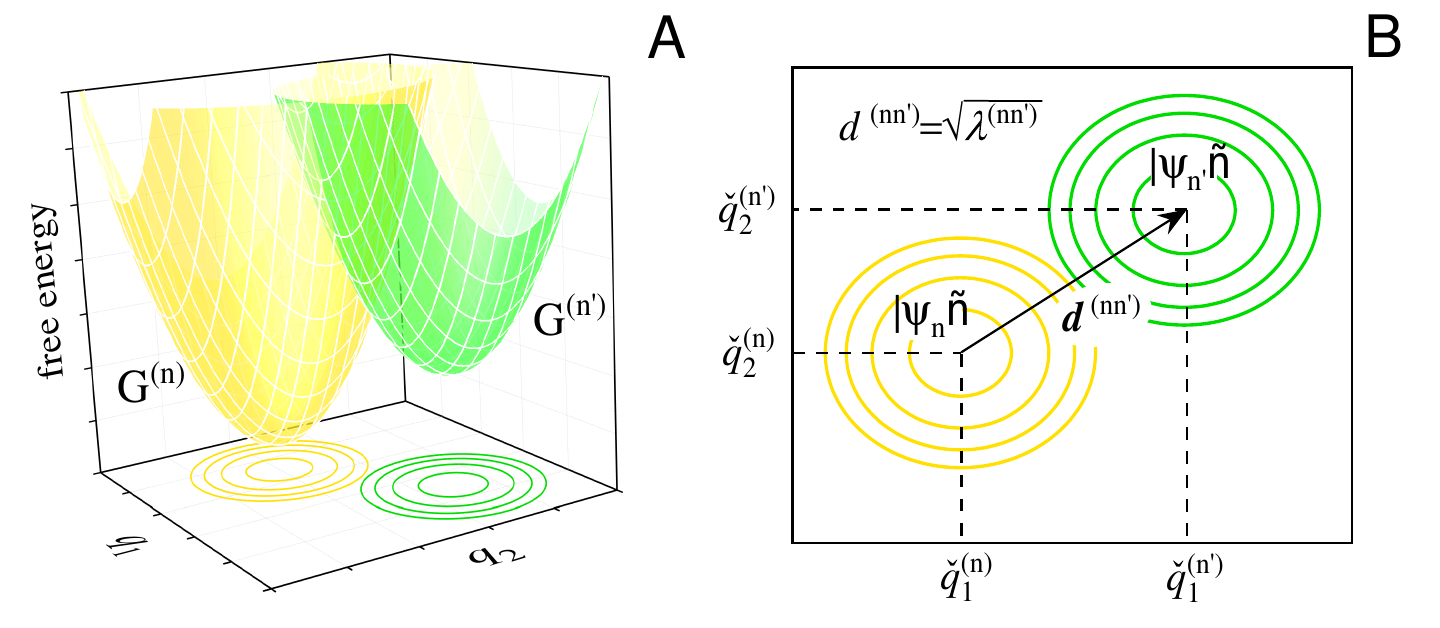}
   \caption{(A) Diabatic free energy surfaces (FESs) for multistage ET, shown in the three-dimensional space of polarization coordinates ($q_1$, $q_2$) and free energy. Each surface $G^{(n)}(\bm{q})$ corresponds to a distinct diabatic state $\vert \psi_n \rangle$. (B) Geometric relationship between the FES minima: the distance between any two equilibrium positions $\check{\bm{q}}^{(n)}$ and $\check{\bm{q}}^{(n^\prime)}$ is equal to $\sqrt{\lambda^{(nn^\prime)}}$, establishing a direct connection between reorganization free energy and spatial configuration of the two FESs.}
   \label{fig:FESs}
\end{figure}

The free energy surface associated with the diabatic electronic state $\vert \psi_n \rangle$ is defined as the diagonal matrix element of the total Hamiltonian, $G^{(n)} \equiv \langle \psi_n \vert \hat{H} \vert \psi_n \rangle$. This matrix element represents the expectation value of the Hamiltonian when the system is constrained to remain in the state $\vert \psi_n \rangle$, and thus provides the energy surface governing polarization coordinates in that electronic state. Substituting Eqs.~\eqref{H_eq} and \eqref{H_ne} yields
\begin{equation}\label{Gn(q)}
  G^{(n)}(\bm{q}) = \check{G}^{(n)} + \sum\limits_{k=1}^{K} \left(q_k - \check{q}_k^{(n)}\right)^2 = \check{G}^{(n)} + \left\vert \bm{q} - \check{\bm{q}}^{(n)} \right\vert^2.
\end{equation}
Each diabatic FES $G^{(n)}$ thus corresponds to a $K$-dimensional paraboloid centered at the equilibrium polarization configuration $\check{\bm{q}}^{(n)}$, with identical curvature along all $q_k$ coordinates and a vertical energy offset $\check{G}^{(n)}$ (see Fig.~\ref{fig:FESs}).

The energetic cost of reorganizing the medium during an electron transfer event $\vert \psi_{n} \rangle \to \vert \psi_{n^\prime} \rangle$ is quantified by the reorganization free energy $\lambda^{(nn^\prime)}$, defined as the energy required to move the system from its equilibrium configuration in state $n$ to the equilibrium configuration of state $n^\prime$, while remaining on the FES of state $n$
\begin{equation}\label{lambda_def}
   \lambda^{(nn^\prime)} \equiv G^{(n)}\!\left(\check{\bm{q}}^{(n^\prime)}\right) - G^{(n)}\!\left(\check{\bm{q}}^{(n)}\right).
\end{equation}
Using Eq.~\eqref{Gn(q)}, this becomes
\begin{align} \nonumber
  \lambda^{(nn^\prime)} = G^{(n)}\!\left(\check{\bm{q}}^{(n^\prime)}\right) - G^{(n)}\!\left(\check{\bm{q}}^{(n)}\right) = \check{G}^{(n)} + \sum\limits_{k=1}^{K} \left(\check{q}_k^{(n^\prime)} - \check{q}_k^{(n)}\right)^2 - \\
  - \left[ \check{G}^{(n)} + \sum\limits_{k=1}^{K} \left(\check{q}_k^{(n)} - \check{q}_k^{(n)}\right)^2 \right]= \sum\limits_{k=1}^{K} \left(\check{q}_k^{(n^\prime)} - \check{q}_k^{(n)}\right)^2 = \left\vert \bm{d}^{(nn^\prime)} \right\vert^2,
  \label{lambda(q)}
\end{align}
where $\bm{d}^{(nn^\prime)} \equiv \check{\bm{q}}^{(n^\prime)} - \check{\bm{q}}^{(n)}$ is the displacement vector between the FES minima corresponding to states $n$ and $n^\prime$. Thus, the reorganization energy $\lambda^{(nn^\prime)}$ is directly related to the squared distance between the equilibrium polarization configurations in $\bm{q}$-space. Given a set of $\lambda^{(nn^\prime)}$ values, one can calculate the pairwise distances $d^{(nn^\prime)}$ between FES minima as $d^{(nn^\prime)} = \sqrt{\lambda^{(nn^\prime)}}$. These distances provide the necessary input for constructing the diabatic FESs.

\begin{figure}[t]
   \includegraphics[scale=0.55]{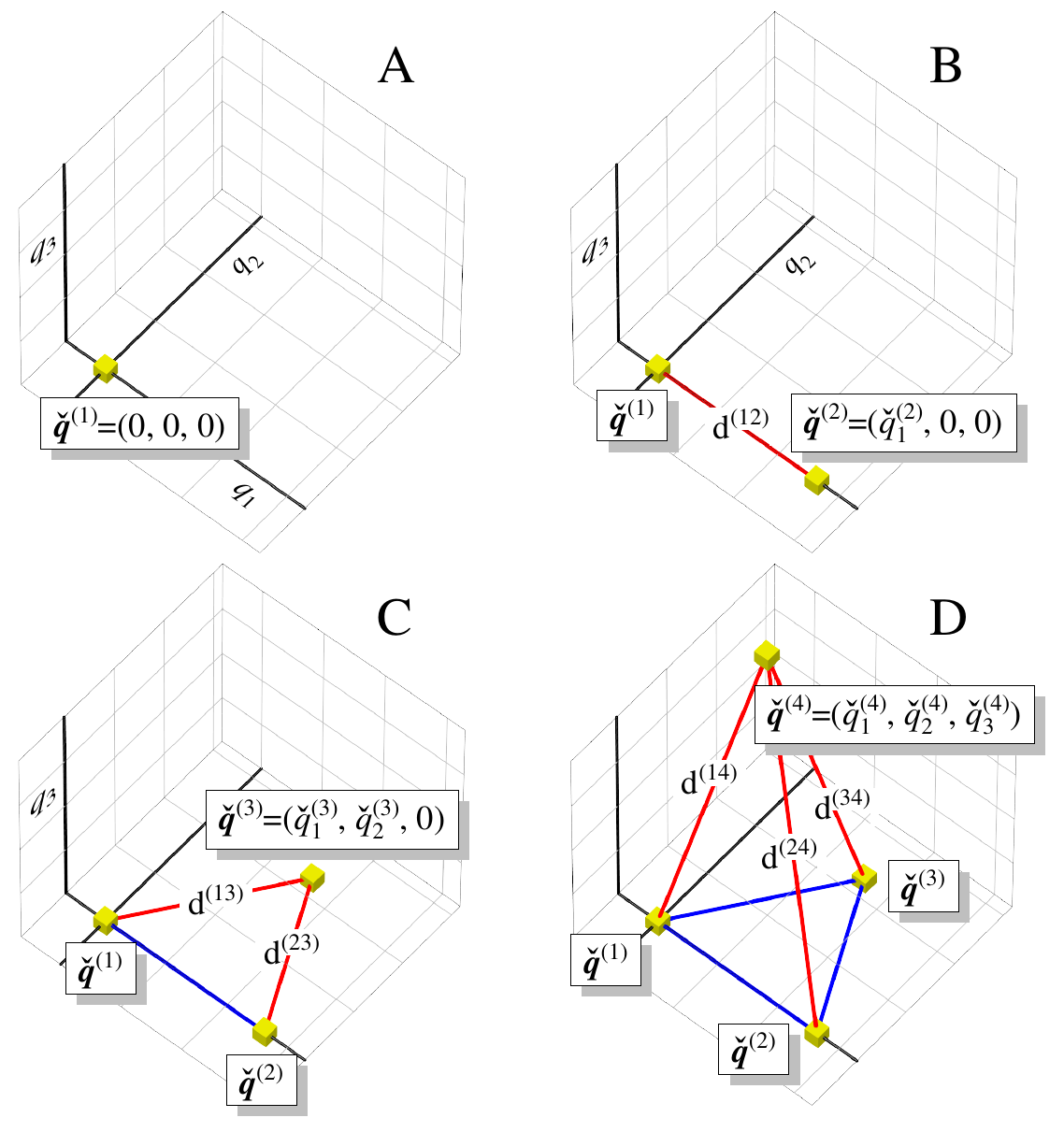}
   \caption{Step-by-step construction of the FES minima $\check{\bm{q}}^{(n)}$ for a macromolecular system with $N=4$ redox centers. Because the dimensionality of the polarization coordinate space is $N-1$, the minima are placed in a three-dimensional space $(q_1,q_2,q_3)$. Panels (A–D) illustrate the sequential placement of the $\check{\bm{q}}^{(n)}$ points. Their spatial arrangement is determined by the inter-state distances $d^{(nn^\prime)}$, which are directly related to the corresponding reorganization free energies $\lambda^{(nn^\prime)}$.}
   \label{fig:algorithm}
\end{figure}

From a geometrical standpoint, constructing the diabatic free energy surfaces $G^{(n)}$ reduces to the problem of embedding $N$ points in an $(N-1)$-dimensional configuration space, such that the pairwise distances correspond to the known $\lambda^{(nn^\prime)}$. When the matrix of reorganization energies, $\hat{\lambda}$, is non-degenerate, this embedding problem has a well-defined solution: a unique set of $N$ distinct points $\check{\bm{q}}^{(n)}$ in $\mathbb{R}^{N-1}$, determined up to isometric (distance-preserving) transformations. These transformations include global translations, rotations, and reflections of the coordinate frame. Importantly, this ambiguity does not influence the ET model, as isometric transformations leave the ET energetics unchanged and therefore have no impact on observable quantities.

The spatial configuration of the FES minima $\check{\bm{q}}^{(n)}$ can be constructed using the iterative embedding algorithm proposed in Ref.~\cite{feskov_jcp_18}. This procedure incrementally builds the multidimensional coordinate system by sequentially introducing each point $\check{\bm{q}}^{(n)}$ and increasing the dimensionality of the $\bm{q}$-space as necessary. Figure~\ref{fig:algorithm} illustrates this algorithm for a representative case involving four redox centers embedded in a three-dimensional space.

The construction begins with the placement of the first FES minimum, $G^{(1)}$, at the origin: $\check{\bm{q}}^{(1)} = \left(0, 0, 0\right)$ (Fig.~\ref{fig:algorithm}A). The second minimum, $\check{\bm{q}}^{(2)}$, is positioned along the $q_1$ axis at a distance $d^{(12)} = \sqrt{\lambda^{(12)}}$, yielding $\check{\bm{q}}^{(2)} = \left(d^{(12)}, 0, 0\right)$ (Fig.~\ref{fig:algorithm}B). The third point, $\check{\bm{q}}^{(3)}$, is placed in the $(q_1, q_2)$ plane such that the distances to the previous points match $d^{(13)} = \sqrt{\lambda^{(13)}}$ and $d^{(23)} = \sqrt{\lambda^{(23)}}$ (Fig.~\ref{fig:algorithm}C). Finally, the fourth point, $\check{\bm{q}}^{(4)}$, is located in full three-dimensional space, consistent with its distances to the first three points: $d^{(14)}$, $d^{(24)}$, and $d^{(34)}$ (Fig.~\ref{fig:algorithm}D).

In low-dimensional cases, the construction of diabatic FESs using the proposed algorithm can be carried out analytically. An explicit analytical solution for the $N = 3$ case is presented in Section 2. For systems with larger numbers of redox centers ($N > 3$), the positions of the FES minima $\check{\bm{q}}^{(n)}$ must typically be obtained numerically. A widely used approach is classical multidimensional scaling (MDS), which embeds a set of points in a Euclidean space such that the pairwise distances match a given dissimilarity matrix — in this context, derived from the reorganization free energies $\lambda^{(nn^\prime)}$. MDS minimizes a stress function that quantifies the discrepancy between the target and realized distances, yielding an optimal configuration in the least-squares sense \cite{borg_book_05}.

\subsection{Relaxation Components and Extended Coordinate Space}

The system's motion along the $q_k$ coordinates is primarily governed by the dynamic properties of the environment, particularly its response to charge redistribution during ET. Key contributors to this response include the electronic and dipolar components of medium polarization, as well as large-scale conformational intramolecular reorganization, such as shifts in the positions of redox centers within the macromolecule. This chapter focuses on ultrafast photochemical CS processes occurring on timescales of up to several tens of picoseconds. At such timescales, large-scale conformational modes can be considered effectively frozen and are therefore excluded from the present model.

The dynamic response of the medium is incorporated into the model through the energy-gap autocorrelation function,  
\[
\Gamma(t) = \frac{\langle \Delta E(0)\Delta E(t)\rangle}{\langle \Delta E(0)^2 \rangle},
\]
where $\Delta E(t)$ denotes the instantaneous energy gap between the ground state with neutral reactants and the charge-separated excited state, and $\langle \dots \rangle$ indicates ensemble averaging. The function $\Gamma(t)$ is an experimentally accessible quantity and is commonly represented as a sum of several exponentials \cite{jimenez_nat_94, maroncelli_jpc_93}:
\begin{equation} \label{Gamma(t)} 
  \Gamma(t) = \sum_{i=1}^{L} \Gamma_i(t) = \sum_{i=1}^{L} \gamma_i e^{-t/\tau_i}.
\end{equation}
Here, $\Gamma_i(t)$ represents the $i$-th relaxation component of the medium, with $\gamma_i$ and $\tau_i$ denoting its weight and relaxation timescale. The weights satisfy the normalization condition $\sum \gamma_i = 1$. Polar solvents typically exhibit two to three relaxation components, with $\tau_i$ values often varying by an order of magnitude. In mixtures and structured environments, the range of $\tau_i$ values can be even wider. The early-time behavior of $\Gamma(t)$ is often studied using specialized experimental approaches that combine ultrafast spectroscopic techniques with computational methods \cite{nazarov_jml_22}.

The exponential form of $\Gamma_i(t)$ in Eq.~\eqref{Gamma(t)} suggests an inertialess diffusive motion of the system along the energy-gap coordinate $\Delta E$. Consequently, the sum of $L$ exponentials corresponds to $L$ distinct energy-diffusion relaxation processes in the environment, each characterized by a diffusion coefficient $D_i = 2k_\mathrm{B}T/\tau_i$ \cite{zusman_cp_88, bagchi_acp_99, feskov_jppc_16}.

Owing to the linearity of the medium response, the decomposition of $\Gamma(t)$ in Eq.~\eqref{Gamma(t)} applies not only to the energy-gap coordinate $\Delta E$, but also to each polarization coordinate of the system. Following the approach of Ref.~\cite{feskov_rjpcb_24}, we generalize the single-coordinate representation of $q_k$ by introducing the $L$-dimensional vector
\begin{equation}\label{q-split}
    \bm{y}_k = \bm{x} q_k = \left( x_1 q_k,\, x_2 q_k,\, \dots,\, x_L q_k \right)^\mathrm{T},
\end{equation}
where $\bm{x} \equiv (x_1, x_2, \dots, x_L)^\mathrm{T}$ is defined as
\begin{equation*}
    \bm{x} = \left(\sqrt{\gamma_1}, \sqrt{\gamma_2}, \dots, \sqrt{\gamma_L}\right)^\mathrm{T}.
\end{equation*}
The vector $\bm{x}$ is normalized to unity, and its components are weighted by the relaxation amplitudes $\gamma_i$ of the medium.  

\begin{figure}[t]
\sidecaption[t]
   \includegraphics[scale=0.32]{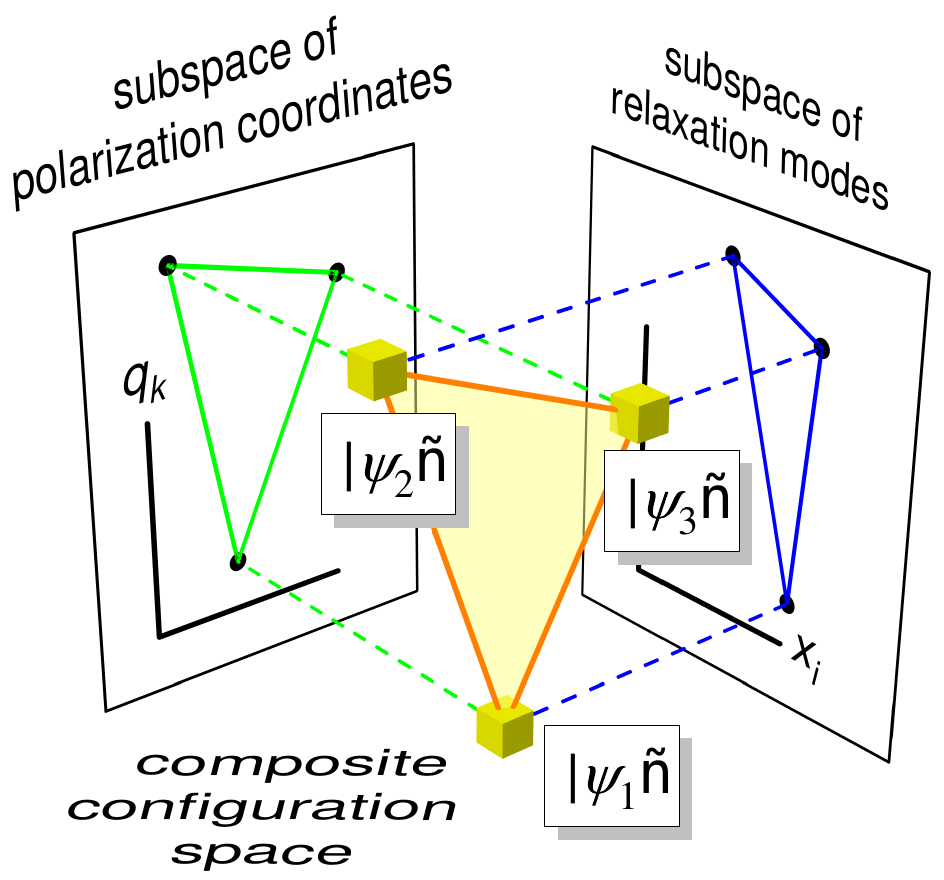}
   \caption{Visualization of the coordinate space architecture used in the extended ET framework. Shown are the polarization coordinate subspace ($\bm{q}$), the relaxation component subspace ($\bm{x}$), and the composite configuration space ($\bm{y}$). Yellow objects indicate equilibrium positions $\bm{\check{y}}^{(n)}$ corresponding to different diabatic states $\vert \psi_n \rangle$.}
   \label{fig:subspaces}
\end{figure}

The transformation in Eq.~\eqref{q-split} maps the polarization coordinates $q_k$ into an extended configuration space $\bm{y}$, constructed as the outer product of the $\bm{x}$ and $\bm{q}$ subspaces (see Fig.~\ref{fig:subspaces}):
\begin{equation} \label{direct_product}
   \bm{y} = \mathrm{vec}\!\left( \bm{x} \otimes \bm{q} \right) 
   = \mathrm{vec} \!\left[
      \begin{array}{cccc}
          x_1 q_1 & \dots & x_1 q_K \\
          \vdots & \ddots & \vdots \\
          x_L q_1 & \dots & x_L q_K
      \end{array}
   \right],
\end{equation}
where $\otimes$ denotes the outer product and $\mathrm{vec}(\cdot)$ is the vectorization operation that reshapes a matrix into a column vector. The resulting vector $\bm{y}$ has dimension
\begin{equation} \label{M}
    M = L K = L (N-1).
\end{equation}
Physically, this construction embeds the system into an enlarged configuration space that combines the $K$ structural polarization coordinates with the $L$ characteristic relaxation modes of the medium, thereby capturing both spatial and temporal aspects of solvent response.

Eq.~\eqref{direct_product} provides a higher-dimensional representation of the system, facilitating the analysis of multi-component relaxation dynamics. As follows from this expression, the $(ik)$-th element of the extended configuration space corresponds to the $i$-th relaxation component of the $q_k$ polarization coordinate. This explicit separation of relaxation components allows for a detailed description of the medium response and provides a framework for modeling complex relaxation processes \cite{feskov_ijms_22}.

Since $\bm{x}$ is a unit vector, $\vert\bm{x}\vert = 1$, it follows directly from Eq.~\eqref{q-split} that $\vert \bm{y} \vert = \vert \bm{q} \vert$, indicating that the coordinate-splitting transformation preserves vector lengths. Consequently, the diabatic free energy surfaces $G^{(n)}$ in the $\bm{y}$-presentation take the form
\begin{equation}\label{Gn(y)}
  G^{(n)}(\bm{y}) = \check{G}^{(n)} + \sum_{k=1}^{K} \sum_{i=1}^{L} \left(y_{ik} - \check{y}_{ik}^{(n)}\right)^2 = \check{G}^{(n)} + \left\vert \bm{y} - \bm{\check{y}}^{(n)} \right\vert^2,
\end{equation}
where $\bm{\check{y}}^{(n)}$ defines the minimum of the FES and is given by 
\begin{equation} \label{y_n}
   \bm{\check{y}}^{(n)} = \mathrm{vec}\left( \bm{x} \otimes \bm{\check{q}}^{(n)} \right) = \mathrm{vec} \left[
      \begin{array}{cccc}
          x_1 \check{q}_1^{(n)} & \dots & x_1 \check{q}_K^{(n)} \\
          \vdots & \ddots & \vdots \\
          x_L \check{q}_1^{(n)} & \dots & x_L \check{q}_K^{(n)}
      \end{array}
   \right].
\end{equation}
By employing Eqs.~\eqref{lambda_def} and \eqref{Gn(y)}, one derives the following expression for the reorganization free energy:
\begin{equation}\label{lambda(y)}
  \lambda^{(nn^\prime)} = \sum_{i,k} \left(\check{y}_{ik}^{(n^\prime)} - \check{y}_{ik}^{(n)}\right)^2 = \left\vert \bm{D}^{(nn^\prime)}\right\vert^2,
\end{equation}
where $\bm{D}^{(nn^\prime)} \equiv \bm{\check{y}}^{(n^\prime)} - \bm{\check{y}}^{(n)}$. This formulation mirrors the structure of Eq.~\eqref{lambda(q)}, but is applicable to the higher-dimensional representation. 

Together with the expansion of the coordinate space defined in Eq.~\eqref{q-split}, we introduce its inverse operation — a projection from the composite space $\bm{y}$ onto the subspaces $\bm{q}$ and $\bm{x}$. Projection onto the $\bm{q}$ subspace is performed by the operator $\hat{P}_q$
\begin{equation} \label{projection_q_def}
    \hat{P}_q \, \bm{y} = \vert \bm{x} \vert \bm{q} = \bm{q}.
\end{equation}
Similarly, projection onto the $\bm{x}$ subspace is defined by the operator $\hat{P}_x$
\begin{equation} \label{projection_x_def}
    \hat{P}_x \, \bm{y} = \vert \bm{q} \vert\, \bm{x}.
\end{equation}
Applying these operators to the displacement vector $\bm{D}^{(nn^\prime)}$ gives
\begin{equation} \label{Dnn_projections}
    \hat{P}_q \, \bm{D}^{(nn^\prime)} = \bm{d}^{(nn^\prime)}, 
    \qquad
    \hat{P}_x \, \bm{D}^{(nn^\prime)} = \sqrt{\lambda^{(nn^\prime)}} \bm{x}.
\end{equation}
These relations establish a direct correspondence between the system parameters — namely, the solvent reorganization energies and the relaxation amplitudes of the medium — and the spatial arrangement of the diabatic FESs in different configuration spaces.

\subsection{ET Energy-Gap Coordinates and Equations of Motion }

\begin{figure}[t]
\sidecaption[t]
   \includegraphics[scale=0.28]{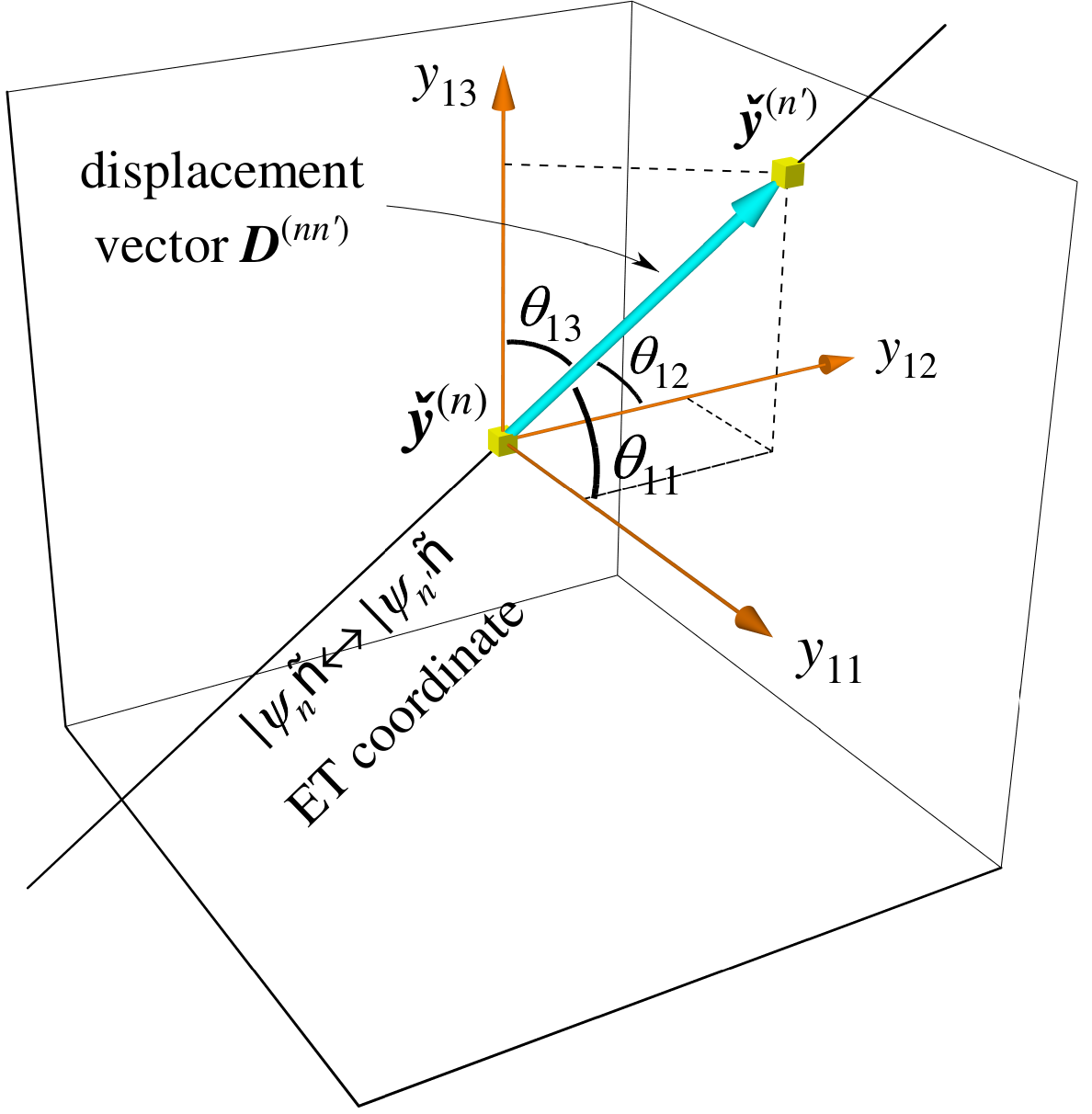}
   \caption{Illustration of the displacement vector $\bm{D}^{(nn^\prime)}$ connecting the equilibrium positions $\check{\bm{y}}^{(n)}$ and $\check{\bm{y}}^{(n^\prime)}$ of two diabatic FESs in the extended configuration space $\bm{y}$. The components of $\bm{D}^{(nn^\prime)}$ along the coordinate axes define the directional cosines $\cos\theta_{ik}^{(nn^\prime)}$, which are used in the construction of the energy-gap reaction coordinate $z^{(nn^\prime)}$.}
   \label{fig:D_cosines}
\end{figure}

Nonadiabatic ET typically occurs when the free energies of the reactant and product states are equal, a condition realized in the vicinity of the intersection region of the corresponding diabatic FESs. The $\vert \psi_n \rangle \to \vert \psi_{n^\prime} \rangle$ transitions can thus be described in terms of the energy gap between the two diabatic states, $\Delta G^{(nn^\prime)}(\bm{y}) \equiv G^{(n^\prime)}(\bm{y}) - G^{(n)}(\bm{y})$. Substituting Eq.~\eqref{Gn(y)}, this gap can be expressed as
\begin{equation*}
    \Delta G^{(nn^\prime)}(\bm{y}) = \Delta \check{G}^{(nn^\prime)} + \lambda^{(nn^\prime)} - 2 \sum\limits_{k,i} \left(y_{ik} - \check{y}_{ik}^{(n)} \right) \left( \check{y}_{ik}^{(n^\prime)} - \check{y}_{ik}^{(n)} \right),
\end{equation*}
where $\Delta \check{G}^{(nn^\prime)} = \check{G}^{(n^\prime)} - \check{G}^{(n)}$ is the driving force of ET. Employing the definition $\bm{D}^{(nn^\prime)} \equiv \bm{\check{y}}^{(n^\prime)} - \bm{\check{y}}^{(n)}$, the expression simplifies to 
\begin{equation*}
    \Delta G^{(nn^\prime)}(\bm{y}) = \Delta \check{G}^{(nn^\prime)} + \lambda^{(nn^\prime)} - 2 \left( \bm{y} - \bm{\check{y}}^{(n)}\right) \cdot \bm{D}^{(nn^\prime)}.
\end{equation*}
The final term here represents a scalar product and can be reformulated as 
\begin{equation} \label{DeltaG_nn(y)}
    \Delta G^{(nn^\prime)}(\bm{y}) = \Delta \check{G}^{(nn^\prime)} + \lambda^{(nn^\prime)} - 2 \sqrt{\lambda^{(nn^\prime)}} z^{(nn^\prime)},
\end{equation}
where $z^{(nn^\prime)}$ is calculated as 
\begin{equation} \label{z_nn_def}
    z^{(nn^\prime)} \equiv \sum\limits_{i,k} \left( y_{ik} - \check{y}_{ik}^{(n)} \right) \cos{\theta_{ik}^{(nn^\prime)}},
\end{equation}
and $\cos{\theta_{ik}^{(nn^\prime)}}$ are the directional cosines of the displacement vector $\bm{D}^{(nn^\prime)}$
\begin{equation} \label{cos_theta_ik}
    \cos{\theta_{ik}^{(nn^\prime)}} = \frac{\check{y}_{ik}^{(n^\prime)} - \check{y}_{ik}^{(n)}}{\sqrt{\lambda^{(nn^\prime)}}}.
\end{equation}
As follows from Eq.~\eqref{DeltaG_nn(y)}, the quantity $z^{(nn^\prime)}$ represents the energy-gap reaction coordinate for the $\vert \psi_n \rangle \to \vert \psi_{n^\prime} \rangle$ ET transition, and Eq.~\eqref{z_nn_def} provides a straightforward computational recipe for evaluating this coordinate in terms of the directional cosines. This geometric construction is illustrated in Fig.~\ref{fig:D_cosines}, where the displacement vector $\bm{D}^{(nn^\prime)}$ is resolved into directional cosines that determine the energy-gap reaction coordinate $z^{(nn^\prime)}$.

We now apply this result to evaluate the intrinsic rates of ET transitions in the macromolecule. In the weak electronic coupling limit, the position-dependent rate $K_\mathrm{cs}^{(nn^\prime)}(\bm{y})$ is given by Fermi’s Golden Rule
\begin{equation} \label{K_nn}
    K_\mathrm{cs}^{(nn^\prime)} = \frac{2\pi}{\hbar} \vert V_\mathrm{cs}^{(nn^\prime)} \vert^2 \, 
    \delta\!\left( G^{(n^\prime)} - G^{(n)}\right) 
    = \frac{\pi\vert V_\mathrm{cs}^{(nn^\prime)} \vert^2}{\hbar\sqrt{\lambda^{(nn^\prime)}}} \, 
    \delta\!\left( z^{(nn^\prime)} - \tilde{z}^{(nn^\prime)} \right),
\end{equation}
where $\delta(z)$ is the Dirac delta function, defined as a generalized function that vanishes everywhere except at $z=0$ and satisfies 
\[
\int_{-\infty}^{\infty} f(z)\,\delta(z)\,dz = f(0)
\]
for any smooth test function $f(z)$. In this context, the delta function enforces the energy-conservation condition at the crossing of diabatic FESs. The quantity 
\begin{equation} \label{tilde_z_nn}
    \tilde{z}^{(nn^\prime)} = \frac{\Delta \check{G}^{(nn^\prime)} + \lambda^{(nn^\prime)}}{2\sqrt{\lambda^{(nn^\prime)}}} 
\end{equation}
is the value of the reaction coordinate $z^{(nn^\prime)}$ at the intersection point of the two diabatic surfaces. Analogous expressions can be written for $K_\mathrm{cr}^{(n0)}$ by replacing $V_\mathrm{cs}^{(nn^\prime)}$ with $V_\mathrm{cr}^{(n0)}$ in Eq.~\eqref{K_nn}.

We now turn to the dynamic description of photochemical processes in macromolecular systems. The kinetics of multistage ET are formulated in terms of time-dependent probability density functions $\varrho_n(\bm{y},t)$, which describe the distribution of the system over nuclear configurations $\bm{y}$ within a given diabatic electronic state $\vert \psi_n \rangle$. The complete state of the system is thus represented by an $(N+1)$-dimensional vector
\begin{equation*}
 \bm{\varrho} = (\varrho_0, \varrho_1, \dots, \varrho_N)^\mathrm{T},   
\end{equation*}
whose temporal evolution is governed by the following kinetic master equation
\begin{equation} \label{eq_of_motion}
    \frac{\partial \bm{\varrho}(\bm{y},t)}{\partial t} = \left( \hat{W} + \hat{T} + \hat{L} \right) \bm{\varrho}(\bm{y},t). 
\end{equation}
In this formulation, $\hat{W}$ denotes the operator describing nonadiabatic electronic transitions, encompassing both charge separation and charge recombination processes, with transition rates $K_\mathrm{cs}^{(nn^\prime)}$ and $K_\mathrm{cr}^{(n0)}$ specified by Fermi’s Golden Rule. The operator $\hat{T}$ accounts for irreversible deactivation of the photoexcited state, such as internal conversion or radiative relaxation. The operator $\hat{L}$ represents Smoluchowski diffusion (i.e., the overdamped limit of the Fokker–Planck equation) and describes relaxation of the nuclear degrees of freedom through interactions with the thermal environment.

The ET operator $\hat{W}$ is written as a summation over all allowed transition channels
\begin{equation} \label{W_operator}
   \hat{W} = \sum\limits_{n > 1} K_\mathrm{cr}^{(n0)} \hat{M}^{(n0)} + \sum\limits_{n \geq 1} \sum\limits_{n^\prime > n} K_\mathrm{cs}^{(nn^\prime)} \hat{M}^{(nn^\prime)},
\end{equation}
where $\hat{M}^{(nn^\prime)}$ is a sparse transition matrix. The matrix $\hat{M}^{(nn^\prime)}$ contains nonzero elements at four positions: the diagonal components $(n,n)$ and $(n^\prime,n^\prime)$ are set to $-1$, while the off-diagonal components $(n,n^\prime)$ and $(n^\prime,n)$ are assigned the value $+1$, thereby ensuring proper accounting of population depletion and accumulation between diabatic states.

The deactivation operator $\hat{T}$ describes the irreversible decay of the excited electronic state $\vert \psi_1 \rangle$ to the ground state $\vert \psi_0 \rangle$ with a first-order rate constant $k_\mathrm{d}$
\begin{equation} \label{T_operator}
  [ \hat{T} ]_{n,n^\prime} = k_\mathrm{d} \, \delta_{n,0} \, \delta_{n^\prime,1} - k_\mathrm{d} \, \delta_{n,1} \, \delta_{n^\prime,1},
\end{equation}
where $\delta_{i,j}$ denotes the Kronecker delta.

The nuclear relaxation operator $\hat{L}$ governs the diffusive dynamics in the extended nuclear configuration space and incorporates the multi-component nature of the environmental response. Given independence of the nuclear coordinates $y_{ik}$, the total operator $\hat{L}$ can be expressed as a sum over one-dimensional Smoluchowski operators
\begin{equation}\label{L_operator}
[ \hat{L} ]_{n,n^\prime} = \delta_{n,n^\prime} \sum\limits_{i,k} \hat{L}_{ik}^{(n)}, \quad
\hat{L}_{ik}^{(n)} = \frac{1}{\tau_{i}}
\left[ 1 + \left(y_{ik} - \check{y}_{ik}^{(n)}\right) \frac{\partial}{\partial y_{ik}} + k_\mathrm{B}T \frac{\partial^2}{\partial y_{ik}^2} \right],
\end{equation}
where $\tau_i$ is the relaxation time associated with the $i$-th mode, $k_\mathrm{B}$ is the Boltzmann constant, and $T$ denotes the temperature.

Equations~\eqref{eq_of_motion}–\eqref{L_operator} constitute a comprehensive kinetic framework for modeling ultrafast multistage electron transfer dynamics. This formulation captures the essential features of nonadiabatic electronic transitions, nuclear relaxation, and irreversible deactivation of excited states. The initial condition for the probability density vector $\bm{\varrho}(\bm{y},t)$ is determined by the physical conditions under which photoexcitation occurs. In particular, when excitation is induced by a femtosecond laser pulse, the initial nonequilibrium distribution reflects not only intrinsic molecular parameters and excitation frequency, but also the spectral width of the excitation pulse \cite{nicolet_jpca_05, fedunov_jcp_04, feskov_jpca_08, barykov_jpcc_15}.

Experimentally relevant observables include the time-dependent populations of diabatic states, defined by the configuration-space integrals of the corresponding probability densities
\begin{equation} \label{P_n(t)}
   P_n(t) = \int \varrho_n(\bm{y},t) , d\bm{y}.
\end{equation}

In the subsequent section, we address numerical techniques for solving the kinetic equation~\eqref{eq_of_motion}, with emphasis on computational approaches applicable to multistage ET cascades in complex molecular environments.

\section{Numerical Method: Brownian Simulations with Surface Hopping} \label{sec:algorithm}

From a computational perspective, a distinctive feature of the kinetic equation~\eqref{eq_of_motion} is the presence of multiple sink and source terms associated with nonadiabatic transitions. These terms are proportional to position-dependent transition rates, $K_\mathrm{cs}^{(nn^\prime)}(\bm{y})$ and $K_\mathrm{cr}^{(n0)}(\bm{y})$, whose sharply localized, delta-function-like structure follows directly from Eq.~\eqref{K_nn}. This singular behavior poses a major challenge for traditional numerical approaches, such as finite-difference schemes and grid-based Monte Carlo methods, which typically require high spatial resolution to achieve sufficient accuracy.

This computational difficulty becomes even more pronounced when intramolecular quantum vibrational modes are included in the model. In these scenarios, the number of coupled electron–vibrational states increases substantially, leading to transition regions that are no longer confined to narrow domains in configuration space. As a result, the sink and source terms become distributed, further diminishing the efficiency of conventional discretization-based methods for solving the kinetic equations \cite{siplivy_jcp_20, feskov_jpca_06, feskov_jpca_11, nazarov_jpcb_16}.

To overcome the numerical challenges posed by delta-localized transition regions, we employ a trajectory-based simulation method that avoids the need for explicit spatial discretization of the coupling terms and has proven effective for solving the kinetic equations~\eqref{eq_of_motion}. The computational cost of this algorithm scales linearly with the dimensionality of the nuclear configuration space, i.e., with the number of independent coordinates $y_{ik}$, making it particularly well-suited for high-dimensional systems. This approach has previously been applied to simulate a variety of ultrafast ET processes, including intervalence charge transfer in mixed-valence metal complexes \cite{feskov_cpl_08}, photoinduced charge separation in zinc–porphyrin/imide dyads and triads \cite{feskov_jpca_13}, and fluorescence quenching via electron transfer in polar solvents \cite{feskov_jpca_09, feskov_jcp_19}.

In the algorithm, the nuclear dynamics are treated as overdamped Brownian motion on diabatic free energy surfaces, representing stochastic fluctuations of the polarization coordinates driven by thermal interactions with the environment. Surface hopping is superimposed on this Brownian propagation: when a trajectory enters a region of near-degeneracy between two FESs, stochastic transition rules determine whether the system hops to the neighboring electronic state. Thus, Brownian motion accounts for the continuous, thermally induced evolution of nuclear coordinates, while surface hopping captures the discrete and probabilistic character of nonadiabatic electronic transitions.

The general Brownian surface-hopping algorithm proceeds according to the following sequence

\begin{enumerate}
  \item Initialize an ensemble of particles representing the initial probability density $\varrho_0(\bm{y})$ on the excited-state free energy surface $G^{(1)}$.
  \item For each particle residing in electronic state $\vert \psi_n \rangle$, propagate a stochastic Brownian trajectory on the corresponding diabatic surface $G^{(n)}$, performing the following operations at each time step:
  \begin{enumerate}
     \item Detect whether the particle enters a transition region between $G^{(n)}$ and a neighboring surface $G^{(n^\prime)}$ by evaluating the energy-gap coordinate $z^{(nn^\prime)}$.
     \item For each potential transition, compute the probability of nonadiabatic transition and determine stochastically whether a surface hop occurs.
     \item If a hop occurs, update the electronic state to $\vert \psi_{n^\prime} \rangle$ and continue trajectory propagation on $G^{(n^\prime)}$.
     \item If the particle remains in the photoexcited state $\vert \psi_1 \rangle$, evaluate the probability of irreversible deactivation within the time step. If deactivation occurs, transfer the particle to the ground-state surface $G^{(0)}$ and continue propagation accordingly.
  \end{enumerate}
  \item Compute the time-dependent populations $P_n(t)$ of each electronic state by averaging over the full ensemble of trajectories
  \begin{equation*}
        P_n(t) = \left\langle \chi_n(t) \right\rangle,
  \end{equation*}
  where $\chi_n(t)$ is a binary indicator function that equals 1 if the particle is in state $\vert \psi_n \rangle$ at time $t$, and 0 otherwise.
\end{enumerate}

In the following subsections, we describe the core components of this algorithm in detail, with particular attention to the stochastic integration of Brownian dynamics and the simulation of surface-hopping events, including both nonadiabatic charge transfer and irreversible excited-state decay.

\subsection{Stochastic Propagation on Diabatic Free Energy Surfaces}

Non-reactive Brownian trajectories on a given diabatic FES $G^{(n)}$ are numerically generated using the Green’s function formalism, which evaluates the propagator corresponding to the Smoluchowski operator $\hat{L}_{ik}^{(n)}$. Due to the orthogonality of the generalized coordinates $y_{ik}$, the multidimensional diffusion process decomposes into a set of independent one-dimensional stochastic processes. For clarity, we suppress the indices $n$, $i$, and $k$, and consider a single representative coordinate $y$. The Green’s function $F(y,t \,|\, y_0)$ is defined as the solution to the equation
\begin{equation}\label{F_equation}
    \left(\frac{\partial}{\partial t} - \hat{L} \right) F(y,t \,|\, y_0) = \delta(t) \delta(y - y_0),
\end{equation}
where $\hat{L}$ is the one-dimensional Smoluchowski operator associated with the harmonic potential $G(y) = (y - \check{y})^2$.

The corresponding Green’s function has a closed-form Gaussian solution
\begin{equation}\label{F_solution}
    F(y, t \,|\, y_0) = \frac{1}{\sqrt{2\pi \sigma^2(t)}} \exp\left[ - \frac{(y - \bar{y}(t))^2}{2\sigma^2(t)} \right],
\end{equation}
with mean and variance given by
\begin{equation}
    \bar{y}(t) = \check{y} - (y_0 - \check{y}) e^{-t/\tau}, \qquad
    \sigma^2(t) = k_\mathrm{B}T \left(1 - e^{-2t/\tau} \right),
\end{equation}
where $\tau$ denotes the relaxation time associated with the coordinate $y$.

As evident from Eq.~\eqref{F_solution}, the propagator remains Gaussian at all times, with the mean relaxing exponentially toward the potential minimum $\check{y}$ and the variance growing from zero to its thermal equilibrium value. In the long-time limit ($t \to \infty$), the distribution approaches the stationary Boltzmann distribution corresponding to the harmonic potential.

Given the analytical form of the Green’s function, non-reactive Brownian trajectories can be generated using the following update rule
\begin{equation}\label{brownian_algorithm}
    y(t + \Delta t) = \check{y} + \left(y(t) - \check{y}\right) e^{-\Delta t/\tau} + N \sqrt{k_\mathrm{B}T \left(1 - e^{-2\Delta t/\tau}\right)},
\end{equation}
where $\Delta t$ is the time step, and $N$ is a normally distributed random variable with zero mean and unit variance, i.e., $\langle N \rangle = 0$ and $\langle N^2 \rangle = 1$. This update rule is exact with respect to the time step $\Delta t$, as it is derived directly from the closed-form solution of the Smoluchowski equation and does not rely on finite-difference approximations.

It is important to note that Eq.~\eqref{brownian_algorithm} does not generate a continuous physical trajectory of the particle on the $G(y)$ FES. In the mathematical sense, Brownian motion is nowhere differentiable, and thus the notion of a smooth trajectory is ill-defined. Rather, the update rule provides a stochastic sequence of positions at discrete time points $t_m = t_0 + m\Delta t$, which accurately reflect the statistical properties of the underlying diffusion process. The particle’s path between consecutive time steps remains undefined, representing the inherent stochasticity of Brownian dynamics at short timescales.

As a consequence of this discretization, certain dynamical quantities cannot be directly obtained from the simulation. These include: (1) the exact time $t^*$ at which a particle crosses the intersection between two diabatic FESs, (2) the instantaneous velocity $v_y$ at the point of crossing, and (3) the number of repeated crossings that may occur within a single time step $\Delta t$. Nevertheless, the algorithm does allow for accurate estimation of the total residence time spent within an intersection region, and therefore yields statistically reliable rates for charge separation and recombination processes, in agreement with Fermi’s Golden Rule.

A key component of the algorithm is the probabilistic evaluation of surface-hopping transitions, which is addressed in detail in the following subsection.

\subsection{Modeling Surface Hops at FESs Intersections}

Within the nonadiabatic framework, electronic transitions between diabatic states are permitted only when the system's nuclear configuration enters a region where two diabatic FESs become degenerate, $G^{(n)} = G^{(n^\prime)}$. Accurately detecting such intersection events is therefore essential for simulating surface hops.

The intersection region between surfaces $G^{(n)}$ and $G^{(n^\prime)}$ corresponds to a hyperplane defined by the condition $z^{(nn^\prime)} = \tilde{z}^{(nn^\prime)}$, where $z^{(nn^\prime)}$ is the scalar projection of the configuration vector $\bm{y}$ onto the displacement vector $\bm{D}^{(nn^\prime)}$ connecting the minima of the two surfaces. This hyperplane is oriented perpendicular to $\bm{D}^{(nn^\prime)}$ in the extended configuration space. To identify potential transitions, the algorithm monitors the sign of the function $z^{(nn^\prime)} - \tilde{z}^{(nn^\prime)}$ along the Brownian trajectory. A change in sign between two successive time steps indicates that the trajectory has crossed the corresponding intersection hyperplane, and thus entered the region where a nonadiabatic electronic transition may occur.

Let us consider a Brownian trajectory that intersects the degeneracy region between $G^{(n)}$ and $G^{(n^\prime)}$ within the time interval $[t, t + \Delta t]$. The probability for the particle to remain in the original diabatic state $\vert \psi_n \rangle$ throughout this interval is given by the survival function
\begin{align}
    S^* &= \exp\left\{ - \int_{t}^{t+\Delta t} K^{(nn^\prime)}\left( \bm{y}(t) \right) dt \right\} \nonumber \\
    &= \exp\left\{ - \frac{\pi \vert V^{(nn^\prime)} \vert^2}{\hbar \sqrt{\lambda^{(nn^\prime)}}} \int_{t^* - 0}^{t^* + 0} \delta\left( z^{(nn^\prime)}(t) - \tilde{z}^{(nn^\prime)} \right) dt \right\}, \label{S*_definition}
\end{align}
where $t^*$ denotes the (unknown) time at which the trajectory intersects the hyperplane, $\bm{y}(t)$ is the particle’s position in configuration space at time $t$, and $z^{(nn^\prime)}(t)$ is its projection onto the corresponding ET reaction coordinate.

This formulation is consistent with Fermi’s Golden Rule for transitions between diabatic states. The presence of the Dirac delta function in Eq.~\eqref{S*_definition} reflects the assumption that electronic transitions occur only at the exact crossing point between the diabatic surfaces, where their energies coincide. In practical simulations, the evaluation of this expression is replaced by a discrete probabilistic criterion for surface hopping.

The integral in Eq.~\eqref{S*_definition} can be evaluated analytically, yielding an explicit expression for the survival probability of the particle in its initial diabatic state \cite{feskov_jcp_18}
\begin{equation} \label{S*_solution}
    S^* = \exp\left\{ -\frac{\pi \vert V^{(nn^\prime)} \vert^2} {\hbar \vert \dot{z}^{(nn^\prime)} \vert \sqrt{\lambda^{(nn^\prime)}}} \right\},
\end{equation}
where $\dot{z}^{(nn^\prime)}$ denotes the average velocity of the particle along the corresponding energy-gap coordinate, computed over the time interval $[t, t + \Delta t]$ as
\begin{equation} \label{z-velocity}
    \dot{z}^{(nn^\prime)} = \frac{z^{(nn^\prime)}(t+\Delta t) - z^{(nn^\prime)}(t)}{\Delta t}.
\end{equation}

Equation~\eqref{S*_solution} forms the basis for implementing electronic transitions within the surface-hopping algorithm. A transition event is simulated by drawing a uniform random number $\xi \in [0,1)$ and comparing it to the calculated survival probability $S^*$. If $\xi > S^*$, the particle undergoes a nonadiabatic transition from $\vert \psi_n \rangle$ to $\vert \psi_{n^\prime} \rangle$. Otherwise, the particle remains on the current FES $G^{(n)}$. This stochastic decision-making procedure ensures consistency with Fermi’s Golden Rule and accurately captures the probabilistic nature of electronic transitions at surface intersections.

\begin{figure}[ht]
   \includegraphics[width=0.99\textwidth]{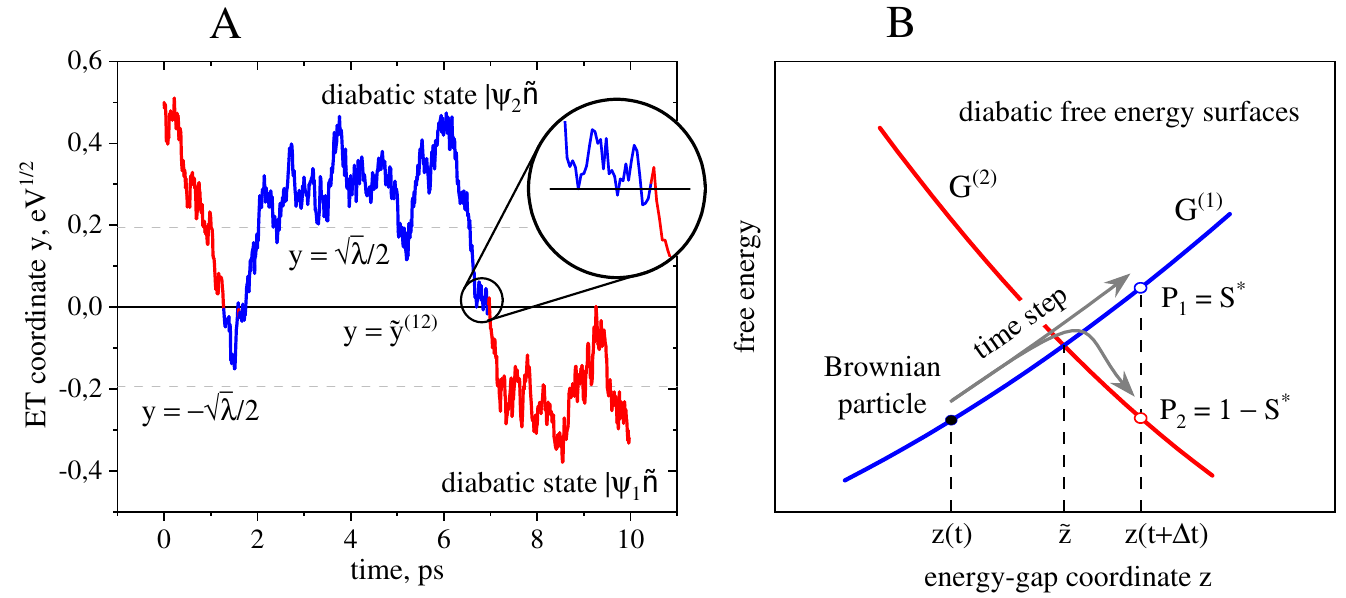}
   \caption{(A) Simulated Brownian trajectory in a one-dimensional model with two intersecting diabatic FESs, $G_1(y)$ and $G_2(y)$. Dashed lines mark the FESs minima, the solid line at $y = 0$ indicates the intersection point. Surface hopping events are indicated by color changes: red corresponds to state $\vert \psi_1 \rangle$, blue to $\vert \psi_2 \rangle$. (B) Schematic illustration of the decision process for probabilistic surface hops during Brownian motion.}
   \label{fig:trajectory}
\end{figure}

Figure~\ref{fig:trajectory} illustrates the surface hopping algorithm for a simple one-dimensional model consisting of two symmetric diabatic FESs, defined by $G_1(y) = (y + \sqrt{\lambda}/2)^2$ and $G_2(y) = (y - \sqrt{\lambda}/2)^2$. These surfaces intersect at $\tilde{y}^{(12)} = 0$. Fig.~\ref{fig:trajectory}A shows a Brownian trajectory calculated using Eq.~\eqref{brownian_algorithm} with a time step $\Delta t = 0.01$~ps. The inset highlights that only a part of the intersection events result in electronic transitions. Fig.~\ref{fig:trajectory}B provides a schematic representation of the surface hopping mechanism. Possible realizations of the Brownian trajectory near the intersection region are depicted, along with the associated survival and transition probabilities computed over a single time step $\Delta t$. The parameters used in the simulation are: $\lambda = 0.15$~eV, $k_\mathrm{B}T = 0.025$~eV, $V = 0.006$~eV, $\tau = 1$~ps.

Unlike the update rule for nonreactive trajectory Eq.~\eqref{brownian_algorithm}, the surface hopping algorithm based on Eq.~\eqref{S*_solution} exhibits sensitivity to the time step $\Delta t$. Due to well-known characteristics of Brownian motion, the mean velocity $\dot{z}^{(nn^\prime)}$, calculated using Eq.~\eqref{z-velocity}, depends on $\Delta t$ and even diverges as $\Delta t \to 0$. This dependence influences the single-crossing survival probability $S^*$ and, consequently, the frequency of reaction events. Nevertheless, ensemble-averaged simulations reveal that the overall ET kinetics remain invariant with respect to the choice of $\Delta t$, provided the following condition is met throughout the simulation \cite{feskov_jcp_18}
\begin{equation}\label{validity}
    \langle T^* \rangle \ll 1,
\end{equation}
where $T^* \equiv 1 - S^*$ represents the probability of a transition during a single crossing event. This criterion ensures that the system operates in the weak-coupling regime and can be satisfied by selecting a sufficiently small time step $\Delta t$. 

Assuming small time steps $\Delta t \ll \tau$, one can derive from Eqs.~\eqref{brownian_algorithm} and \eqref{z-velocity} the following estimate for the particle’s velocity along the energy-gap coordinate
\begin{equation*}\label{velocity_z}
  \dot{z}^{(nn^\prime)} = - \sum\limits_{i,k} \frac{y_{ik} - \check{y}_{ik}^{(n)}}{\tau_i}\cos{\theta_{ik}^{(nn^\prime)}} + N \sqrt{\frac{2k_\mathrm{B}T}{\Delta t}} \sum\limits_{i,k} \frac{\cos{\theta_{ik}^{(nn^\prime)}}}{\sqrt{\tau_i}}.
\end{equation*}
The first term in this expression describes a regular drift of the particle toward the FES minimum, while the second term represents stochastic fluctuations arising from thermal noise. At $\Delta t \ll \tau k_\mathrm{B}T/\lambda^{(nn^\prime)}$, the stochastic component dominates. In this regime, the mean absolute velocity $\langle \vert \dot{z}^{(nn^\prime)} \vert \rangle$ can be estimated as
\begin{equation}\label{mean_z-velocity1}
    \langle \vert \dot{z}^{(nn^\prime)} \vert \rangle \approx
    \frac{2}{\sqrt{\pi}} \sqrt{\frac{k_\mathrm{B}T}{\Delta t}} \zeta^{(nn^\prime)},
    \qquad  
    \zeta^{(nn^\prime)} \equiv \sum\limits_{i,k} \frac{\cos{\theta_{ik}^{(nn^\prime)}}}{\sqrt{\tau_i}}.    
\end{equation}
The quantity $\zeta^{(nn^\prime)}$ characterizes the effective diffusive mobility of a particle along the energy-gap coordinate $z^{(nn^\prime)}$ on short time scales. For convenience, this result can be recast in terms of an effective relaxation time
\begin{equation}\label{mean_z-velocity2}
    \langle \vert \dot{z}^{(nn^\prime)} \vert \rangle \approx
    \frac{2}{\sqrt{\pi}} \sqrt{\frac{k_\mathrm{B}T}{\tau_\mathrm{eff}^{(nn^\prime)} \Delta t}},    
\end{equation}
where the effective relaxation time $\tau_\mathrm{eff}^{(nn^\prime)}$ is defined as 
\begin{equation}\label{tau_eff}
    \tau_\mathrm{eff}^{(nn^\prime)} = \left( \zeta^{(nn^\prime)} \right)^{-2}. 
\end{equation}

By inserting Eq.~\eqref{mean_z-velocity2} into Eq.~\eqref{S*_solution}, one obtains an explicit relation between the time step $\Delta t$ and the average single-crossing transition probability $\langle T^* \rangle$. The condition $\langle T^* \rangle \ll 1$, required for the validity of the algorithm, is satisfied if the time step obeys the inequality 
\begin{equation}\label{dt_choice}
    \Delta t \ll \frac{4 \hbar^2 \lambda^{(nn^\prime)}k_\mathrm{B}T}{\pi^3 \tau_\mathrm{eff}^{(nn^\prime)} \left| V^{(nn^\prime)} \right|^4}.
\end{equation}
This expression highlights the strong dependence of the allowed time step on the electronic coupling $V^{(nn^\prime)}$. As a result, the algorithm is particularly efficient for simulating nonadiabatic (weak-coupling) ET processes, where $V^{(nn^\prime)}$ is small. 

It is important to note that Eq.~\eqref{dt_choice} is a local condition, meaning it only needs to be satisfied in the vicinity of the FESs intersection region where transitions occur. This locality enables the use of efficient adaptive time-stepping schemes, in which smaller time steps are applied near term-crossing regions, while larger steps can be used elsewhere to reduce computational cost \cite{feskov_jcp_18}.

Although the above approach is particularly useful in the weak-coupling regime, alternative methods exist for stronger electronic coupling, as discussed in the following subsection.

\subsection{Comparison with Reactive Green’s Function Method}

As an alternative to the surface hopping algorithm described above, a distinct numerical technique was developed in Refs.~\cite{feskov_cpl_07, feskov_cpl_08, feskov_ctc_18} for treating electron transfer in the strong-coupling (solvent-controlled) regime, where $V^{(nn^\prime)}$ is relatively large. This method is based on the evaluation of Green's functions for coupled diffusive equations that include delta-function sink-source terms to model ET transitions.

In contrast to the nonadiabatic surface hopping algorithm, the trajectories generated by the reactive Green’s function method are inherently reactive — electronic transitions are incorporated directly into the propagation of each Brownian path. This intrinsic reactivity makes the method especially efficient for simulating solvent-controlled ET, as confirmed by prior applications \cite{feskov_ctc_18}. However, in the weak-coupling regime, the approach becomes less computationally efficient due to the increased numerical effort required to resolve rare transition events. Table~\ref{table:comparison} summarizes the key characteristics of both simulation strategies, highlighting their respective strengths and domains of applicability.

\begin{table*}[ht]
  \begin{tabularx}{\textwidth}{ 
   >{\raggedright\arraybackslash}X 
   >{\raggedright\arraybackslash}X 
   >{\raggedright\arraybackslash}X }
   \hline
   \textbf{Feature} & \textbf{Surface Hopping Algorithm} & \textbf{Reactive Green’s Function Method} \\ [15pt]
   \hline 
   \textbf{ET regime applicability} 
   & Best suited for nonadiabatic (weak coupling) regime
   & Best suited for diffusion-controlled (strong coupling) regime \\ [25pt]
   \textbf{Trajectory type} 
   & Nonreactive; surface hops occur probabilistically at the FESs intersection regions 
   & Reactive; transitions are incorporated within the trajectory \\ [25pt]
   \textbf{Time step sensitivity} 
   & High sensitivity; requires small $\Delta t$ to maintain accuracy 
   & Less sensitive; transitions handled within Green’s function \\ [25pt]
   \textbf{Treatment of FESs crossings} 
   & Requires detection and evaluation of survival probability 
   & Handled analytically through delta-localized terms \\ [25pt]
   \textbf{Computational cost per step} 
   & Low (simple updates and random number generation) 
   & Higher (requires evaluation of the inverse Laplace transforms) \\ [5pt]
   \hline
  \end{tabularx}
  \caption{Comparison of the two numerical schemes for simulating ET dynamics in the diabatic representation: the nonadiabatic surface hopping algorithm and the reactive Green’s function method from Ref.~\cite{feskov_ctc_18}.}
  \label{table:comparison}
\end{table*}

We now apply the general theoretical and computational framework to a representative three-center molecular system. Specifically, we consider D–A$_1$–A$_2$ triads, which serve as minimal models for investigating competitive ultrafast ET processes in multiredox architectures.

\section{Competitive ET in Donor–Acceptor(1)–Acceptor(2) Compounds: General Formulation of the Model} \label{sec:3centers}

Building on the theoretical framework developed in Section~\ref{sec:theory}, we now consider a model molecular triad of the form DA$_1$A$_2$, where D is a photoactive electron donor, and A$_1$ and A$_2$ are electron acceptors. This DA$_1$A$_2$ motif serves as a minimal yet representative architecture for a broad class of molecular assemblies employed in optoelectronics, photocatalysis, and solar energy conversion, where the efficiency of photoinduced charge separation is essential \cite{scattergood_dt_14, bottari_ccr_21, machin2023, wang_as_24, souza_cc_09, guldi_csr_02, wallin_jpca_10, wrobel_ccr_11, kirner_cs_15, phelan_jacs_19, loong_jpcb_22, brown_jpca_24, wang_jpcc_23}. Moreover, natural photosynthetic reaction centers in plants and bacteria embody analogous multicenter systems that exploit ultrafast ET to achieve highly efficient conversion of solar energy into long-lived charge-separated states \cite{cherepanov2022, nguyen_sci_23}.

\begin{figure}[ht]
\sidecaption[t]
   \includegraphics[width=0.55\textwidth]{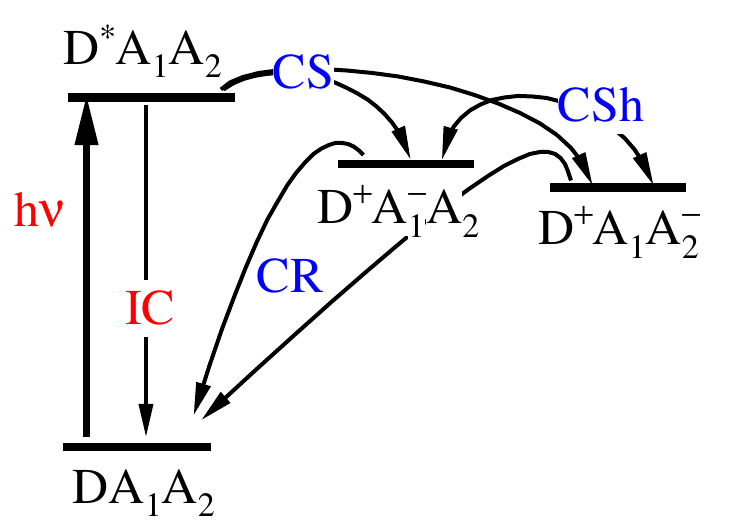}
   \caption{Photochemical processes in the DA$_1$A$_2$ compound. CS, CR, CSh and IC denote charge separation, charge recombination, charge shift and internal conversion, respectively.}
   \label{fig:triad}
\end{figure}

Photoexcitation of the DA$_1$A$_2$ triad initiates a sequence of nonadiabatic electronic transitions, as depicted schematically in Fig.~\ref{fig:triad}. Some of these transitions may take place under nonequilibrium conditions, during the system's evolution on diabatic FESs prior to full thermal relaxation. To account for these effects, we employ a general model incorporating a multicomponent relaxation function $\Gamma(t)$, as introduced in Eq.~\eqref{Gamma(t)}. A key advantage of this three-center/multicomponent formulation is that the corresponding diabatic FESs can be derived in closed analytical form. This property is important for validation of the theoretical framework developed in this work.

We assume that photoexcitation does not produce a substantial redistribution of electron density within the triad, thereby justifying the neglect of environmental reorganization at the excitation stage. Accordingly, the minima of the ground- and excited-state FESs are taken to coincide. The system is described in a diabatic basis of states with fixed electronic configurations
\begin{equation*}
  | \varphi_0 \rangle = | \mathrm{D} \mathrm{A}_1\mathrm{A}_2 \rangle, \quad
  | \varphi_1 \rangle = | \mathrm{D}^\ast \mathrm{A}_1\mathrm{A}_2 \rangle, \quad
  | \varphi_2 \rangle = | \mathrm{D}^+\mathrm{A}_1^-\mathrm{A}_2 \rangle, \quad
  | \varphi_3 \rangle = | \mathrm{D}^+\mathrm{A}_1\mathrm{A}_2^- \rangle.
\end{equation*}
The equilibrium free energies $\check{G}^{(n)}$ associated with diabatic states $|\varphi_n\rangle$, as well as the corresponding reorganization free energies $\lambda^{(nn^\prime)}$, are treated as known parameters of the model.

For a three-center molecular system ($N = 3$) embedded in an $L$-mode environment, the extended configuration space $\bm{y}$ contains $(N - 1)L = 2L$ independent nuclear coordinates (see Eq.~\eqref{M}). It is convenient to represent the vector $\bm{y}$ in matrix form as
\begin{equation}\label{y_matrix}
   \bm{y} = \bm{x}^{\mathrm{T}} \otimes
   \begin{pmatrix}
      q_{1} \\
      q_{2}
   \end{pmatrix} =
   \begin{pmatrix}
      y_{11} & y_{21} & \dots & y_{L1} \\
      y_{12} & y_{22} & \dots & y_{L2}
   \end{pmatrix},
\end{equation}
where $(q_1, q_2)$ are the polarization coordinates that span the two-dimensional configuration space, and $\bm{x}^{\mathrm{T}} = (x_1, x_2, \dots, x_L)$ contains the corresponding relaxation amplitudes. The coefficients $x_i = \sqrt{\gamma_i}$ are normalized such that $\sum x_i^2 = 1$.

To compute the coordinates of the FES minima $\check{\bm{q}}^{(n)}$ (in polarization space) and $\check{\bm{y}}^{(n)}$ (in the full configuration space), we apply the general algorithm described in Section~\ref{sec:theory}. Since arrangement of the $\check{\bm{q}}^{(n)}$ points depends solely on the relative displacement vectors $\bm{d}^{(nn^\prime)}$ (see Eq.~\eqref{lambda(q)}), the absolute positions of the FES minima in the $\bm{q}$-space are defined only up to isometric transformations, i.e., global translations and rotations. This invariance allows us to fix one of the FES minima arbitrarily. For convenience, we place the minimum of the excited state at the origin
\begin{equation}\label{q1_vector}
    \check{\bm{q}}^{(1)} =
    \begin{pmatrix}
        0 \\
        0
    \end{pmatrix}.
\end{equation}
This choice simplifies the construction of the remaining minima.

The corresponding minimum in the extended space $\bm{y}$ is then obtained by applying the outer product structure
\begin{equation}\label{y1_vector}
    \check{\bm{y}}^{(1)} = \bm{x}^{\mathrm{T}} \otimes \check{\bm{q}}^{(1)} =
    \begin{pmatrix}
        0 & 0 & \dots & 0 \\
        0 & 0 & \dots & 0
    \end{pmatrix}.
\end{equation}

To determine $\check{\bm{y}}^{(2)}$, we use the known reorganization energy $\lambda^{(12)}$ associated with the $|\varphi_1\rangle \to |\varphi_2\rangle$ transition. According to Eq.~\eqref{lambda(q)}, the displacement between $\check{\bm{q}}^{(1)}$ and $\check{\bm{q}}^{(2)}$ is $\sqrt{\lambda^{(12)}}$. At this stage, it is sufficient to consider a displacement along a single polarization coordinate, such as $q_1$. This yields
\begin{equation}\label{q2_vector}
  \check{\bm{q}}^{(2)} =  
  \left(
    \begin{matrix} 
    \sqrt{\lambda^{(12)}} \\ 
    0 \\ 
    \end{matrix}
  \right),
\end{equation}
and the corresponding FES minimum in the composite space
\begin{equation}\label{y2_vector}
  \check{\bm{y}}^{(2)} = \bm{x}^{\mathrm{T}} \otimes \check{\bm{q}}^{(2)} = 
  \sqrt{\lambda^{(12)}}
    \left( 
       \begin{matrix} 
         x_1 & x_2 & \dots & x_L \\ 
         0 & 0 & \dots & 0 \\ 
        \end{matrix}
    \right).
\end{equation}

Similar geometric approach is used to evaluate $\check{\bm{y}}^{(3)}$. Given the reorganization energies $\lambda^{(13)}$ and $\lambda^{(23)}$, the distances from $\check{\bm{q}}^{(3)}$ to $\check{\bm{q}}^{(1)}$ and $\check{\bm{q}}^{(2)}$ are $d^{(13)} = \sqrt{\lambda^{(13)}}$ and $d^{(23)} = \sqrt{\lambda^{(23)}}$, respectively. The point $\check{\bm{q}}^{(3)}$ therefore lies in the plane spanned by the two polarization coordinates $q_1$ and $q_2$. The explicit form is given by
\begin{equation}\label{q3_vector}
  \check{\bm{q}}^{(3)} =  
  \left(
    \begin{matrix} 
    \sqrt{\lambda^{(13)}} \cos{\theta} \\ 
    \sqrt{\lambda^{(13)}} \sin{\theta} \\ 
    \end{matrix}
  \right),
\end{equation}
with the angle $\theta$ determined by the law of cosines
\begin{equation}\label{cos(theta)}
  \cos{\theta} = \frac{\lambda^{(12)} + \lambda^{(13)} - \lambda^{(23)}}{2\sqrt{\lambda^{(12)}\lambda^{(13)}}}.
\end{equation}
This result can be verified by direct evaluation of $d^{(nn^\prime)}$ from Eqs.~\eqref{q1_vector}--\eqref{q3_vector}, and comparing them to the expected values $\sqrt{\lambda^{(nn^\prime)}}$.

Using Eq.~\eqref{q3_vector} one obtains the coordinates of the $G^{(3)}$ FES minimum in $\bm{y}$ space
\begin{equation}\label{y3_vector}
  \check{\bm{y}}^{(3)} = \bm{x}^{\mathrm{T}} \otimes \check{\bm{q}}^{(3)} = 
  \sqrt{\lambda^{(13)}} \left( 
    \begin{matrix} 
      x_1 \cos{\theta}\ & x_2\cos{\theta}\ & \dots\ & x_L \cos{\theta}\ \\ x_1 \sin{\theta}\ & x_2 \sin{\theta}\ & \dots\ & x_L \sin{\theta}\ \\ 
    \end{matrix} \right).
\end{equation}

From Eqs.~\eqref{Gn(y)} and \eqref{y1_vector}--\eqref{y3_vector}, the diabatic FESs of the DA$_1$A$_2$ triad are expressed in terms of the $y_{ik}$ coordinates as follows
\begin{align} \label{triad_Gn(y)}
    G^{(0)}(\bm{y}) &= \sum_i \left( y_{i1}^2 + y_{i2}^2 \right) + \check{G}^{(0)}, \\ \nonumber
    G^{(1)}(\bm{y}) &= \sum_i \left( y_{i1}^2 + y_{i2}^2 \right) + \check{G}^{(1)}, \\ \nonumber
    G^{(2)}(\bm{y}) &= \sum_i \left[ \left( y_{i1} - x_i \sqrt{\lambda^{(12)}} \right)^2 + y_{i2}^2 \right] + \check{G}^{(2)}, \\ \nonumber
    G^{(3)}(\bm{y}) &= \sum_i \left[ \left( y_{i1} - x_i \cos{\theta} \sqrt{\lambda^{(13)}} \right)^2 + \left( y_{i2} - x_i \sin{\theta} \sqrt{\lambda^{(13)}} \right)^2 \right] + \check{G}^{(3)},
\end{align}
where $\check{G}^{(n)}$ denotes the equilibrium free energy of diabatic state $|\varphi_n \rangle$.

This set of FESs fully specifies the energetic landscape for all ET processes within the triad, including activation barriers for both charge separation and recombination transitions. As an illustrative example, consider electron transfer from the photoexcited donor D$^\ast$ to the primary acceptor A$_1$, corresponding to the transition $|\varphi_1 \rangle \rightarrow |\varphi_2 \rangle$. The relevant diabatic FESs, $G^{(1)}$ and $G^{(2)}$, intersect along a $(2L - 1)$-dimensional hyperplane, defined in terms of the energy-gap coordinate by
\begin{equation}\label{intersection_12}
    z^{(12)} = \sum\limits_{i=1}^L y_{i1} x_i = \frac{\lambda^{(12)} + \Delta \check{G}^{(21)}}{2\sqrt{\lambda^{(12)}}} \equiv \tilde{z}^{(12)}.
\end{equation}
This result gives the ET activation free energy
\begin{equation}\label{G_activation_12}
   G^\sharp = \frac{\left( \lambda^{(12)} + \Delta \check{G}^{(21)} \right)^2}{4\lambda^{(12)}},
\end{equation}
which exactly recovers the classical Marcus result for nonadiabatic electron transfer \cite{marcus_bba_85}.

Similar expressions can be derived for the activation barriers of other elementary transitions in the triad, confirming that the proposed model reproduces the correct equilibrium-limit behavior for all quasi-thermal ET reactions.

In the following section, we extend the analysis beyond the equilibrium framework to demonstrate that the model also captures nonequilibrium effects in ultrafast ET. Specifically, we examine two limiting cases of the DA$_1$A$_2$ model for which exact analytical solutions are available in the literature, providing benchmarks for validating the theoretical approach.

\subsection{Model Validation via Subspace Projection Analysis} \label{sec:validation}

The general theoretical framework for DA$_1$A$_2$ systems developed in the preceding subsection can be validated by comparison with two well-established models that represent limiting cases of our formulation. Specifically, we consider: (1) the Najbar–Tachiya model, which describes sequential ET in a three-center system coupled to a single-mode Debye solvent \cite{najbar_jpc_94}, and (2) the extended Sumi–Marcus model, which treats single-step ET in a two-center system embedded in a multicomponent environment \cite{sumi_jcp_86, bicout_jcp_97}.

These models differ in both system dimensionality and the structure of their configuration spaces. The Najbar–Tachiya model incorporates three redox centers ($N = 3$) but restricts the environmental dynamics to a single collective relaxation mode ($L = 1$). In contrast, the extended Sumi–Marcus model accommodates an arbitrary number of relaxation modes ($L$), while limiting the electronic subsystem to two redox centers ($N = 2$).

A central distinction between the two lies in the choice of coordinate representation: the Najbar–Tachiya model is formulated in polarization coordinates, whereas the extended Sumi–Marcus model adopts relaxation coordinates. These coordinate sets span different subspaces of the full multidimensional configuration space employed in our generalized approach, as schematically illustrated in Fig.~\ref{fig:subspaces}.  As our framework generalizes both limits, it should correctly reproduce their results under appropriate parameter constraints, consistent with the correspondence principle.

To establish agreement with the Najbar–Tachiya model, we impose uniform solvent relaxation times, $\tau_1 = \tau_2 = \dots = \tau_L = \tau$, thereby collapsing the relaxation dynamics to a single effective mode with a time scale $\tau$. We then project the diabatic FESs from the full $\bm{y}$-space onto the reduced polarization subspace $\bm{q}$, effectively eliminating the relaxation coordinates $\bm{x}$ by folding the $\bm{x}$-space. This yields the following coordinates for the FES minima
\begin{align}\label{triad_q_minima} \nonumber
    \check{\bm{q}}^{(0)} &= \check{\bm{q}}^{(1)} = \hat{P}_q\, \check{\bm{y}}^{(1)} = \left(\begin{matrix} 0 \\ 0 \end{matrix}\right), & & \text{($|$DA$_1$A$_2\rangle$ and $|$D$^*$A$_1$A$_2\rangle$ states)} \\
    \check{\bm{q}}^{(2)} &= \hat{P}_q\, \check{\bm{y}}^{(2)} = \left(\begin{matrix} \sqrt{\lambda^{(12)}} \\ 0 \end{matrix} \right), & & \text{($|$D$^+$A$_1^-$A$_2\rangle$ state)} \\
    \check{\bm{q}}^{(3)} &= \hat{P}_q\, \check{\bm{y}}^{(3)} = \left(\begin{matrix} \sqrt{\lambda^{(13)}} \cos{\theta} \\ \sqrt{\lambda^{(13)}} \sin{\theta} \end{matrix}\right), & & \text{($|$D$^+$A$_1$A$_2^-\rangle$ state)} \nonumber
\end{align}
where $\hat{P}_q$ denotes the projection operator defined by Eq.~\eqref{projection_q_def}.

The corresponding free energy surfaces in the polarization coordinate space $(q_1, q_2)$ are then given by
\begin{align} \nonumber
    G^{(0)}(q_1, q_2) &= q_1^2 + q_2^2 + \check{G}^{(0)}, \qquad
    G^{(1)}(q_1, q_2) = q_1^2 + q_2^2 + \check{G}^{(1)}, \\ \label{triad_Gn(q)}
    G^{(2)}(q_1, q_2) &= \left(q_1 - \sqrt{\lambda^{(12)}}\right)^2 + q_2^2 + \check{G}^{(2)}, \\
    G^{(3)}(q_1, q_2) &= \left(q_1 - \cos{\theta} \sqrt{\lambda^{(13)}}\right)^2 + \left(q_2 - \sin{\theta} \sqrt{\lambda^{(13)}}\right)^2 + \check{G}^{(3)}. \nonumber
\end{align}

The system’s time evolution in this representation is governed by a set of coupled Smoluchowski equations describing diffusion over the two-dimensional parabolic potentials $G^{(n)}(q_1, q_2)$. The diffusion coefficients along both polarization coordinates are equal, $D_1 = D_2 = k_BT/\tau$, reflecting the single-mode relaxation assumption. This formulation recovers the structure of the Najbar–Tachiya model for three-center electron transfer in a Debye solvent \cite{najbar_jpc_94, zusman_jcp_99, newton_ijc_04}. The two-dimensional description of diabatic FESs in terms of polarization coordinates, as expressed in Eqs.~\eqref{triad_Gn(q)}, has been widely employed in models that incorporate electron–vibrational interactions in DA$_1$A$_2$ systems (e.g., \cite{feskov_jppc_16, feskov_jpca_13, feskov_rjpca_16, feskov_cp_16}).

\begin{figure}[t]
  \sidecaption[t]
  \includegraphics[scale=0.32]{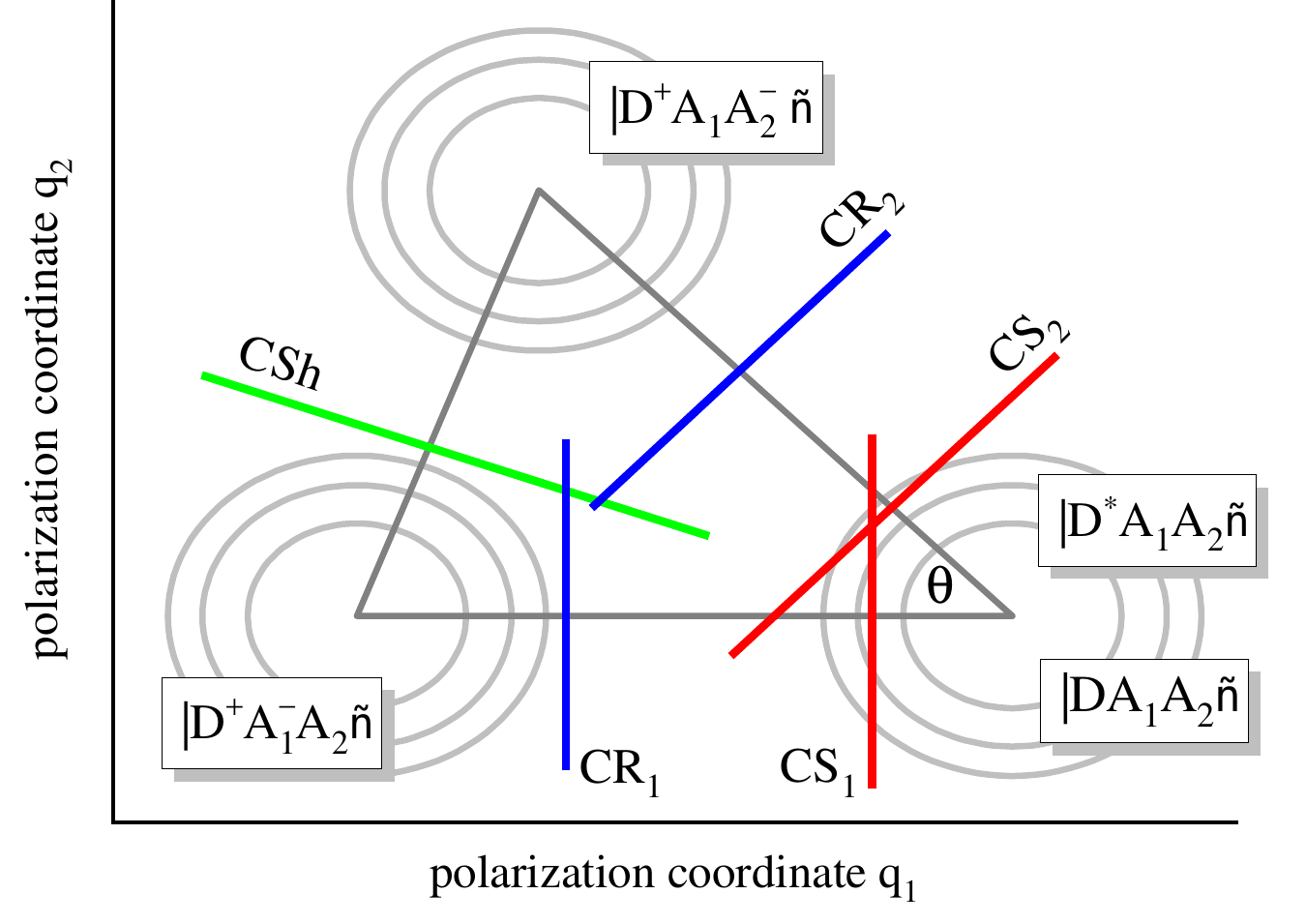}
   \caption{Diabatic FESs of the DA$_1$A$_2$ compound in the polarization coordinate space $\bm{q}$. Colored lines indicate electron transfer regions: red (CS), blue (CR), and green (CSh), corresponding to charge separation, recombination, and shift, respectively. The position and orientation of each intersection line relative to the FES minima are governed by the values of $\Delta \check{G}^{(nn^\prime)}$ and $\lambda^{(nn^\prime)}$ for the corresponding ET transition.}
\label{fig:triad_in_q-space}
\end{figure}

Figure~\ref{fig:triad_in_q-space} depicts the diabatic free energy surfaces projected into the $\bm{q}$-space, constructed using Eqs.~\eqref{triad_Gn(q)}. The parameters are set to: $\lambda^{(12)} = 0.9$ eV, $\lambda^{(13)} = 1.0$ eV, and $\lambda^{(23)} = 0.4$ eV, resulting in an inter-coordinate angle $\theta \approx 40^\circ$ between the energy-gap directions $z^{(12)}$ and $z^{(13)}$. Colored lines indicate regions of FES intersection corresponding to different ET channels: red denotes charge separation (CS), blue marks charge recombination (CR), and green represents the charge shift (CSh) between ionic configurations. These processes compete during the early stages following photoexcitation. Within this reduced, two-dimensional representation, the FES topology fully captures the energetics and competition between ultrafast ET pathways in the DA$_1$A$_2$ triad embedded in a single-mode polar environment.

We now examine the correspondence between the general DA$_1$A$_2$ framework and the extended Sumi–Marcus model by projecting the diabatic FESs from the full configuration space $\bm{y}$ onto the relaxation subspace $\bm{x}$. The resulting FESs take the form 
\begin{equation} \label{triad_Gn(X)_def}
    G^{(n)}(\bm{X}) = \sum_i \left( X_i - \check{X}_i^{(n)} \right)^2  + \check{G}^{(0)},
\end{equation}
where the new variables are defined as $X_i \equiv (\hat{P}_x \bm{y})_i = |\bm{q}| x_i$, and $\hat{P}_x$ denotes the projection operator onto the relaxation coordinate subspace. The equilibrium positions of the FES minima in this representation are given by 
\begin{equation} \label{triad_Xn_def}
    \check{X}_i^{(n)} \equiv \left( \hat{P}_x\, \check{\bm{y}}^{(n)} \right)_i = \vert \check{\bm{q}}^{(n)} \vert x_i,
\end{equation}
with $\vert \check{\bm{q}}^{(n)} \vert$ representing the Euclidean norm of the polarization coordinate vector associated with state $|\varphi_n \rangle$.

Applying this projection to the state vectors $\check{\bm{y}}^{(n)}$ derived previously (Eqs.~\eqref{y1_vector}, \eqref{y2_vector}, \eqref{y3_vector}) yields
\begin{equation*} 
  \check{\bm{X}}^{(0)} = \check{\bm{X}}^{(1)} = \bm{0}, \quad
  \check{\bm{X}}^{(2)} = \sqrt{\lambda^{(12)}} \bm{x}, \quad
  \check{\bm{X}}^{(3)} = \sqrt{\lambda^{(13)}} \bm{x}, 
\end{equation*}
where $\bm{0}$ is the $L$-dimensional zero vector. The corresponding free energy surfaces, expressed in terms of the $X_i$ coordinates, are
\begin{align} \label{triad_Gn(X)} \nonumber
    G^{(0)}(\bm{X}) &=\sum_i X_i^2 + \check{G}^{(0)}, \qquad 
    G^{(1)}(\bm{X}) =\sum_i X_i^2 + \check{G}^{(1)}, \\
    G^{(2)}(\bm{X}) &= \sum_i \left( X_i - \sqrt{\lambda^{(12)}} x_i \right)^2 + \check{G}^{(2)}, \\ \nonumber
    G^{(3)}(\bm{X}) &= \sum_i \left( X_i - \sqrt{\lambda^{(13)}} x_i \right)^2 + \check{G}^{(3)}.
\end{align}

We now demonstrate that the present framework reproduces the well-established multicomponent model of electron transfer for any pair of diabatic states. As a representative example, consider ET between the photoexcited and charge-separated configurations, $|\mathrm{D}^*\mathrm{A}_1\mathrm{A}_2\rangle$ and $|\mathrm{D}^+\mathrm{A}_1^-\mathrm{A}_2\rangle$, associated with the reorganization free energy $\lambda^{(12)}$. To facilitate comparison with standard models, we introduce a rescaled set of relaxation coordinates defined as $Q_i = X_i / 2\sqrt{\lambda^{(12)}}$. In terms of these variables, the diabatic FESs corresponding to the excited and charge-separated states are given by
\begin{equation} \label{triad_validation2}
    G^{(1)}(\bm{Q}) = \sum_i \frac{Q_i^2}{4\lambda^{(12)}_i} + \check{G}^{(1)}, \qquad 
    G^{(2)}(\bm{Q}) = \sum_i \frac{(Q_i - 2\lambda^{(12)}_i)^2}{4\lambda^{(12)}_i} + \check{G}^{(2)},
\end{equation}
where $\lambda^{(12)}_i = \lambda^{(12)} x_i^2$ denotes the contribution of the $i$-th environmental relaxation mode to the total reorganization energy. These expressions exactly reproduce the canonical form of the two-state, multicomponent ET model frequently employed in the theoretical treatment of single-step ET processes in non-Debye solvents (see, e.g., \cite{bagchi_acp_99, zusman_cp_88, feskov_jpcb_20}).

\begin{figure}[t]
  \sidecaption[t]
  \includegraphics[scale=0.32]{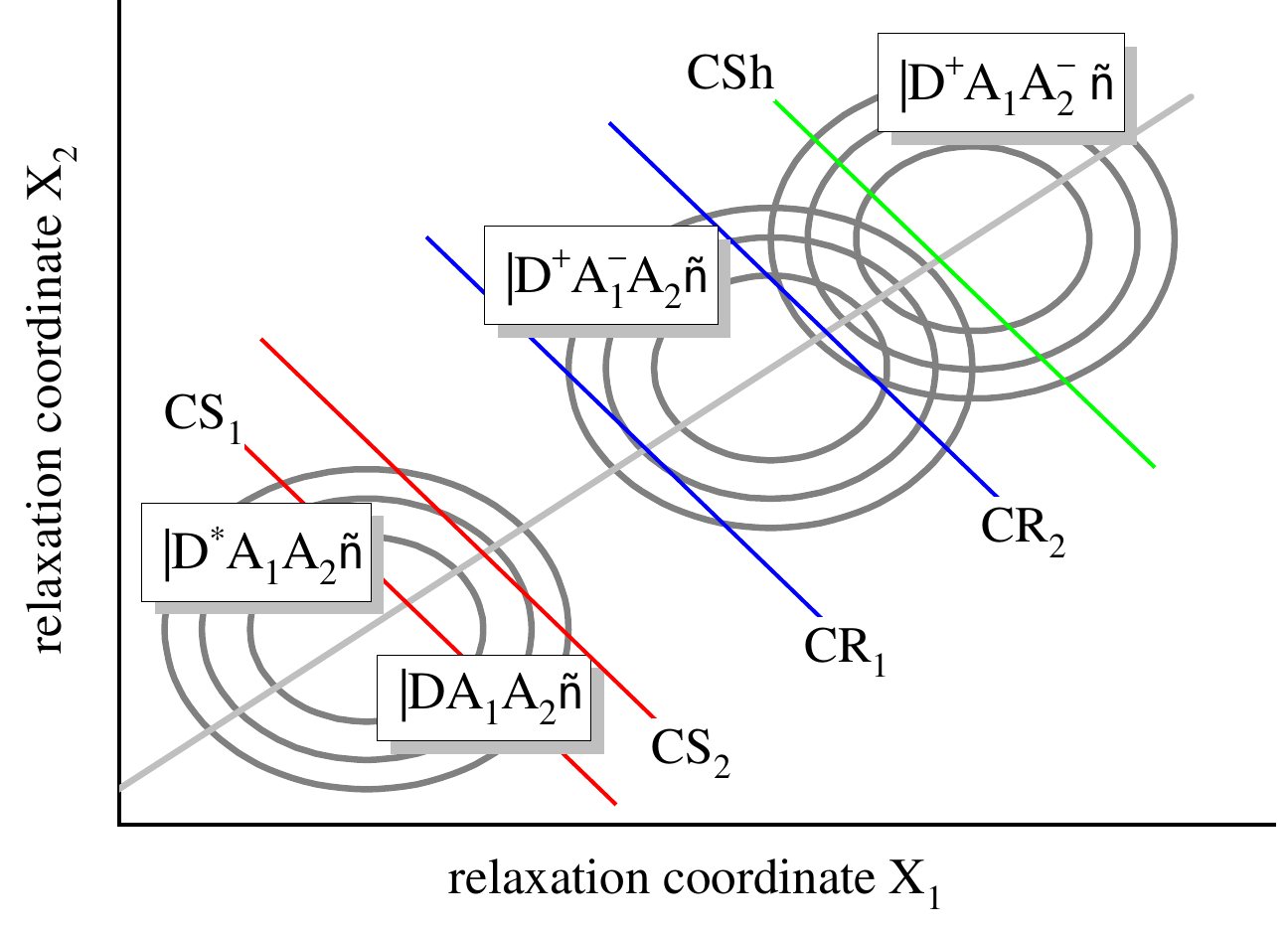}
   \caption{Diabatic FESs of the DA$_1$A$_2$ triad in the relaxation coordinate space $\bm{X}$, evaluated using Eqs.~\eqref{triad_Gn(X)}. The gray line, defined by $X_2 = \sqrt{\gamma_2/\gamma_1} X_1$, represents the direction of the energy-gap ET coordinate. Intersection lines for CS, CR, and CSh processes are highlighted in color.}
  \label{fig:triad_in_X-space}
\end{figure}

Figure~\ref{fig:triad_in_X-space} illustrates the arrangement of the diabatic FESs $G^{(n)}$ in the relaxation coordinate space $\bm{X}$, computed according to Eqs.~\eqref{triad_Gn(X)}. The model parameters used are: $\lambda^{(12)} = 0.4$~eV, $\lambda^{(13)} = 0.9$~eV, with relaxation mode weights $\gamma_1 = 0.6$, $\gamma_2 = 0.4$, which define the respective contributions of the two environmental modes to the total reorganization energies. As implied by Eq.~\eqref{triad_Xn_def}, the minima of the diabatic FESs in $\bm{X}$-space lie along a common line directed along the unit vector $\bm{x}$. Explicitly, these positions satisfy the linear relation $X_2 = \sqrt{\gamma_2/\gamma_1} X_1$, reflecting the anisotropic relaxation contributions encoded by $\gamma_1$ and $\gamma_2$.

The intersection hyperplane between any two diabatic surfaces $G^{(n)}(\bm{X})$ and $G^{(n^\prime)}(\bm{X})$ can be derived analytically by equating the corresponding expressions in Eq.~\eqref{triad_Gn(X)}. This yields the condition
\begin{equation} \label{triad_X_intersection}
    \sum_i x_i X_i = \bm{x} \cdot \bm{X} = - \frac{\lambda^{(nn^\prime)} + \Delta\check{G}^{(nn^\prime)}}{2\sqrt{\lambda^{(nn^\prime)}}},
\end{equation}
where $\Delta\check{G}^{(nn^\prime)} \equiv \check{G}^{(n^\prime)} - \check{G}^{(n)}$ is the equilibrium free energy gap between the diabatic states.  The resulting hyperplane, orthogonal to $\bm{x}$, defines the reaction zone for each ET process (see Fig.~\ref{fig:triad_in_X-space}).

Equation~\eqref{triad_X_intersection} provides the condition for resonance between two diabatic states: it specifies the set of points $\bm{X}$ in the relaxation coordinate space where the free energies of $| \psi_n \rangle$ and $| \psi_{n^\prime} \rangle$ are equal. Geometrically, this relation defines a hyperplane perpendicular to the vector $\bm{x}$, which encodes the relative contributions of the environmental relaxation modes. The position of the hyperplane is determined by two energetic quantities: (i) the reorganization free energy $\lambda^{(nn^\prime)}$, which sets the distance between the minima of the diabatic surfaces, and (ii) the thermodynamic driving force $\Delta\check{G}^{(nn^\prime)}$, which shifts the relative vertical alignment of the surfaces. Thus, Eq.~\eqref{triad_X_intersection} directly links the system’s energetic parameters to the spatial location of the ET reaction zone in $\bm{X}$-space.

In nonequilibrium ET, the system does not propagate from the minimum of the donor FES but instead starts from a displaced configuration, typically generated by direct photoexcitation or a preceding ET step. As a result, the system’s trajectory in $\bm{X}$-space may enter the reaction region before reaching thermal equilibrium — a process known as non-thermal or hot electronic transition \cite{feskov_jppc_16, fedunov_jcp_04, canton_nc_15, gladkikh_jcp_05}. In the following section, we explore how hot transitions in molecular triads can suppress hot charge recombination, thereby enhancing the efficiency of photoinduced charge separation in DA$_1$A$_2$ compounds.

\section{Efficiency of Charge Separation From the Second
Excited State in DA$_1$A$_2$ Compounds: Impact of Hot Electron Transfer} \label{sec:CS_from_S2}

A promising direction in the development of photovoltaic and optoelectronic materials involves designing macromolecular systems that facilitate efficient photochemical charge separation through cascaded ultrafast ET steps, leading to the formation of long-lived charge-separated states. A broad class of such systems has been explored, frequently employing porphyrins as photoactive electron donors in combination with fullerenes or imide-based moieties as electron acceptors \cite{rizzi_jpca_08, wallin_jpca_10, rego_jpcc_14, manna_jpcl_15, sun_jpcc_18, hou_jmcc_19, hu_jpcb_20}. These donor–acceptor assemblies emulate the architecture of natural photosynthetic reaction centers, where multistep ET enables spatial separation of charges and suppression of recombination, thereby stabilizing the CS state.

Recent experimental investigations have shown that molecular triads of the type A$_\mathrm{L}$–D–A$_\mathrm{R}$, in which the electron donor possesses two distinct locally excited states, can exhibit wavelength-dependent charge separation. In such systems, the direction of ET can be selectively modulated by tuning the excitation frequency. A representative example is the NDI–ZnP–NI triad, in which ZnP (zinc porphyrin) functions as the central photoactive donor, while NI (naphthaleneimide) and NDI (naphthalenediimide) serve as electron-accepting units on opposite sides of the porphyrin ring \cite{wallin_jpca_10}. Upon excitation to the lowest singlet excited state (S$_1$, Q-band), electron transfer predominantly proceeds toward the NDI acceptor (A$_\mathrm{L}$). Conversely, excitation to the higher-lying singlet state (S$_2$, Soret band) initiates ET to the NI acceptor (A$_\mathrm{R}$), enabling photoselective control over the CS direction.

\begin{figure}[ht]
  \includegraphics[width=0.95\textwidth]{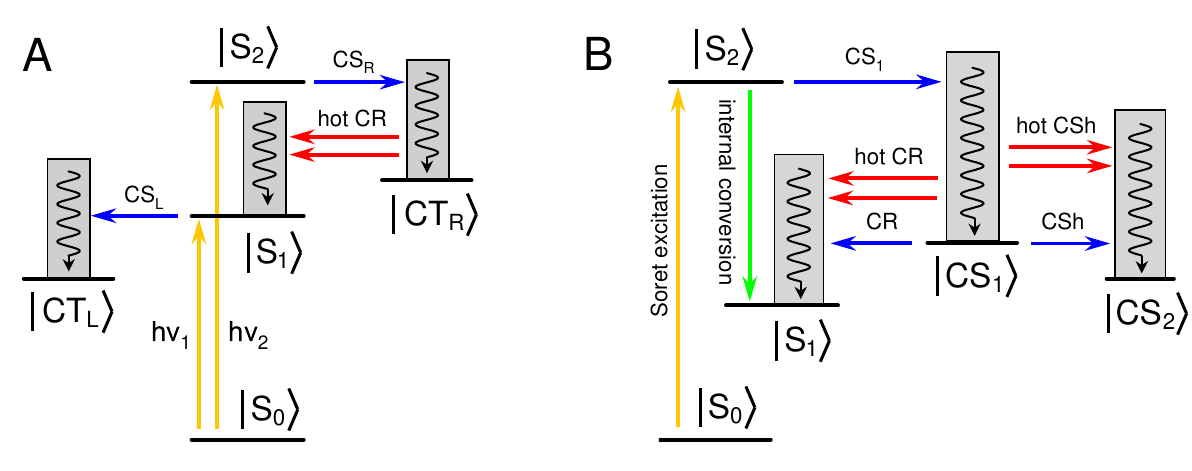}
   \caption{(A) Wavelength-selective photoinduced charge separation in the NDI–ZnP–NI triad, as reported in Ref.~\cite{wallin_jpca_10}. The two charge-separated states, $|\text{CT}_\mathrm{L} \rangle$ and $|\text{CT}_\mathrm{R} \rangle$, correspond to NDI$^-$–ZnP$^+$–NI and NDI–ZnP$^+$–NI$^-$ configurations, respectively. Red arrows denote nonequilibrium back electron transfer (hot charge recombination), which significantly reduces the CS efficiency in the right-hand ET branch. (B) Schematic representation of photochemical processes in DA$_1$A$_2$ compound, which models the right-hand ET branch of the NDI–ZnP–NI triad but includes an additional secondary acceptor covalently attached to the NI moiety. This design aims to suppress hot CR by enabling ultrafast charge shift (CSh) to A$_2$.}
  \label{fig:switch}
\end{figure}

Such wavelength-dependent ET behavior opens opportunities for optoelectronic applications, including optical molecular switches whose dipole orientation and CS character depend on the excitation frequency. However, the CS efficiency of ZnP-based molecular switches remains limited, primarily due to ultrafast charge recombination in the D$^+$A$_\mathrm{R}^-$ state generated from the S$_2$ excitation \cite{wallin_jpca_10}. The electronic structure of these systems promotes fast back electron transfer from the CS state $|\text{CT}_\mathrm{R} \rangle$ to the lower-lying neutral excited state $|\text{S}_1 \rangle$, driven by solvent and vibrational relaxation (see Fig.~\ref{fig:switch}A). This recombination process, which occurs before full thermal equilibration, constitutes hot electron transfer \cite{ivanov_cp_99}. Time-resolved spectroscopic studies reveal that only 10–20\% of the initially formed D$^+$A$_\mathrm{R}^-$ pairs avoid hot recombination within the first few picoseconds following excitation.

The issue of hot charge recombination in the right-hand ET branch of the NDI–ZnP–NI triad may be addressed by incorporating an additional electron acceptor covalently linked to the NI moiety. In this extended four-center configuration, A$_\mathrm{L}$–D–A$_\mathrm{R1}$–A$_\mathrm{R2}$, the initially formed charge-separated state D$^+$A$_\mathrm{R1}^-$ can undergo ultrafast subsequent electron transfer to A$_\mathrm{R2}$. This sequential process competes kinetically with charge recombination and effectively diverts the system away from back electron transfer to the lower-lying neutral state $|\text{S}_1 \rangle$. Such intramolecular competition offers a strategy for suppressing hot recombination and enhancing the overall yield of photoinduced charge separation.

In this section, we examine the potential of suppressing hot CR via ultrafast sequential ET in DA$_1$A$_2$ triads. We assume that photoinduced charge separation is initiated from the second locally excited state $|\text{S}_2 \rangle$, consistent with the mechanism illustrated in Fig.~\ref{fig:switch}B. To focus on the relevant dynamics, we restrict our analysis to the “right” ET branch of the system, neglecting the “left” branch on the premise that it exerts minimal influence on the overall photochemical behavior under the conditions considered. The primary objective of this analysis is to quantify the effect of the secondary acceptor A$_2$ on the CS efficiency in triads.

\subsection{Model Equations}

To describe the photochemical dynamics of the DA$_1$A$_2$ system following excitation to the Soret band, we employ the general theoretical framework presented in Section~\ref{sec:3centers}. For notational clarity, we define the relevant diabatic electronic states as follows
\begin{align} \nonumber
  &|\psi_0 \rangle = |\text{S}_1 \rangle = |\text{D}^{*}\text{A}_1\text{A}_2\rangle, \quad &&|\psi_1 \rangle = |\text{S}_2 \rangle = |\text{D}^{**}\text{A}_1\text{A}_2\rangle, \\
  &|\psi_2 \rangle = |\text{CS}_1\rangle = |\text{D}^{+} \text{A}_1^-\text{A}_2 \rangle, \quad &&|\psi_3 \rangle = |\text{CS}_2 \rangle = |\text{D}^{+} \text{A}_1 \text{A}_2^- \rangle.
\end{align}

The electronic ground state, $|\text{S}_0 \rangle = |\text{D} \text{A}_1 \text{A}_2 \rangle$, is excluded from subsequent analysis, as recombination to this state is negligible on the timescale considered. This simplification is justified by the substantial energy gap between the $|\text{CS}_1 \rangle$ and $|\text{S}_0 \rangle$ states in ZnP–imide systems, expressed as $\Delta G_\mathrm{cr0} = \check{G}_\mathrm{CS1} - \check{G}_\mathrm{S0} \ll - \lambda^{(12)}$, which renders this recombination pathway significantly slower than the picosecond-scale dynamics of interest.

In a single-component polar environment, ultrafast competitive CS and CR processes in DA$_1$A$_2$ triads are frequently modeled using a two-dimensional polarization coordinate space \cite{cho_jcp_95, ando_jpcb_98, najbar_jpc_94, tachiya_cp_96}. Since the present analysis focuses on the sequential two-step ET pathway $|\text{S}_2\rangle \rightarrow |\text{CS}_1\rangle \rightarrow |\text{CS}_2\rangle$, it is convenient to characterize the geometry of the diabatic FESs in terms of the angle between the displacement vectors $\bm{d}^{(21)}$ and $\bm{d}^{(23)}$ corresponding to the respective ET transitions. This angle, denoted $\theta^{(23)}$, is given by the geometric relation
\begin{equation}\label{ch5_theta^(23)}
    \cos{\theta^{(23)}} = \frac{\lambda^{(12)} + \lambda^{(23)} - \lambda^{(13)}}{2\sqrt{\lambda^{(12)}\lambda^{(23)}}},
\end{equation}
where $\lambda^{(12)}$, $\lambda^{(23)}$, and $\lambda^{(13)}$ are the reorganization free energies associated with ET between the D–A$_1$, A$_1$–A$_2$, and D–A$_2$ pairs, respectively.

Using the notation introduced in Section~\ref{sec:theory}, the diabatic FESs in the two-dimensional polarization coordinate space $(q_1, q_2)$ are expressed as
\begin{eqnarray} \label{ch5_FESs} \nonumber
    &&G_{\mathrm{S2}}^{(0)} = (q_1 - \sqrt{\lambda^{(12)}})^2 + q_2^2, \\ \nonumber
    &&G_{\mathrm{CS1}}^{(n)} = q_1^2 + q_2^2 + n\hbar\Omega_\mathrm{v} + \Delta G_\mathrm{CS1}, \\ \nonumber
    &&G_{\mathrm{CS2}}^{(m)} = (q_1 - \sqrt{\lambda^{(23)}}\cos \theta^{(23)})^2 + (q_2 - \sqrt{\lambda^{(23)}}\sin\theta^{(23)})^2 + m\hbar\Omega_\mathrm{v} + \Delta G_\mathrm{CS2}, \\
    &&G_{\mathrm{S1}}^{(l)} = (q_1 - \sqrt{\lambda^{(12)}})^2 + q_2^2 + l\hbar\Omega_\mathrm{v} + \Delta G_\mathrm{S1},
\end{eqnarray}
where $n$, $m$, and $l$ denote vibrational quantum numbers, and $\hbar\Omega_\mathrm{v}$ represents the characteristic vibrational energy. The free energy gaps $\Delta G_\mathrm{CS1}$, $\Delta G_\mathrm{CS2}$, and $\Delta G_\mathrm{S1}$ are defined relative to the reference state $|\text{S}_2 \rangle$ as
\begin{equation*}
    \Delta G_\mathrm{CS1} = \check{G}_\mathrm{CS1} - \check{G}_\mathrm{S2}, \quad
    \Delta G_\mathrm{CS2} = \check{G}_\mathrm{CS2} - \check{G}_\mathrm{S2}, \quad
    \Delta G_\mathrm{S1} = \check{G}_\mathrm{S1} - \check{G}_\mathrm{S2}.
\end{equation*}

In Eqs.~\eqref{ch5_FESs}, the influence of ET-active high-frequency intramolecular vibrations is incorporated by introducing vibrational sublevels for each electronic state. It is assumed that these vibrations possess the same frequency $\Omega_\mathrm{v}$ across all electronic states \cite{feskov_jpca_13}. The notation $G_{i}^{(n)} = G_{i}^{(n)}(q_1, q_2)$ denotes the diabatic FES associated with the $n$-th vibrational sublevel of the $i$-th electronic state.

The time evolution of the system is governed by a set of coupled kinetic equations describing the probability density functions $\rho_{i}^{(n)} = \rho_{i}^{(n)}(q_1, q_2, t)$ corresponding to each electron-vibrational state. The governing equations are
\begin{eqnarray}\label{ch5_kinetic_eqs}
     \frac{\partial\rho^{(0)}_\mathrm{S2}}{\partial t} &=& \hat L_{\mathrm{S2}} \rho^{(0)}_\mathrm{S2} - \sum_n K_\mathrm{cs}^{(0n)}\! \left(\rho^{(0)}_\mathrm{S2} -\rho^{(n)}_{\mathrm{CS1}}\right) - k_{\mathrm{d}}\rho^{(0)}_\mathrm{S2}, \\ \nonumber
     \frac{\partial\rho^{(n)}_{\mathrm{CS1}}}{\partial t} &=& \hat L_{\mathrm{CS1}} \rho^{(n)}_{\mathrm{CS1}} + K_\mathrm{cs}^{(0n)}\! \left(\rho^{(0)}_\mathrm{S2} - \rho^{(n)}_{\mathrm{CS1}}\right) - \sum_m K_\mathrm{csh}^{(nm)}\! \left(\rho^{(n)}_{\mathrm{CS1}} -\rho^{(m)}_{\mathrm{CS2}}\right) \\ \nonumber
     &-& \sum_l K_\mathrm{cr}^{(nl)}\! \left(\rho^{(n)}_{\mathrm{CS1}} -\rho^{(l)}_{\mathrm{S1}}\right) - \frac{1}{\tau^{(n)}_\mathrm{v}} \rho^{(n)}_{\mathrm{CS1}} + \frac{1}{\tau^{(n+1)}_\mathrm{v}} \rho^{(n+1)}_{\mathrm{CS1}}, \\ \nonumber
     \frac{\partial\rho^{(m)}_{\mathrm{CS2}}}{\partial t} &=& \hat L_{\mathrm{CS2}} \rho^{(m)}_{\mathrm{CS2}} + \sum_n K_\mathrm{csh}^{(nm)}\! \left(\rho^{(n)}_{\mathrm{CS1}} - \rho^{(m)}_{\mathrm{CS2}}\right) - \frac{1}{\tau^{(m)}_\mathrm{v}} \rho^{(m)}_{\mathrm{CS2}} + \frac{1}{\tau^{(m+1)}_\mathrm{v}} \rho^{(m+1)}_{\mathrm{CS2}}, \\ \nonumber
     \frac{\partial\rho^{(l)}_{\mathrm{S1}}}{\partial t} &=& \hat L_{\mathrm{S1}} \rho^{(l)}_{\mathrm{S1}} + \sum_n K_\mathrm{cr}^{(nl)}\! \left(\rho^{(n)}_{\mathrm{CS1}} - \rho^{(l)}_{\mathrm{S1}}\right) - \frac{1}{\tau^{(l)}_\mathrm{v}} \rho^{(l)}_{\mathrm{S1}} + \frac{1}{\tau^{(l+1)}_\mathrm{v}} \rho^{(l+1)}_{\mathrm{S1}} + \delta_{ll_0} k_\mathrm{d}\rho^{(0)}_{\mathrm{S2}}.
\end{eqnarray}
Here, $\hat{L}_{i}$ denotes the Smoluchowski diffusion operator for the $i$-th electronic state, defined by Eq.~\eqref{L_operator}, and $k_\mathrm{d}$ is the rate constant for irreversible internal conversion from the second to the first singlet excited state, $|\text{S}_2 \rangle \rightarrow |\text{S}_1 \rangle$. The functions $K_\mathrm{cs}^{(0n)}$, $K_\mathrm{csh}^{(nm)}$, and $K_\mathrm{cr}^{(nl)}$ are position-dependent rate constants for nonadiabatic electron-vibrational transitions associated with charge separation, charge shift, and charge recombination, respectively. These rates are evaluated as
\begin{align} \label{ch5_K_rates} \nonumber
   K_\mathrm{cs}^{(0n)}(q_1, q_2) &= \frac{2\pi}{\hbar} |V_{\mathrm{cs}}|^2 F_{0n} \, \delta\!\left(G^{(0)}_\mathrm{S2} - G^{(n)}_\mathrm{CS1} \right), \\ \nonumber
   K_\mathrm{cr}^{(nl)}(q_1, q_2) &= \frac{2\pi}{\hbar} |V_{\mathrm{cr}}|^2 F_{nl} \, \delta\! \left(G^{(n)}_\mathrm{CS1} - G^{(l)}_\mathrm{S1} \right), \\
   K_\mathrm{csh}^{(nm)}(q_1, q_2) &= \frac{2\pi}{\hbar} |V_{\mathrm{csh}}|^2 F_{nm} \, \delta\! \left(G^{(n)}_\mathrm{CS1} - G^{(m)}_\mathrm{CS2} \right),
\end{align}
where $V_\mathrm{cs}$, $V_\mathrm{cr}$, and $V_\mathrm{csh}$ are the corresponding electronic coupling elements.

The Franck–Condon factors $F_{nm}$ in Eq.~\eqref{ch5_K_rates} quantify the overlap between the $n$-th and $m$-th vibrational levels and are calculated using the standard expression
\begin{equation}\label{ch5_fc_factors}
  F_{nm} = e^{-S}n!m! \left[\sum^{\min(n,m)}_{r=0} \frac{(-1)^{n-r} (\sqrt{S})^{n+m-2r}}{r!(n-r)!(m-r)!} \right]^2,
\end{equation}
where $S = \lambda_\mathrm{vib} / \hbar\Omega_\mathrm{v}$ is the Huang–Rhys factor representing the strength of electron–vibrational coupling, and $\lambda_\mathrm{vib}$ is the reorganization energy associated with a high-frequency intramolecular mode.

The Franck–Condon factors $F_{nm}$ quantify the vibrational overlap between the nuclear wavefunctions of the initial and final electronic states. Physically, they represent the probability that an electronic transition between two diabatic states is accompanied by a change in vibrational quantum number from $n$ to $m$. In the displaced harmonic oscillator model adopted here, $F_{nm}$ depends on the Huang–Rhys factor $S$, which measures the strength of electron–vibrational coupling. Large values of $S$ correspond to strong vibronic displacement and hence significant population of higher vibrational levels after an electronic transition, while small $S$ values indicate that transitions predominantly occur without vibrational excitation (the so-called zero-phonon line). Thus, the Franck–Condon factors provide the link between molecular geometry changes upon electronic excitation and the distribution of accessible vibronic states during charge separation and recombination.

In Eqs.~\eqref{ch5_kinetic_eqs}, vibrational relaxation is modeled as a single-quantum cascade process, wherein each transition from level $n$ to $n-1$ occurs with a rate constant $1/\tau_\mathrm{v}^{(n)}$. The relaxation times are assumed to follow the relation $\tau_\mathrm{v}^{(n)} = \tau_\mathrm{v}^{(1)}/n$, consistent with previously established treatments of high-frequency intramolecular vibrational dissipation.

The initial conditions for the kinetic system \eqref{ch5_kinetic_eqs} are specified by assuming that, immediately following excitation, the system is thermally equilibrated on the lowest vibrational sublevel of the excited-state FES $G_\mathrm{S2}^{(0)}(q_1, q_2)$. This leads to the initial probability distributions
\begin{eqnarray}\label{ch5_initial}
    \rho_{\mathrm{S2}}^{(0)}(q_1, q_2, t = 0) &=& \frac{1}{\pi k_BT} \exp\left( - \frac{q_1^2 + q_2^2}{k_BT} \right), \\
    \rho_{\mathrm{CS1}}^{(n)}(q_1, q_2, t = 0) &=& \rho_{\mathrm{CS2}}^{(m)}(q_1, q_2, t = 0) = \rho_{\mathrm{S1}}^{(l)}(q_1, q_2, t = 0) = 0. \nonumber
\end{eqnarray}

The time-dependent population $P_i(t)$ of each diabatic electronic state $|\psi_i \rangle$ is obtained by integrating the corresponding vibrationally resolved probability densities over the configuration space and summing over all vibrational sublevels
\begin{equation}\label{ch5_population}
    P_i(t) = \sum_n \int \rho_i^{(n)}(q_1, q_2, t) \, dq_1 \, dq_2.
\end{equation}
These populations provide a quantitative description of the kinetics of photoinduced CS and CR under nonequilibrium conditions. They also offer insight into the role of vibrational relaxation and solvent dynamics in modulating the efficiency of ultrafast ET processes.

To quantitatively characterize the efficiency of charge separation in DA$_1$A$_2$ triads during the nonequilibrium (hot) stage of the photoreaction, we define the hot product yields $Y_i$ for all electronic states ($i = $ S2, CS1, CS2, S1) as the population values $Y_i = P_i(t')$ at a characteristic time $t'$ that marks the end of the transient regime and precedes the onset of quasi-equilibrium dynamics. This time point should be chosen such that fast hot transitions are complete, but thermally activated ET processes have not yet significantly occurred.

\begin{figure}[t]
    \sidecaption[t]
    \includegraphics[scale = 0.3]{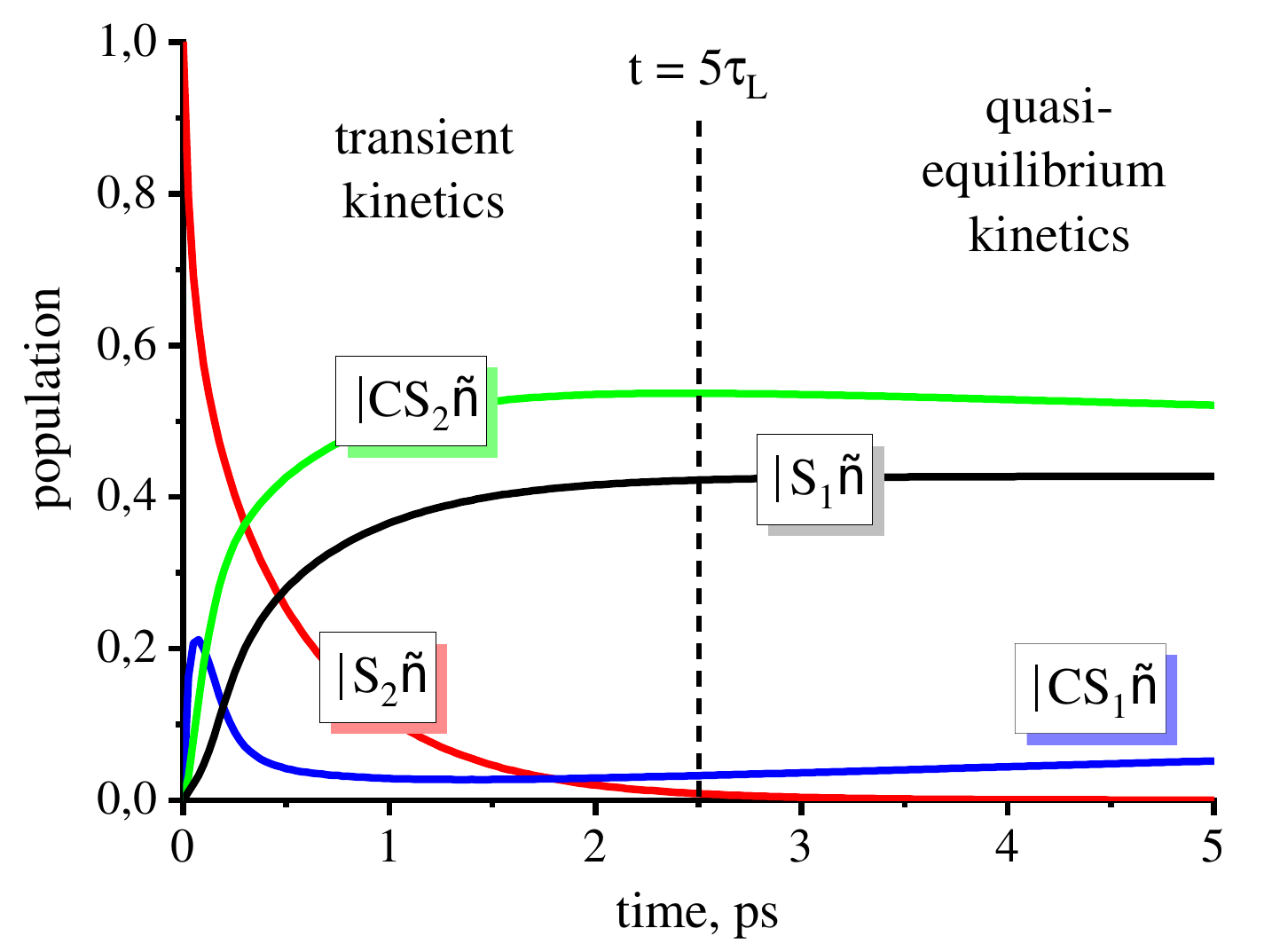}
    \caption{Time evolution of diabatic state populations $P_i(t)$ in DA$_1$A$_2$ triad following Soret-band excitation. The ultrafast photochemical processes are complete by $t' = 2.5~\mathrm{ps} = 5\tau_\mathrm{L}$ (dashed vertical line). The populations at this time, $P_i(t')$, define the hot product yields $Y_i$.}
    \label{fig:kinetics}
\end{figure}

Figure~\ref{fig:kinetics} illustrates the time evolution of the state populations $P_i(t)$ over a few picoseconds following photoexcitation. The early-time dynamics ($t < 2.5$ ps) are dominated by nonequilibrium transitions: the population of the initially excited state $|\text{S}_2\rangle$ (red curve) decays sharply within the first 1–2 ps, primarily due to charge separation into the $|\text{CS}_1\rangle$ state (blue curve), which quickly rises and subsequently feeds into the $|\text{CS}_2\rangle$ state (green curve) through hot charge shift. Simultaneously, the $|\text{S}_1\rangle$ population (black curve) increases due to internal conversion and hot charge recombination from $|\text{CS}_1\rangle$.

These processes are essentially complete by $t' = 5\tau_\mathrm{L} = 2.5$ ps, as indicated in the figure by the vertical line. Beyond this point, population changes proceed slowly, governed by thermally activated transitions within the quasi-equilibrium regime. Accordingly, we adopt $t' = 2.5$ ps as the cutoff time for evaluating the hot product yields $Y_i$ and analyzing the efficiency of ultrafast, nonequilibrium charge separation in the triad. 

The numerical simulations shown in Fig.~\ref{fig:kinetics} were performed using the following model parameters. The lifetime of the second excited state was set to $\tau_\mathrm{d} = k^{-1}_\mathrm{d} = 2$~ps, a representative value for ZnP-based systems \cite{mataga_cp_03}. The characteristic medium relaxation time was chosen as $\tau_\mathrm{L} = 0.5$~ps, corresponding to low-viscosity polar solvents such as acetonitrile. The thermal energy was fixed at $k_\mathrm{B}T = 0.025$~eV. Reorganization energies for low-frequency (solvent) modes were set to $\lambda^{(12)} = 0.9$~eV, $\lambda^{(23)} = 0.4$~eV, and $\lambda^{(13)} = 0.8$~eV, while the contribution from high-frequency intramolecular vibrations was neglected ($\lambda_\mathrm{vib} = 0$). Free energy gaps $\Delta G_\mathrm{CS1}$, $\Delta G_\mathrm{CS2}$ and $\Delta G_\mathrm{S1}$ were taken equal to $-0.8$~eV. Electronic coupling parameters for charge separation, charge shift, and charge recombination transitions were uniformly set to $V_\mathrm{cs} = V_\mathrm{csh} = V_\mathrm{cr} = 0.05$~eV.

To characterize the sensitivity of CS efficiency to the underlying energetic parameters of the DA$_1$A$_2$ system, the following section presents numerical simulations analyzing how the hot product yields $Y_i$ depend on key factors, including (i) the angle $\theta^{(23)}$ between the energy-gap coordinates associated with the two sequential ET steps (CS and CSh), (ii) the low-frequency and high-frequency reorganization free energies, $\lambda^{(nn^\prime)}$ and $\lambda_\mathrm{vib}$, respectively, and (iii) the electronic coupling parameter $V_\mathrm{cr}$, which governs the rate of hot charge recombination from the intermediate charge-separated state $|\text{CS}_1 \rangle$ to the first excited state $|\text{S}_1 \rangle$.

\subsection{Numerical Results: Effect of the System Energetics}

\begin{figure}[ht]
   \includegraphics[scale=0.43]{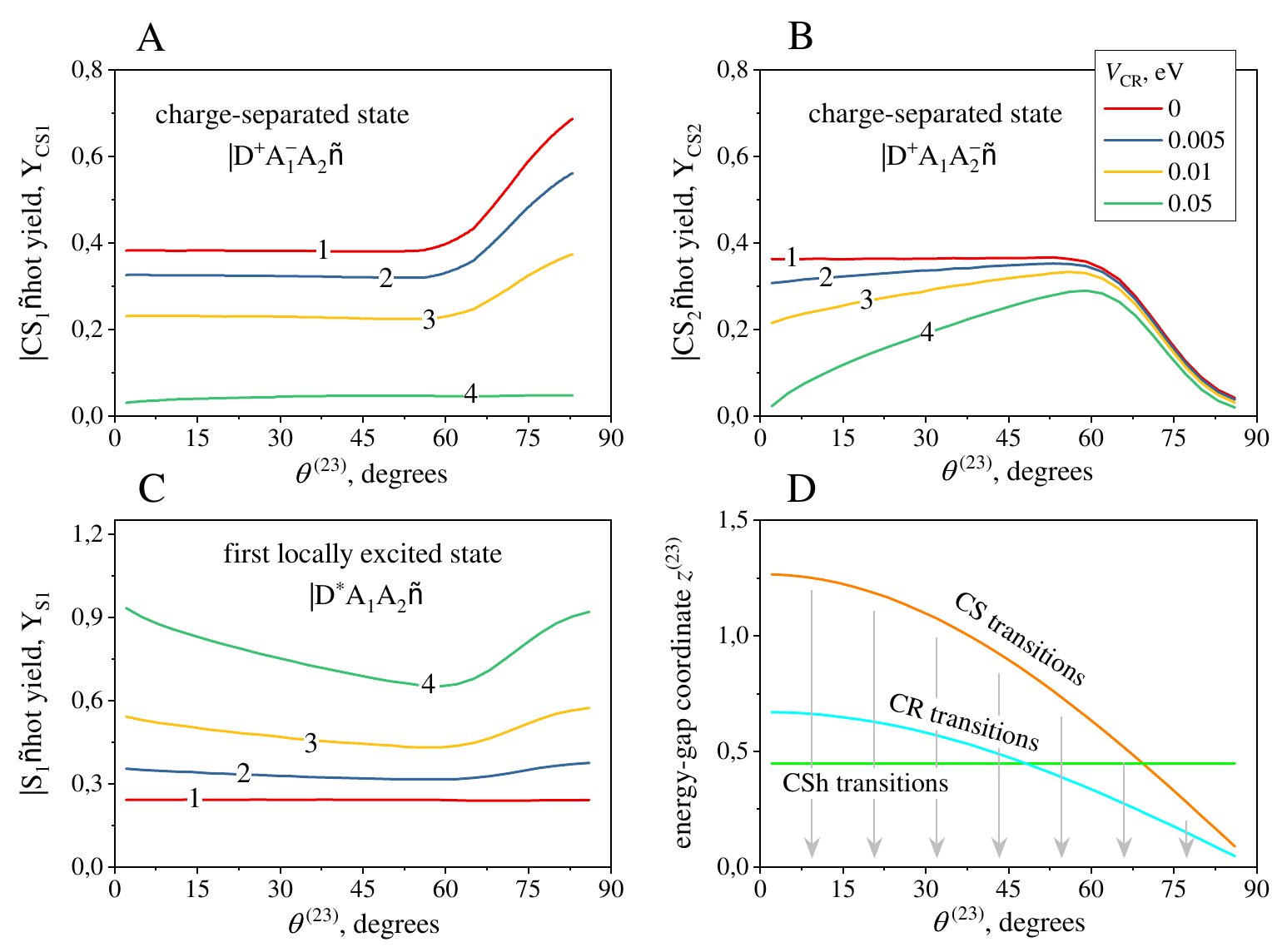}
   \caption{Influence of nonequilibrium (hot) charge recombination on CS efficiency in DA$_1$A$_2$ triads. (A--C) Hot product yields for the charge-separated states $|\mathrm{CS}_1 \rangle$, $|\mathrm{CS}_2 \rangle$ and the first excited state $|\mathrm{S}_1 \rangle$ are shown as functions of the angle $\theta^{(23)}$ between the CS and CSh energy-gap coordinates. Results are presented for several values of the electronic coupling $V_\mathrm{cr}$, which governs the rate of hot recombination: 0 (1), 0.005 (2), 0.01 (3), and 0.05~eV (4). The CR driving force is fixed at $\Delta G_\mathrm{cr} = -0.3$~eV. All other parameters are as given in the caption to Fig.~\ref{fig:kinetics}. (D) Geometric interpretation of the transition regions in $\bm{q}$-space: intersection points $\tilde{z}^{(23)}$ for CS, CR, and CSh processes, plotted as functions of $\theta^{(23)}$. Vertical arrows illustrate the trajectory of a nonequilibrium wave packet on the $G_\mathrm{CS1}$ surface, highlighting how spatial overlap with different transition regions is controlled by $\theta^{(23)}$.}
   \label{fig:hot_yield1}
\end{figure}

To analyze the influence of nonequilibrium charge recombination and charge shift processes on the photochemical dynamics of DA$_1$A$_2$ triads, we performed a series of numerical simulations under the condition of zero high-frequency vibrational reorganization energy ($\lambda_\mathrm{vib} = 0$). The electronic coupling strength for the CR transition, $V_\mathrm{cr}$, was systematically varied to examine its impact on the CS efficiency. The resulting hot product yields $Y_i$ for the $|\text{CS}_1 \rangle$, $|\text{CS}_2 \rangle$, and $|\text{S}_1 \rangle$ states are shown in Figs.~\ref{fig:hot_yield1}A–\ref{fig:hot_yield1}C as functions of the angle $\theta^{(23)}$ between the two sequential reaction coordinates.

To facilitate interpretation of the simulation results, Fig.~\ref{fig:hot_yield1}D presents a geometric visualization of the intersection regions between diabatic FESs associated with charge separation, charge recombination, and charge shift transitions. As previously illustrated in Fig.~\ref{fig:triad_in_q-space}, the spatial arrangement of these intersections, serving as reactive sinks, along the relaxation trajectory of the nuclear wave packet on the $G_\mathrm{CS1}$ surface is a key determinant of the photoreaction outcome.

Notably, the relative locations of the CR and CSh sinks with respect to the initial CS region vary systematically with the angle $\theta^{(23)}$, thereby modulating the extent of their dynamic overlap with the evolving wave packet. This geometric sensitivity underscores an important principle for optimizing the CS efficiency: by tuning the angle $\theta^{(23)}$ through rational molecular design, one may reduce the undesired recombination and selectively enhance the forward ET pathway.

To evaluate the role of geometric factors, we first examine the red curves in Figs.~\ref{fig:hot_yield1}A, \ref{fig:hot_yield1}B, and \ref{fig:hot_yield1}C, which correspond to vanishing charge recombination, $V_\mathrm{cr} = 0$. In this limit, Fig.~\ref{fig:hot_yield1}C shows that the hot yield of the $|\text{S}_1 \rangle$ state, $Y_\mathrm{S1}$, remains constant across all values of $\theta^{(23)}$. This outcome has a clear physical interpretation: in the absence of recombination, $Y_\mathrm{S1}$ is solely determined by the competition between charge separation $|\text{S}_2\rangle \to |\text{CS}_1\rangle$ and internal conversion $|\text{S}_2\rangle \to |\text{S}_1\rangle$ processes that are independent of $\theta^{(23)}$.

A similar trend is observed for the yields of the $|\text{CS}_1 \rangle$ and $|\text{CS}_2 \rangle$ states, which exhibit no dependence on $\theta^{(23)}$ within a certain angular range. Specifically, this independence persists as long as the nonequilibrium wave packet formed on the $G_\mathrm{CS1}$ surface lies above the CSh transition region. This condition is fulfilled for $\theta^{(23)} < \theta_c^{(23)}$, where $\theta_c^{(23)}$ denotes the critical angle at which the CSh transition sink intersects the path of the CS-generated wave packet (see the red and green lines in Fig.~\ref{fig:hot_yield1}D).

This insensitivity to geometry arises because the rate of a hot transition is primarily determined by the gradient (slope) of the diabatic FESs at their intersection, which remains independent of $\theta^{(23)}$ under the assumptions of the model \cite{ivanov_cp_99}. When $\theta^{(23)}$ exceeds this critical value, $Y_\mathrm{CS2}$ decreases significantly: the CS-generated wave packets then fall below the CSh reaction sink, reducing the likelihood of hot charge shift to the secondary acceptor. This behavior emphasizes the geometric control that $\theta^{(23)}$ exerts over ET pathway selectivity.

As the electronic coupling $V_\mathrm{cr}$ increases, the population yield of the $|\text{S}_1 \rangle$ state correspondingly rises, while the yields of the charge-separated states $|\text{CS}_1 \rangle$ and $|\text{CS}2 \rangle$ decrease, reflecting the enhanced efficiency of charge recombination. At higher values of $V_\mathrm{cr}$, the $|\text{CS}_2 \rangle$ yield displays a pronounced peak near the critical angle $\theta^{(23)}_c$, accompanied by a corresponding dip in the $|\text{S}_1 \rangle$ yield. This behavior is attributable to the interaction between the CR and CSh pathways. This interplay between competitive ET channels has been thoroughly analyzed in Ref.~\cite{feskov_jpca_13}.

\begin{figure}[ht]
   \includegraphics[scale=0.43]{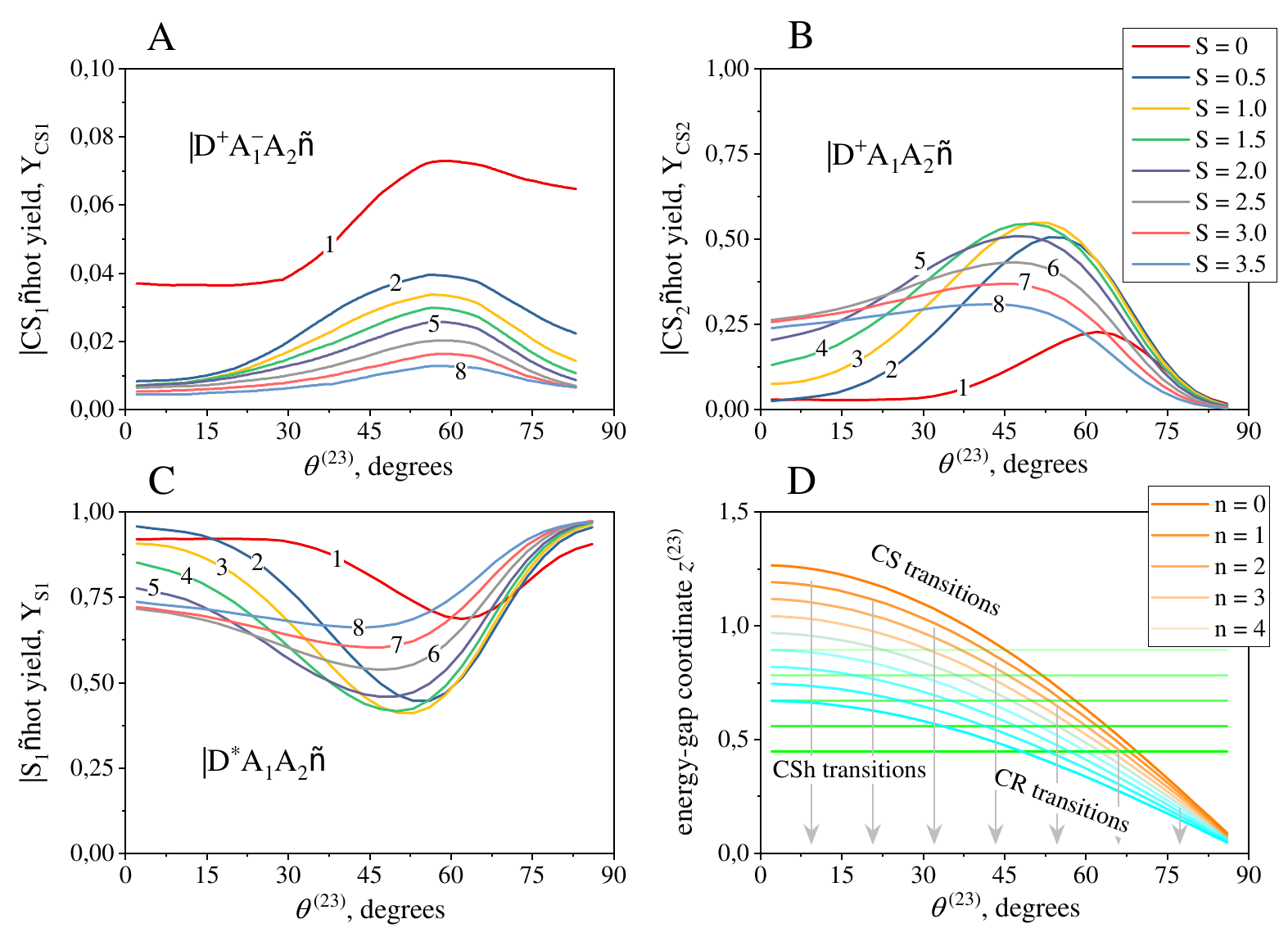}
   \caption{Effect of ET-active high-frequency intramolecular vibrations on the efficiency of charge separation in DA$_1$A$_2$ triads. (A–C) Hot product yields for $|\text{CS}_1 \rangle$, $|\text{CS}_2 \rangle$, and $|\text{S}_1 \rangle$ states plotted as functions of the angle $\theta^{(23)}$, shown for different values of the Huang–Rhys factor $S$. Curve labels correspond to $S = 0$ (1), 0.5 (2), 1.0 (3), 1.5 (4), 2.0 (5), 2.5 (6), 3.0 (7), and 3.5 (8). (D) Arrangement of the CS, CSh, and CR transition regions in the energy-gap coordinate $z^{(23)}$, taking into account vibrational structure of diabatic FESs.}
   \label{fig:hot_yield2}
\end{figure}

The effect of high-frequency vibrational reorganization on hot product yields is illustrated in Fig.~\ref{fig:hot_yield2}, which presents the calculated dependencies $Y_i(\theta^{(23)})$ for several values of the Huang–Rhys parameter, $S = \lambda_\mathrm{vib}/\hbar\Omega_\mathrm{v}$. At small $\theta^{(23)}$, the hot yield of the $|\text{CS}_2 \rangle$ state increases monotonically with $S$, reflecting the growing contribution of vibrationally assisted charge shift transitions.

As $\theta^{(23)}$ approaches the region associated with maximum yield, the population of $|\text{CS}_2 \rangle$ exhibits a pronounced peak whose magnitude increases with $S$ up to approximately $S \approx 1.0$–$1.5$, beyond which a gradual decline is observed. Furthermore, the position of this maximum shifts systematically toward smaller values of $\theta^{(23)}$ as $S$ increases, indicating that vibrational reorganization facilitates more efficient charge separation at reduced angular displacements between the CS and CSh transition vectors.

Reorganization of intramolecular high-frequency vibrational modes opens additional reaction channels by enabling transitions into vibrationally excited product states. This effect initially enhances the efficiency of hot electronic transitions, as previously discussed in Ref.~\cite{feskov_jpca_06}. As a result, the hot yields of both the $|\text{S}_1 \rangle$ and $|\text{CS}_2 \rangle$ states are expected to increase with growing vibrational reorganization energy $\lambda_\mathrm{vib}$, or equivalently, with the Huang–Rhys factor $S$. This trend is clearly reflected in Fig.~\ref{fig:hot_yield2}. However, for sufficiently large values of $\lambda_\mathrm{vib}$, this enhancement becomes non-monotonic and eventually declines. This reversal in behavior has been explained in earlier studies \cite{Ivanov_rusjpcb_08} and is attributed to the redistribution of transition probability among an increasing number of vibrational levels, which effectively weakens the strength of individual reaction sinks. Despite this limitation, ET-active high-frequency vibrational modes can significantly enhance the overall yield of the final charge-separated state, $|\text{CS}_2 \rangle$, with values reaching up to 0.6 under optimal conditions. In contrast, the hot yield of the intermediate charge-separated state $|\text{CS}_1 \rangle$ remains consistently low (below 0.04) when $\lambda_\mathrm{vib}$ is nonzero, indicating fast ET progression toward the terminal acceptor.

\begin{figure}[ht]
   \includegraphics[scale=0.48]{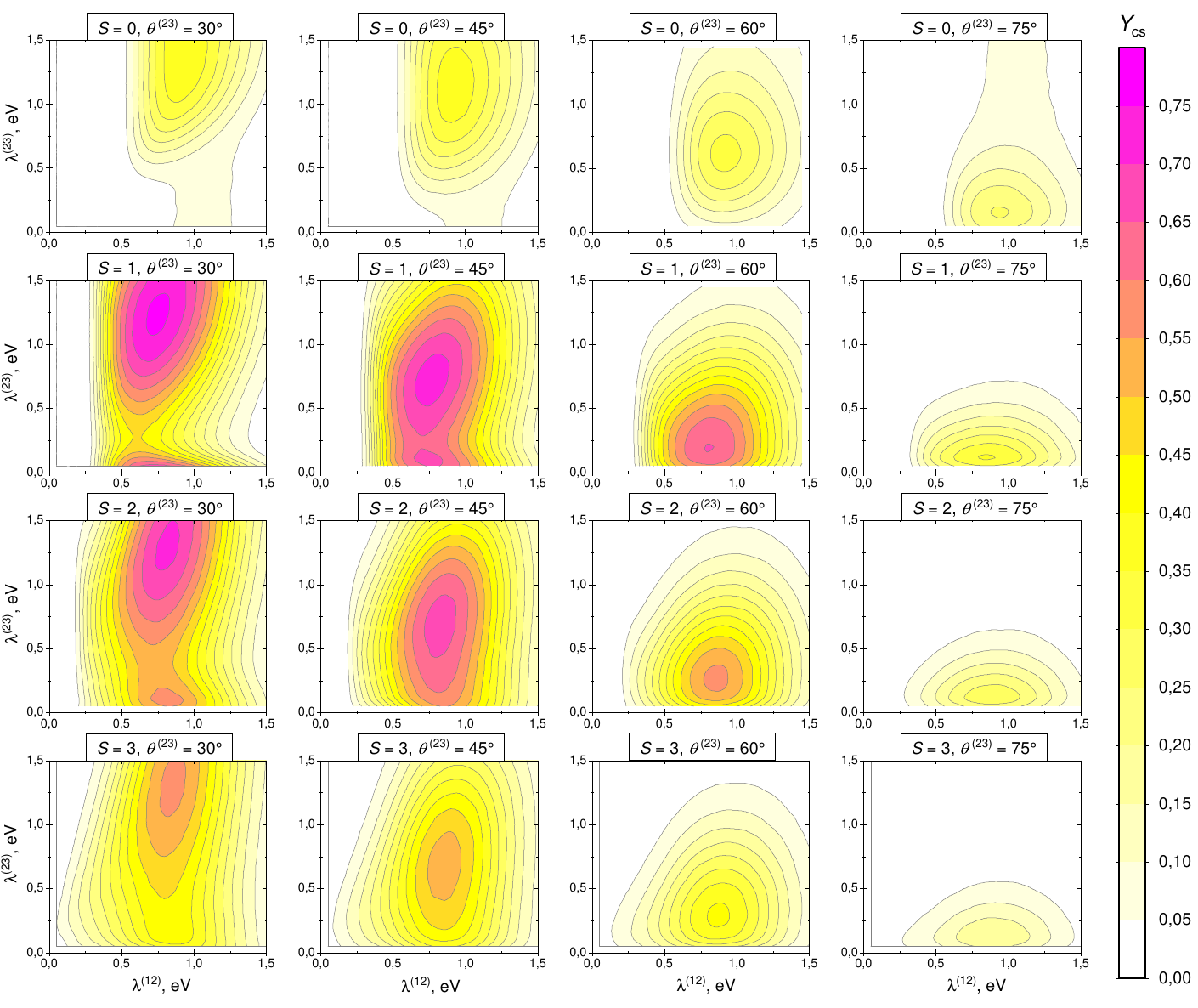} 
   \caption{Quantum yield of ultrafast charge separation, $Y_\mathrm{cs}$, as a function of the reorganization energies $\lambda^{(12)}$ and $\lambda^{(23)}$, for different values of the Huang–Rhys factor $S$ and angular displacement $\theta^{(23)}$. Each panel corresponds to a specific $(S, \theta^{(23)})$ pair, as indicated. Other model parameters are consistent with those used in Figs.~\ref{fig:hot_yield1} and \ref{fig:hot_yield2}.}
   \label{fig:Y_cs_maps}
\end{figure}

To analyze the efficiency of ultrafast charge separation in DA$_1$A$_2$ triads, we define the total yield of charge-separated states as
\begin{equation*}
    Y_\mathrm{cs} = Y_\mathrm{CS1} + Y_\mathrm{CS2},
\end{equation*}
where $Y_\mathrm{CS1}$ and $Y_\mathrm{CS2}$ denote the hot yields of the intermediate and terminal charge-separated states, respectively. Figure~\ref{fig:Y_cs_maps} presents the computed $Y_\mathrm{cs}$ values on the $(\lambda^{(12)}, \lambda^{(23)})$ plane for several combinations of Huang–Rhys parameter $S$ and angular displacement $\theta^{(23)}$.

The results reveal that high quantum yields ($Y_\mathrm{cs} > 0.7$) are predominantly observed when two conditions are met: (i) the angle $\theta^{(23)}$ is relatively small ($\lesssim 45^\circ$), and (ii) the intramolecular vibrational reorganization is strong ($S \geq 1$). These findings are consistent with previous analyses (e.g., \cite{feskov_jpca_13}), which identified that maximum CS efficiency occurs when the relaxation trajectory of the nuclear wave packet on the $G_\mathrm{CS1}$ surface first intersects the CSh transition region before reaching the CR transition region.

Furthermore, the location of the $Y_\mathrm{cs}$ maximum on the $(\lambda^{(12)}, \lambda^{(23)})$ landscape is sensitive to $\theta^{(23)}$: for $\theta^{(23)} = 30^\circ$, the yield is maximized at $\lambda^{(23)} \approx 1.2$~eV, whereas at $\theta^{(23)} = 75^\circ$, the optimal value shifts to $\lambda^{(23)} \approx 0.15$~eV. This shift reflects the changing geometry of the FES intersection regions with respect to the relaxation direction. In contrast, the position of the maximum is relatively insensitive to the Huang–Rhys parameter $S$, although the absolute yield increases with $S$ in most cases.

\subsection{Numerical Results: Effect of Molecular Geometry}

We now examine how the geometric structure of the DA$_1$A$_2$ triad influences the efficiency of ultrafast charge separation. Specifically, we investigate the impact of donor and acceptor effective radii, center-to-center ET distances, and the triad bending angle on the hot CS yield $Y_\mathrm{cs}$.

The sensitivity of $Y_\mathrm{cs}$ to structural parameters arises from their direct influence on key quantities governing electron transfer: the reorganization free energies $\lambda^{(12)}$, $\lambda^{(23)}$, and $\lambda^{(13)}$, as well as the electronic coupling strengths $V_\mathrm{cs}$, $V_\mathrm{cr}$, and $V_\mathrm{csh}$. Previous studies have demonstrated that $Y_\mathrm{cs}$ in DA$_1$A$_2$ systems is particularly responsive to ET energetics, especially during ultrafast charge shift from A$_1$ to A$_2$ \cite{feskov_jpca_13, feskov_cp_16, feskov_rjpca_16, bazlov_rjpcb_17}.

To quantify these effects, we adopt a classical electrostatic model in which the donor (D) and acceptors (A$_1$, A$_2$) are represented as spherical cavities embedded in a dielectric medium. The respective radii of the redox centers are denoted by $R_1$, $R_2$, and $R_3$, and the center-to-center distances between sites $n$ and $n^\prime$ are indicated as $R_{nn^\prime}$ ($n, n^\prime = 1,2,3$). The bending angle of the triad is designated by $\Phi$.

Within this spherical-cavity model, the solvent reorganization energies can be evaluated using the classical Marcus formula \cite{marcus_jcp_56}
\begin{equation} \label{marcus_lambda}
    \lambda^{(nn^\prime)} = \frac{c_0 e^2}{2} \left( \frac{1}{R_n} + \frac{1}{R_{n^\prime}} - \frac{2}{R_{nn^\prime}}\right),
\end{equation}
where $e$ is the elementary charge, and $c_0$ is the Pekar factor, given by $c_0 = 1/\epsilon_{\infty} - 1/\epsilon_0$. For acetonitrile, a prototypical polar solvent, we adopt $\epsilon_{\infty} = 1.806$ and $\epsilon_0 = 36.64$.

By substituting Eq.~\eqref{marcus_lambda} into Eq.~\eqref{ch5_theta^(23)}, one obtains an analytical expression for the angle $\theta^{(23)}$ between the ET energy-gap coordinates associated with the $|\mathrm{S}_2 \rangle \to |\mathrm{CS}_1 \rangle$ and $|\mathrm{CS}_1 \rangle \to |\mathrm{CS}_2 \rangle$ transitions \cite{feskov_cp_16}
\begin{equation}\label{cos_theta^(23)_R}
    \cos\theta^{(23)} = \frac{R_{2}^{-1} - R_{23}^{-1} - R_{12}^{-1} + R_{13}^{-1}}{\sqrt{\left( R_{1}^{-1} + R_{2}^{-1} - R_{12}^{-1}\right) \left( R_{2}^{-1} + R_{3}^{-1} - R_{23}^{-1} \right)}}.
\end{equation}
This result demonstrates that, within the Marcus formalism, the angular correlation $\theta^{(23)}$ is determined exclusively by the molecular geometry and is independent of the dielectric properties of the medium. Numerical evaluations indicate that physically reasonable variations in the radii and inter-site distances yield a wide range of $\theta^{(23)}$ values, typically between $40^\circ$ and $85^\circ$.

In addition to the geometric dependence of the reorganization free energies, the electronic coupling elements $V_\mathrm{cs}$, $V_\mathrm{cr}$, and $V_\mathrm{csh}$ are also strongly influenced by molecular structure. A widely employed approach to estimate these couplings is the exponential distance-dependence model, which assumes that the electronic interaction between two redox centers decays exponentially with their center-to-center separation. Within this model, the coupling constant $V_i$ for the $i$-th ET process ($i = \text{cs}, \text{cr}, \text{csh}$) is expressed as
\begin{equation}\label{V(r)}
    V_i = V^{(0)}_{i} \exp\left(-\frac{R_{nn^\prime} - R_n - R_{n^\prime}}{L_{i}}\right),
\end{equation}
where $V_{i}^{(0)}$ denotes the maximum coupling at van der Waals contact ($R_{nn^\prime} = R_n + R_{n^\prime}$), and $L_i$ is the characteristic tunneling decay length for the corresponding ET process \cite{bazlov_rjpcb_17}.

To identify structural motifs that enhance ultrafast charge separation in DA$_1$A$_2$ triads, we adopt the following two-step strategy. First, we analyze a simpler DA$_1$ dyad to determine the geometric parameters that maximize the initial CS efficiency. Subsequently, we introduce a secondary acceptor (A$_2$), and systematically vary its spatial configuration to estimate its effect on the overall CS efficiency. This methodology provides a clearer understanding of how individual geometrical features influence the dynamics of multistep ET.

As a starting point, we consider the photophysical behavior of the DA$_1$ system following Soret-band excitation at $t = 0$. The resulting CS efficiency is determined by competition between two primary processes: (1) internal conversion from the second excited state $|\text{S}_2\rangle$ to the lower-lying state $|\text{S}_1\rangle$, and (2) hot charge recombination from the intermediate CS state $|\text{D}^+ \text{A}_1^- \rangle$ back to $|\text{S}_1\rangle$ during solvent and vibrational relaxation. High efficiency of charge separation requires the following conditions to be met:
\begin{enumerate}
   \item Forward ET from $|\text{S}_2 \rangle$ to $|\text{D}^+ \text{A}_1^- \rangle$ must be faster than internal conversion to $|\text{S}_1\rangle$.
   \item Hot CR from $|\text{D}^+ \text{A}_1^- \rangle$ to $|\text{S}_1\rangle$ must be minimized.
\end{enumerate}
Efficient forward ET is promoted by strong donor–acceptor coupling, which typically requires a short inter-site distance $R_{12}$. Additionally, ultrafast charge separation is most favorable when the reaction is nearly activationless, i.e., when the total reorganization energy satisfies $\lambda^{(12)} + \lambda_\mathrm{vib} \approx -\Delta G_\mathrm{cs}$. In contrast, suppression of hot CR favors larger donor–acceptor separations, which reduce wavefunction overlap and diminish the back-transfer coupling.

\begin{figure}[ht]
   \includegraphics[scale=0.45]{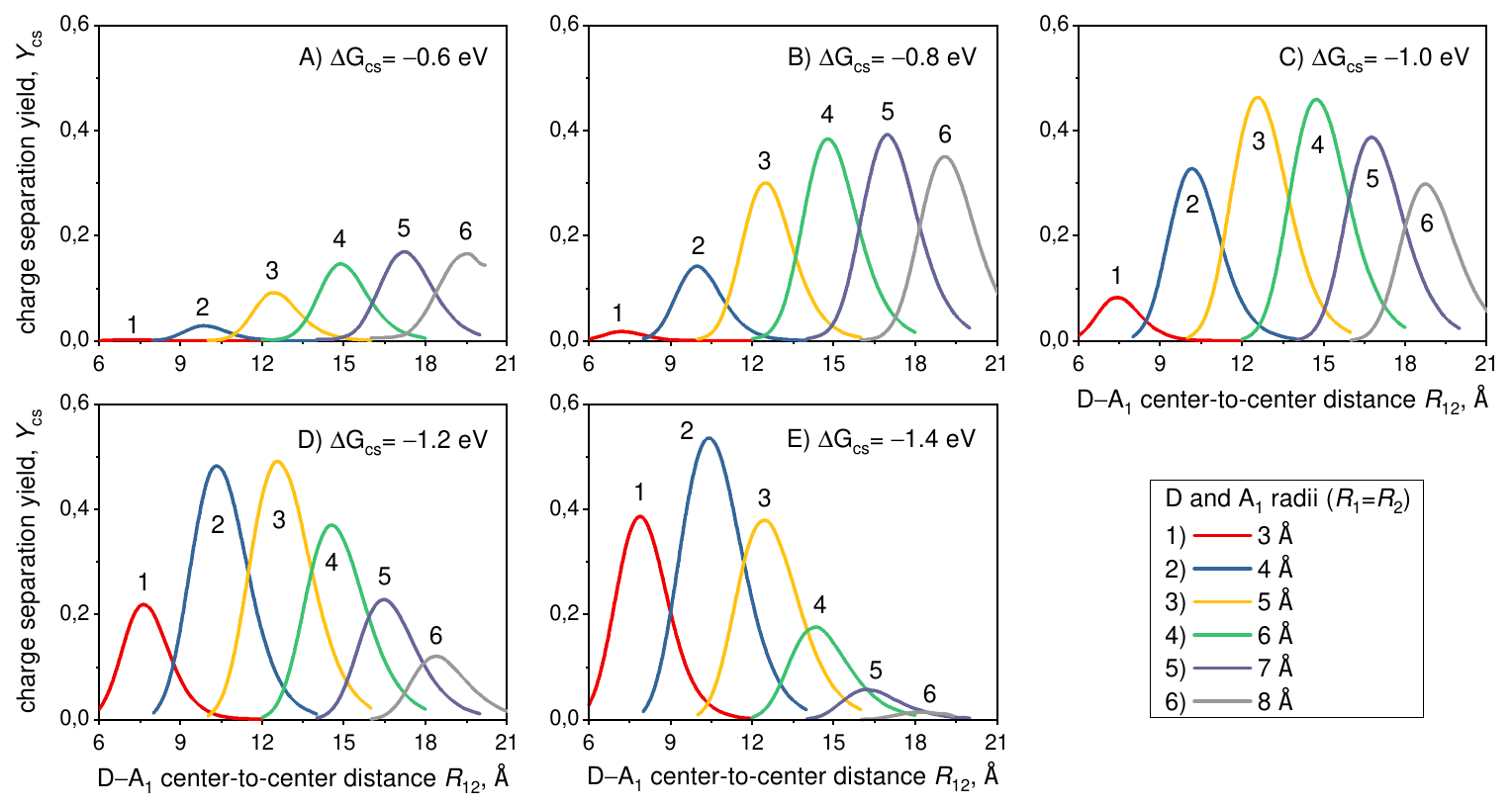}
   \caption{Hot charge separation yield $Y_\mathrm{cs}$ in DA$_1$ dyads as a function of the donor–acceptor center-to-center distance $R_{12}$, for selected values of the driving force $\Delta G_\mathrm{cs}$ and donor/acceptor radii $R_1 = R_2$ (indicated in each panel). A pronounced non-monotonic dependence $Y_\mathrm{cs}(R_{12})$ reflects the competition between efficient forward ET and suppression of ultrafast CR.}
   \label{fig:dyads_R12}
\end{figure}

These competing requirements give rise to a non-monotonic dependence of $Y_\mathrm{cs}$ on $R_{12}$. Specifically, there exists an optimal intermediate distance that balances forward ET efficiency and recombination suppression. This behavior is confirmed by numerical simulations, shown in Fig.~\ref{fig:dyads_R12}, where $Y_\mathrm{cs}$ is plotted as a function of $R_{12}$ for various values of the effective radii ($R_1 = R_2$) and the driving force $\Delta G_\mathrm{cs}$. In all scenarios, $Y_\mathrm{cs}$ exhibits a distinct maximum at a separation exceeding the contact radius. The position of this maximum depends on the system parameters, confirming the existence of a geometry-dependent CS efficiency.

A summary of the optimal yields and corresponding structural parameters is provided in Table~\ref{table:Y_cs(R12)}, which compiles peak $Y_\mathrm{cs}$ values and associated geometric configurations across a range of representative cases.

\begin{table}[ht]
    \centering
    \begin{tabularx}{0.9\textwidth}{ 
    >{\centering\arraybackslash}X 
    >{\centering\arraybackslash}X 
    >{\centering\arraybackslash}X 
    >{\centering\arraybackslash}X } 
      \hline
      $\Delta G_{\mathrm{cs}}$, eV & $R_1$ and $R_2$, \AA & $R_{12}$, \AA & $Y_\mathrm{cs}^{\mathrm{(max)}}$ \\ [3pt] \hline
      $-1.4$ & $4$ & $10.4$ & $0.54$ \\  [3pt]
      $-1.2$ & $5$ & $12.5$ & $0.50$ \\  [3pt]
      $-1.0$ & $6$ & $14.7$ & $0.47$ \\  [3pt]
      $-0.8$ & $7$ & $17.0$ & $0.40$ \\  [3pt]
      \hline
    \end{tabularx}
    \caption{Optimized geometrical parameters for donor–acceptor dyads obtained from numerical simulations. $Y_\mathrm{cs}^{(\mathrm{max})}$ denotes the maximum quantum yield of ultrafast charge separation, corresponding to the free energy gap $\Delta G_\mathrm{cs}$. $R_1$, $R_2$, and $R_{12}$ represent the effective radii and the donor–acceptor center-to-center distance, respectively.}
    \label{table:Y_cs(R12)}
\end{table}

The plots presented in Fig.~\ref{fig:dyads_R12} also demonstrate a pronounced dependence of $Y_\mathrm{cs}$ on the driving force $\Delta G_\mathrm{cs}$ associated with the $|\text{S}_2\rangle \to |\text{CS}_1\rangle$ transition. This sensitivity arises from the requirement that the activation barrier for forward electron transfer remain sufficiently small to ensure fast charge separation. The corresponding activation free energy, derived from the Marcus theory, is given by
\begin{equation}\label{ch5_activation}
    G^{\sharp}_\mathrm{cs} \equiv \frac{\left( \lambda^{(12)} + \lambda_\mathrm{vib} + \Delta G_\mathrm{cs} \right)^2}{4\lambda^{(12)}} \lesssim k_\mathrm{B}T.
\end{equation}
As follows from this expression, larger exergonicity $|\Delta G_\mathrm{cs}|$ of the CS transition requires a proportionally larger reorganization energy $\lambda^{(12)}$ to maintain a low activation barrier. Given that $\lambda^{(12)}$ is inversely related to the effective donor and acceptor radii via Eq.~\eqref{marcus_lambda}, this implies that smaller sizes $R_1$, $R_2$ are advantageous under strongly exergonic conditions. Consequently, the optimal molecular geometry for maximizing $Y_\mathrm{cs}$ depends not only on spatial arrangement but also on the underlying thermodynamics of the ET process.

Despite optimization of donor–acceptor distance, the calculated hot CS yields in Figure~\ref{fig:dyads_R12} do not exceed approximately 50\% for the parameters considered. This upper limit primarily reflects the competition from internal conversion of $|\text{S}_2\rangle$ to $|\text{S}_1\rangle$, occurring on a characteristic timescale of $\tau_\mathrm{d} = 2$~ps in zinc–porphyrin-based systems. Within the current model, this deactivation pathway cannot be readily suppressed.

Nevertheless, there exist molecular systems in which the $|\text{S}_2\rangle$ lifetime is substantially longer. Notably, xanthione-derivative compounds have been shown to possess the $|\text{S}_2\rangle$ state lifetimes on the order of 100~ps and longer \cite{Burdzinski_05}. In such systems, the extended excited-state lifetime may allow for significantly more efficient forward ET and reduced probability of hot charge recombination, thereby enhancing overall ET efficiency.

\begin{figure}[ht]
   \includegraphics[scale=0.45]{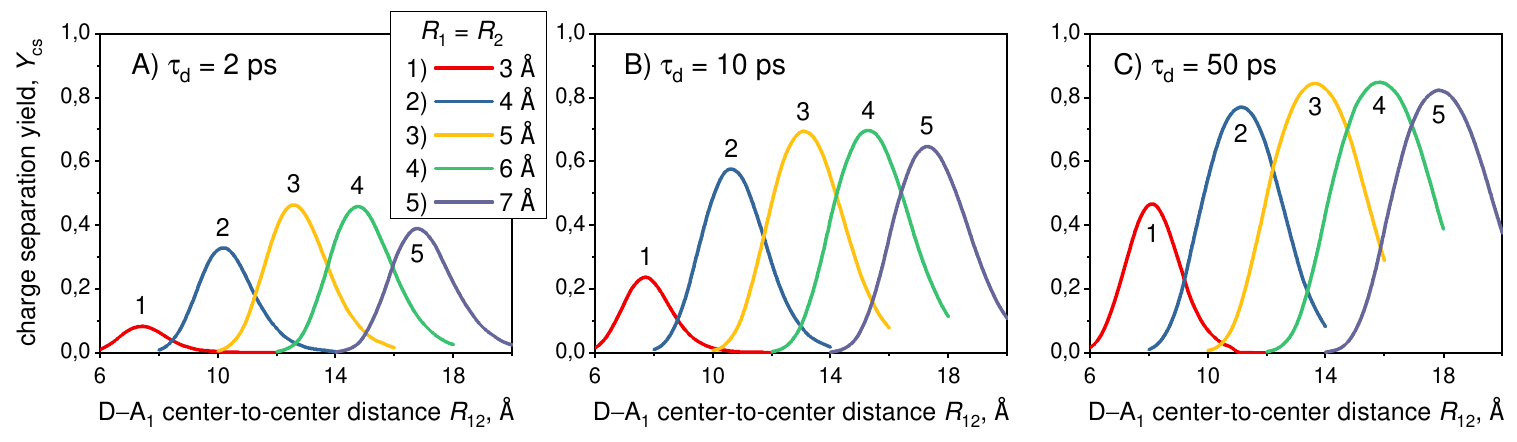}
   \caption{
   Influence of the second excited-state lifetime $\tau_\mathrm{d}$ on the efficiency of ultrafast charge separation in DA$_1$ dyads. The hot CS yield $Y\mathrm{cs}$ is shown as a function of donor–acceptor distance $R_{12}$ for different values of the effective radii $R_1 = R_2 = R$ (indicated in each panel). Panels A–C correspond to $\tau_\mathrm{d} = 2$, 10, and 50 ps, respectively, demonstrating how slower internal conversion enhances the CS efficiency.}
   \label{fig:dyads_tau_ic}
\end{figure}

Figure~\ref{fig:dyads_tau_ic} illustrates the dependence of $Y_\mathrm{cs}$ on the timescale $\tau_\mathrm{d}$ of internal conversion from the second to the first excited singlet state. The results are presented for $\tau_\mathrm{d} = 2$, 10, and 50 ps, and for several values of the effective donor and acceptor radii, $R_1 = R_2 = R$. In each case, $Y_\mathrm{cs}$ is plotted as a function of the donor–acceptor center-to-center distance $R_{12}$. As shown in panels A–C, an increase in $\tau_\mathrm{d}$ leads to a marked enhancement in CS efficiency. For $\tau_\mathrm{d} = 50$ ps, the quantum yield approaches 0.85 under optimal geometric conditions, as compared to less than 50\% for $\tau_\mathrm{d} = 2$ ps. This enhancement reflects the increased probability that the $|\text{S}_2 \rangle$ state undergoes charge separation before internal conversion to $|\text{S}_1\rangle$ occurs.

Having established the structural and kinetic factors that govern CS efficiency in DA$_1$ dyads, we now extend the analysis to DA$_1$A$_2$ triads. In this context, the inclusion of a secondary acceptor A$_2$ can significantly improve the overall $Y_\mathrm{cs}$ yield if two conditions are satisfied: (i) charge shift from A$_1$ to A$_2$ proceeds as ultrafast (hot) electron transfer; and (ii) the charge shift occurs predominantly before the system undergoes hot charge recombination. These requirements impose specific constraints on the relationship between reorganization energies and free energy gaps of the relevant ET steps, as previously discussed in Ref.~\cite{feskov_rjpca_16}.

To quantitatively assess the influence of the triad geometry on quantum yield of hot CS, we systematically explore how variations in the effective radius of the secondary acceptor A$_2$ and the distance between the primary and secondary acceptors affect $Y_\mathrm{cs}$. This analysis is based on the optimized donor–acceptor configurations identified for DA$_1$ dyads (see Table~\ref{table:Y_cs(R12)}), wherein $R_1$, $R_2$, and $R_{12}$ are fixed according to the CS driving force $\Delta G_\mathrm{cs}$. Focusing on the representative case $\Delta G_\mathrm{cs} = -1.4$~eV, we perform simulations over a two-dimensional parameter space defined by the effective radius of the secondary acceptor, $R_3$, and the surface-to-surface separation $\Delta R_{23} \equiv R_{23} - R_2 - R_3$.

\begin{figure}[ht]
  \sidecaption[t]
  \includegraphics[scale = 0.30]{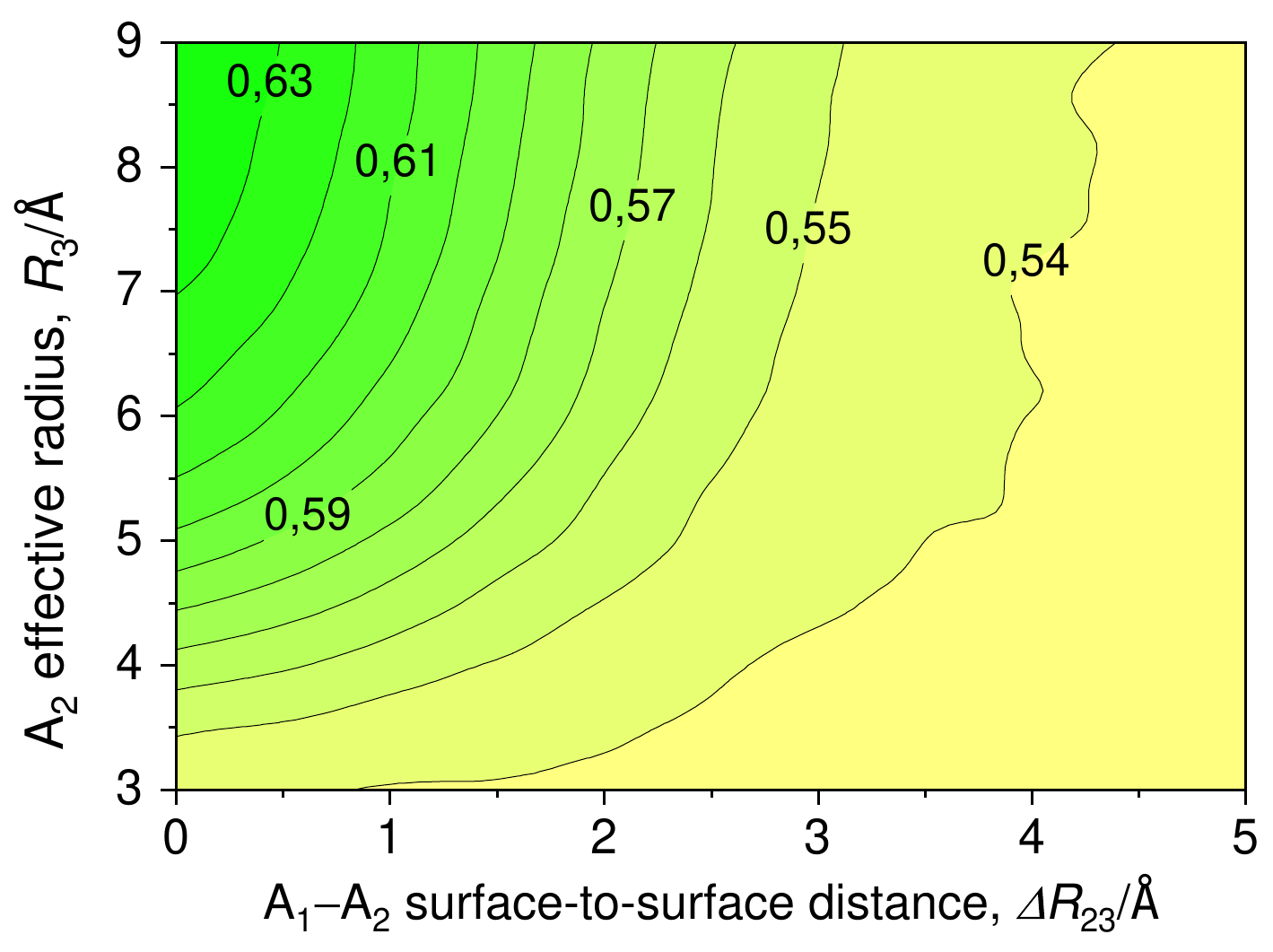}
  \caption{Influence of the secondary acceptor’s geometric configuration on efficiency of ultrafast charge separation in DA$_1$A$_2$ triads. Contour plots show the dependence of $Y_\mathrm{cs}$ on the effective radius of the secondary acceptor A$_2$ and the surface-to-surface distance between A$_1$ and A$_2$. Simulations were performed for $\Delta G_\mathrm{cs} = -1.4$~eV. Parameters $R_1$, $R_2$, and $R_{12}$ are fixed according to the optimized DA$_1$ geometry (Table~\ref{table:Y_cs(R12)}).}
  \label{fig:R3_R23_map}
\end{figure}

As shown in Fig.~\ref{fig:R3_R23_map}, the incorporation of A$_2$ can substantially enhance the CS yield, provided that the secondary acceptor is both sufficiently large in size ($R_3 \sim 8$–$9$~\AA) and spatially proximate to the primary acceptor ($\Delta R_{23} \lesssim 1$~\AA). Under these conditions, $Y_\mathrm{cs}$ increases by approximately 0.12 relative to the dyad baseline, yielding a total CS efficiency of approximately 0.64 for the full triad.

These results highlight a clear structural optimization strategy: high-efficiency ultrafast CS requires compact donor and primary acceptor units (small $R_1$, $R_2$) with large reorganization energies ($\lambda^{(12)} \gtrsim 1$~eV) conducive to ultrafast forward ET. Simultaneously, the reorganization free energy associated with the secondary ET step ($\lambda^{(23)}$) should remain relatively small ($\lambda^{(23)} \lesssim 0.5$~eV), a requirement that can be satisfied by selecting a large-radius A$_2$ moiety and minimizing its separation from A$_1$. This interplay between the structural and energetic parameters underscores the importance of precise molecular design in engineering multistep ET systems for high-efficiency photochemical applications.

Another important structural parameter in donor–acceptor triads is the bending angle $\Phi$, defined as the angle between the D–A$_1$ and A$_1$–A$_2$ segments of the molecular framework. This parameter significantly influences the dielectric polarization response of the environment during the two sequential ET events, namely, the initial charge separation and the subsequent charge shift. The effect arises from the fact that changes in $\Phi$ modify the spatial separation between the donor D and the secondary acceptor A$_2$, thereby altering the $R_{13}$ distance. As indicated by Eqs.~\eqref{marcus_lambda} and \eqref{cos_theta^(23)_R}, such variations impact both the total reorganization energy $\lambda^{(13)}$ and the angle $\theta^{(23)}$ between the energy-gap vectors associated with the two ET steps.

Sensitivity of CS efficiency to triad geometry, particularly to the bending angle, has also been reported in systems exhibiting symmetry-breaking charge separation in excited states. For instance, recent studies of perylenediimide dimers revealed that the CS yield depends strongly on conformational flexibility and the twist angle between the constituent subunits \cite{dereka_jpcl_24, Hariharan25}.

\begin{figure}[ht]
  \sidecaption[t]
  \includegraphics[scale = 0.33]{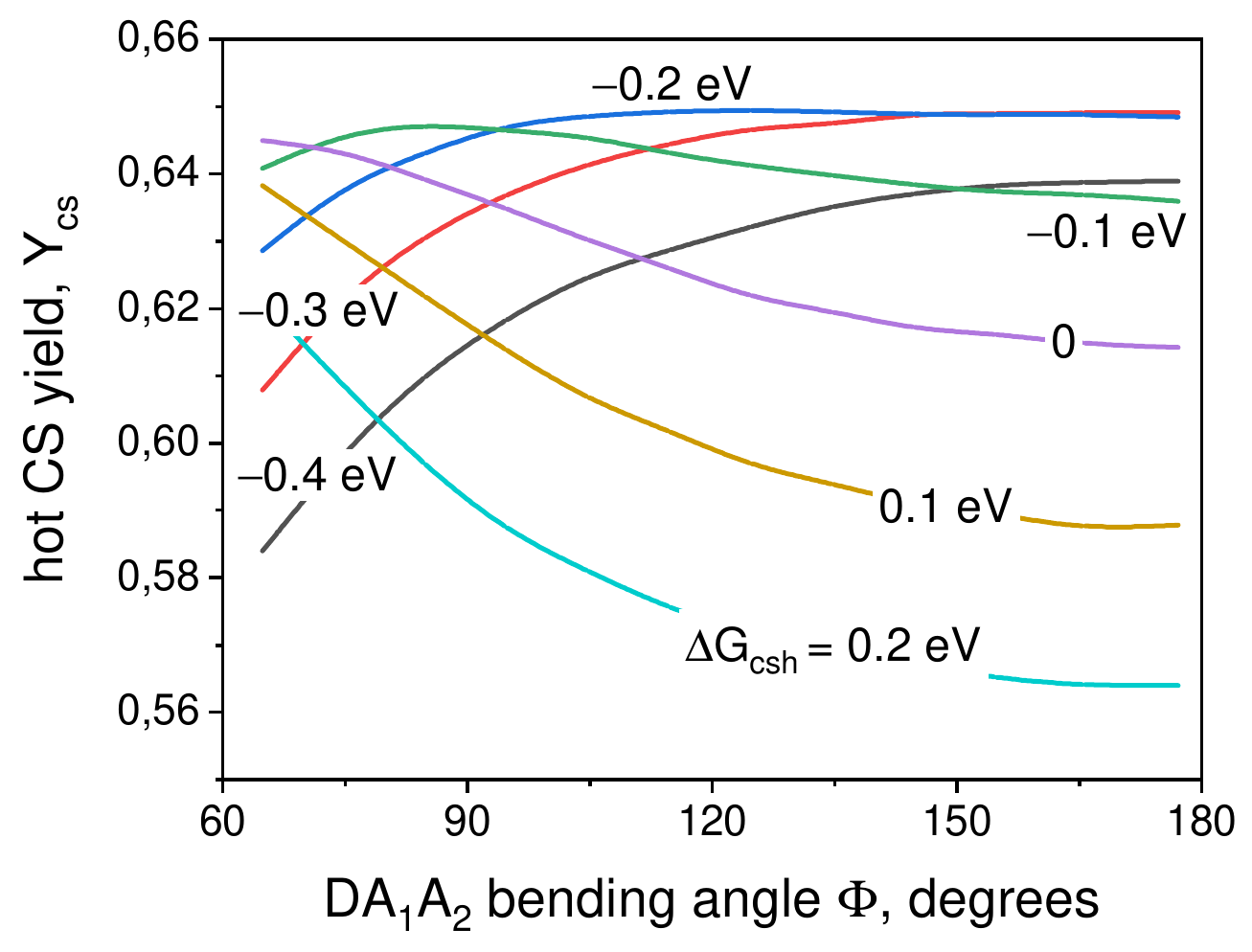}
  \caption{Dependence of the ultrafast charge separation yield $Y_\mathrm{cs}$ on the bending angle $\Phi$ of the DA$_1$A$_2$ triad. Curves correspond to different values of the charge shift free energy gap $\Delta G_\mathrm{csh}$, as indicated. Fixed parameters: $R_1 = R_2 = 4$~\AA, $R_3 = 8$~\AA, $R_{12} = 10.4$~\AA, $R_{23} = 12$~\AA.}
  \label{fig:Phi_dependence}
\end{figure}

To quantitatively evaluate the influence of the bending angle on CS efficiency, a series of numerical simulations was carried out. Figure~\ref{fig:Phi_dependence} presents the resulting dependencies of $Y_\mathrm{cs}$ on $\Phi$ for several values of the charge shift driving force $\Delta G_\mathrm{csh}$. The simulations were performed using the following fixed parameters: donor and primary acceptor radii $R_1 = R_2 = 4$~\AA, secondary acceptor radius $R_3 = 8$~\AA, donor–acceptor distance $R_{12} = 10.4$~\AA, and primary–secondary acceptor distance $R_{23} = 12$~\AA. All other model parameters are consistent with those used in the preceding sections.

As shown in Fig.~\ref{fig:Phi_dependence}, the dependence of $Y_\mathrm{cs}$ on the bending angle $\Phi$ is strongly influenced by the thermodynamic driving force for the charge shift (CSh) process:
\begin{itemize}
   \item For strongly exergonic CSh transitions ($\Delta G_\mathrm{csh} = -0.4$~eV), $Y_\mathrm{cs}$ increases monotonically with increasing $\Phi$, indicating that in this limit more linear DA$_1$A$_2$ geometries promote more efficient sequential ET to the secondary acceptor.
   \item For endergonic charge shifts ($\Delta G_\mathrm{csh} = +0.2$~eV), $Y_\mathrm{cs}$ exhibits a monotonic decrease with $\Phi$, reflecting the diminished probability of uphill CSh transitions at larger bending angles.
   \item For intermediate driving forces ($\Delta G_\mathrm{csh} = -0.2$~eV), the $Y_\mathrm{cs}(\Phi)$ curve is relatively insensitive to variations in $\Phi$, indicating that the system maintains robust CS performance across a wide conformational range.
\end{itemize}
These findings suggest that triad architectures with moderately exergonic charge shift energetics and intermediate bending angles can represent an effective balance between efficient secondary electron transfer and suppression of hot charge recombination. Such configurations offer a practical design strategy for enhancing the photochemical efficiency of multiredox molecular assemblies.

\section{Concluding Remarks}

\subsection{General Formalism}

Many distinctive features of ultrafast photoinduced electron transfer (ET) in macromolecular systems originate from the nonequilibrium character of the nuclear environment. This environment — comprising both solvent and intramolecular vibrational degrees of freedom — is driven out of equilibrium either by optical excitation or by rapid charge redistribution through electronic tunneling between redox centers. A consistent theoretical description of multistage ET processes must therefore incorporate the time-dependent evolution of the nuclear subsystem, including its dynamics during and between successive ET steps.

In this chapter, we have summarized the main results of our recent work on the development of a general semiclassical framework for modeling ultrafast ET in multiredox molecular assemblies. The proposed formalism describes complex ET pathways in macromolecules embedded in polar environments with multicomponent relaxation by employing a multidimensional coordinate space that explicitly incorporates both polarization and relaxation coordinates. This approach captures essential features of nonequilibrium ET dynamics and provides a unified theoretical basis for a broad range of ET regimes, spanning from single-step relaxation-mediated processes to cascaded and hot electron transfer events.

\subsection{Applications}

Application of the framework to donor–acceptor–acceptor (DA$_1$A$_2$) triads has revealed the crucial role of hot electron transfer pathways in enhancing charge-separation efficiency. Numerical simulations demonstrated how molecular and environmental parameters — including reorganization energies, electron–vibrational couplings, donor–acceptor separations, and triad bending angles — influence ultrafast ET kinetics and ultimately determine the quantum yield of the charge-separated state.

\subsection{Outlook}

Taken together, these results elucidate how molecular architecture and environmental relaxation dynamics shape the efficiency and directionality of photoinduced electron transfer. The developed framework provides not only a conceptual foundation but also a predictive tool for guiding the rational design of photochemical systems, including artificial photosynthetic complexes, molecular optoelectronic devices, and organic photovoltaic materials.

\bigskip \bigskip
\textbf{Acknowledgments} This work was supported by Russian Science Foundation (project 22-13-00180-P, https://rscf.ru/en/project/22-13-00180/).

\bibliographystyle{spphys}
\bibliography{biblio_full}

@Article{marcus_jcp_56,
  Title                    = {On the theory of oxidation-reduction reactions involving electron transfer},
  Author                   = {R. A. Marcus},
  Journal                  = {J. Chem. Phys.},
  Year                     = {1956},
  Number                   = {5},
  Pages                    = {966--978},
  Volume                   = {24},
  Doi                      = {10.1063/1.1742723}
}

@Article{zusman_cp_80,
  Title                    = {Outer-sphere electron transfer in polar solvents},
  Author                   = {L. D. Zusman},
  Journal                  = {Chem. Phys.},
  Year                     = {1980},
  Number                   = {2},
  Pages                    = {295 - 304},
  Volume                   = {49},
  DOI                      = {10.1016/0301-0104(80)85267-0}
}

@Article{marcus_bba_85,
  Title                    = {Electron Transfers in Chemistry and Biology},
  Author                   = {R. A. Marcus and N. Sutin},
  Journal                  = {Biochim. Biophys. Acta},
  Year                     = {1985},
  Number                   = {3},
  Pages                    = {265--322},
  Volume                   = {811},
  Doi                      = {10.1016/0304-4173(85)90014-X}
}

@Article{sumi_jcp_86,
  Title                    = {Dynamical effects in electron transfer reactions},
  Author                   = {H. Sumi and R. A. Marcus},
  Journal                  = {J. Chem. Phys.},
  Year                     = {1986},
  Pages                    = {4894--4914},
  Volume                   = {84},
  Doi                      = {10.1063/1.449978}
}

@Article{jortner_jcp_88,
  Title                    = {Intramolecular vibrational excitations accompanying solvent-controlled electron transfer reactions},
  Author                   = {J. Jortner and M. Bixon},
  Journal                  = {J. Chem. Phys.},
  Year                     = {1988},
  Number                   = {1},
  Pages                    = {167-170},
  Volume                   = {88},
  doi                      = {10.1063/1.454632}
}

@Article{zusman_cp_88,
  Title                    = {The theory of electron transfer reactions in solvents with two characteristic relaxation times},
  Author                   = {L. D. Zusman},
  Journal                  = {Chem. Phys.},
  Year                     = {1988},
  Number                   = {1},
  Pages                    = {51 - 61},
  Volume                   = {119},
  ISSN                     = {0301-0104},
  Doi                      = {10.1016/0301-0104(88)80005-3}
}

@Article{akesson_jcp_92,
  Title                    = {Temperature dependence of the inverted regime electron transfer kinetics of betaine-30 and the role of molecular modes},
  Author                   = {E. Akesson and A. E. Johnson and N. E. Levinger and G. C. Walker and T. F. DuBruil and P. F. Barbara},
  Journal                  = {J. Chem. Phys.},
  Year                     = {1992},
  Pages                    = {7859--7862},
  Volume                   = {96}
}

@Article{barbara_sci_92,
  Title                    = {Vibrational Modes and the Dynamic Solvent Effect in Electron and Proton Transfer},
  Author                   = {P. F. Barbara and G. C. Walker and T. P. Smith},
  Journal                  = {Science},
  Year                     = {1992},
  Pages                    = {975--981},
  Volume                   = {256}
}

@Article{marchi_jacs_93,
  Title                    = {Diabatic surfaces and the pathway for primary electron transfer in a photosynthetic reaction center},
  Author                   = {Massimo Marchi and John N. Gehlen and David Chandler and Marshall Newton},
  Journal                  = {J. Am. Chem. Soc.},
  Year                     = {1993},
  Number                   = {},
  Pages                    = {4178-4190},
  Volume                   = {115},
  Doi                      = {10.1021/ja00063a041}
}

@Article{maroncelli_jpc_93,
  Title                    = {A simple interpretation of polar solvation dynamics},
  Author                   = {M. Maroncelli and V. P. Kumar and A. Papazyan},
  Journal                  = {J. Phys. Chem.},
  Year                     = {1993},
  Pages                    = {13--17},
  Volume                   = {97}
}

@Article{jimenez_nat_94,
  Title                    = {Femtosecond solvation dynamics of water},
  Author                   = {R. Jimenez and G. R. Fleming and P. V. Kumar and M. Maroncelli},
  Journal                  = {Nature},
  Year                     = {1994},
  Pages                    = {471--473},
  Volume                   = {369}
}

@Article{najbar_jpc_94,
  Title                    = {Potential energy surfaces for electron transfer in a supramolecular triad system {A1}-{D}-{A2} in a polar solvent},
  Author                   = {J. Najbar and M. Tachiya},
  Journal                  = {J. Phys. Chem.},
  Year                     = {1994},
  Pages                    = {199--205},
  Volume                   = {98},
  Doi                      = {10.1021/j100052a033}
}

@Article{tang_jcp_94,
  Title                    = {On superexchange electron-transfer reactions involving three paraboloidal potential surfaces in a two-dimensional reaction coordinate},
  Author                   = {Jau Tang and James R. Norris},
  Journal                  = {J. Chem. Phys.},
  Year                     = {1994},
  Number                   = {},
  Pages                    = {5615-5622},
  Volume                   = {101},
  Doi                      = {10.1063/1.467348}
}

@Article{cho_jcp_95,
  Title                    = {Nonequilibrium photoinduced electron transfer},
  Author                   = {M. Cho and R. J. Silbey},
  Journal                  = {J. Chem. Phys.},
  Year                     = {1995},
  Pages                    = {595--606},
  Volume                   = {103},
  Doi                      = {10.1063/1.470094}
}

@Article{tachiya_cp_96,
  Title                    = {Competitive electron transfers in model triad systems: continuum model approach},
  Author                   = {T. Motylewski and J. Najbar and M. Tachiya},
  Journal                  = {Chem. Phys.},
  Year                     = {1996},
  Pages                    = {193--206},
  Volume                   = {212}
}

@Article{bicout_jcp_97,
  Title                    = {Electron transfer reactions dynamics in non-{Debye} solvents},
  Author                   = {D. J. Bicout and A. Szabo},
  Journal                  = {J. Chem. Phys.},
  Year                     = {1997},
  Pages                    = {2325--2338},
  Volume                   = {109},
  Doi                      = {10.1063/1.476800}
}

@Article{ando_jpcb_98,
  Title                    = {Nonequilibrium Oscillatory Electron Transfer in Bacterial Photosynthesis},
  Author                   = {K. Ando and H. Sumi},
  Journal                  = {J. Phys. Chem. B},
  Year                     = {1998},
  Pages                    = {10991--11000},
  Volume                   = {102}
}

@Article{bagchi_acp_99,
  Title                    = {Interplay between ultrafast polar solvation and vibrational dynamics in electron transfer reactions: role of high-frequency vibrational modes},
  Author                   = {B. Bagchi and N. Gayathri},
  Journal                  = {Adv. Chem. Phys.},
  Year                     = {1999},
  Pages                    = {1--81},
  Volume                   = {107},
  Doi                      = {10.1002/9780470141663.ch1}
}

@Article{bixon_acp_99,
  Title                    = {Electron Transfer: From Isolated Molecules to Biomolecules},
  Author                   = {M. Bixon and J. Jortner},
  Journal                  = {Adv. Chem. Phys.},
  Year                     = {1999},
  Pages                    = {35--202},
  Volume                   = {106},
  Doi                      = {10.1002/9780470141656.ch3}
}

@Article{ivanov_cp_99,
  Title                    = {Theory of non-thermal electron transfer},
  Author                   = {A. I. Ivanov and V. V. Potovoi},
  Journal                  = {Chem. Phys.},
  Year                     = {1999},
  Number                   = {2},
  Pages                    = {245--259},
  Volume                   = {247},
  Doi                      = {10.1016/S0301-0104(99)00197-4}
}

@Article{zimmermann_jpcb_01,
  Title                    = {Experimental Fingerprints of Vibrational Wave-Packet Motion During Ultrafast Heterogeneous Electron Transfer},
  Author                   = {Zimmermann, C. and Willig, F. and Ramakrishna, S. and Burfeindt, B. and Pettinger, B. and Eichberger, R. and Storck, W.},
  Journal                  = {J. Phys. Chem. B},
  Year                     = {2001},
  Pages                    = {9245--9253},
  Volume                   = {105},
  Doi                      = {10.1021/jp011106z}
}

@Article{barzykin_acp_02,
  Title                    = {Solvent effects in nonadiabatic electron-transfer reactions: theoretical aspects},
  Author                   = {A. V. Barzykin and P. A. Frantsuzov and K. Seki and M. Tachiya},
  Journal                  = {Adv. Chem. Phys.},
  Year                     = {2002},
  Pages                    = {511--616},
  Volume                   = {123}
}

@Article{mataga_cp_03,
  Title                    = {Ultrafast charge transfer and radiationless relaxations from higher excited state (S2) of directly linked Zn-porphyrin (ZP)-acceptor dyads: investigations into fundamental problems of exciplex chemistry},
  Author                   = {N. Mataga and S. Taniguchi and H. Chosrowjan and A. Osuka and N. Yoshida},
  Journal                  = {Chem. Phys.},
  Year                     = {2003},
  Pages                    = {215--228},
  Volume                   = {295}
}

@Article{fedunov_jcp_04,
  Title                    = {Effect of the excitation pulse carrier frequency on the ultrafast charge recombination dynamics of donor-acceptor complexes: Stochastic simulations and experiments},
  Author                   = {R. G. Fedunov and S. V. Feskov and A. I. Ivanov and O. Nicolet and S. Pag\`{e}s and E. Vauthey},
  Journal                  = {J. Chem. Phys.},
  Year                     = {2004},
  Number                   = {8},
  Pages                    = {3643--3656},
  Volume                   = {121},
  Doi                      = {10.1063/1.1772362}
}

@Article{newton_ijc_04,
  Title                    = {Bridge-Mediated Electron Transfer and Multiple Reaction Coordinates},
  Author                   = {Marshall D. Newton},
  Journal                  = {Isr. J. Chem.},
  Year                     = {2004},
  Pages                    = {83--88},
  Volume                   = {44},
  Doi                      = {10.1560/LQ06-T9HQ-MTLM-2VC1}
}

@Article{gladkikh_jcp_05,
  Title                    = {Hot recombination of photogenerated ion pairs},
  Author                   = {V. Gladkikh and A. I. Burshtein and S. V. Feskov and A. I. Ivanov and E. Vauthey},
  Journal                  = {J. Chem. Phys.},
  Year                     = {2005},
  Pages                    = {244510},
  Volume                   = {123},
  Doi                      = {10.1063/1.2140279}
}

@Article{nicolet_jpca_05,
  Title                    = {Effect of the Excitation Wavelength on the Ultrafast Charge Recombination Dynamics of Donor-Acceptor Complexes in Polar Solvents},
  Author                   = {O. Nicolet and N. Banerji and S. Pag\`{e}s and E. Vauthey},
  Journal                  = {J. Phys. Chem. A},
  Year                     = {2005},
  Pages                    = {8236--8245},
  Volume                   = {109},
  Doi                      = {10.1021/jp0532216}
}

@Article{wan_cpl_05,
  Title                    = {Ultrafast Unequilibrated Charge Transfer: a New Channel in the Quenching of Fluorescent Biological Probes},
  Author                   = {Wan, C. and Xia, T. and Becker, H. C. and Zewail, A. H.},
  Journal                  = {Chem. Phys. Lett},
  Year                     = {2005},
  Pages                    = {158--163},
  Volume                   = {412},
  Doi                      = {10.1016/j.cplett.2005.06.101}
}

@book{borg_book_05,
  author                   = {Borg, Ingwer and Groenen, Patrick J. F.},
  title                    = {Modern Multidimensional Scaling: Theory and Applications},
  year                     = 2005,
  address                  = {New York},
  doi                      = {10.1007/0-387-28981-X},
  isbn                     = {038728981X},
  publisher                = {Springer}
}

@Article{feskov_jpca_06,
  Title                    = {Effect of High Frequency Modes and Hot Transitions on Free Energy Gap Dependence of Charge Recombination Rate},
  Author                   = {S. V. Feskov and V. N. Ionkin and A. I. Ivanov},
  Journal                  = {J. Phys. Chem. A},
  Year                     = {2006},
  Number                   = {43},
  Pages                    = {11919--11925},
  Volume                   = {110},
  Doi                      = {10.1021/jp063280z}
}

@Article{vauthey_jppa_06,
  Title                    = {Investigations of bimolecular photoinduced electron transfer reactions in polar solvents using ultrafast spectroscopy},
  Author                   = {E. Vauthey},
  Journal                  = {J. Photochem. Photobiol. A},
  Year                     = {2006},
  Number                   = {1--2},
  Pages                    = {1--12},
  Volume                   = {179},
  Doi                      = {10.1016/j.jphotochem.2005.12.019}
}

@Article{feskov_cpl_07,
  Title                    = {Activationless dynamic solvent effect},
  Author                   = {S. V. Feskov and V. Gladkikh and A. I. Burshtein},
  Journal                  = {Chem. Phys. Lett.},
  Year                     = {2007},
  Pages                    = {162--167},
  Volume                   = {447}
}

@Article{feskov_cpl_08,
  Title                    = {Kinetics of non-thermal electron transfer controlled by the dynamic solvent effect},
  Author                   = {S. V. Feskov and V. Gladkikh and A. I. Burshtein},
  Journal                  = {Chem. Phys. Lett.},
  Year                     = {2008},
  Number                   = {1-3},
  Pages                    = {71--75},
  Volume                   = {458},
  Doi                      = {10.1016/j.cplett.2008.04.063}
}

@Article{feskov_jpca_08,
  Title                    = {Solvent and Spectral Effects in the Ultrafast Charge Recombination Dynamics of Excited Donor-Acceptor Complexes},
  Author                   = {S. V. Feskov and V.N. Ionkin and A. I. Ivanov and H. Hagemann and E. Vauthey},
  Journal                  = {J. Phys. Chem. A},
  Year                     = {2008},
  Number                   = {4},
  Pages                    = {594--601},
  Volume                   = {112},
  Doi                      = {10.1021/jp709587x}
}

@Article{feskov_jpca_09,
  Title                    = {Double-Channel Photoionization Followed by Geminate Charge Recombination/Separation},
  Author                   = {S. V. Feskov and A. I. Burshtein},
  Journal                  = {J. Phys. Chem. A},
  Year                     = {2009},
  Number                   = {48},
  Pages                    = {13528--13540},
  Volume                   = {113},
  DOI                      = {10.1021/jp901863t}
}

@Article{lebard_jpcb_09,
  Title                    = {Energetics of Bacterial Photosynthesis},
  Author                   = {David N. LeBard and Dmitry V. Matyushov},
  Journal                  = {J. Phys. Chem. B},
  Year                     = {2009},
  Number                   = {36},
  Pages                    = {12424-12437},
  Volume                   = {113},
  Doi                      = {10.1021/jp904647m}
}

@Article{ivanov_rcr_10,
  Title                    = {Kinetics of fast photochemical charge separation and charge recombination reactions},
  Author                   = {A. I. Ivanov and V. A. Mikhailova},
  Journal                  = {Russ. Chem. Rev.},
  Year                     = {2010},
  Number                   = {12},
  Pages                    = {1047--1070},
  Volume                   = {79},
  Doi                      = {10.1070/RC2010v079n12ABEH004167}
}

@Article{wallin_jpca_10,
  Title                    = {State-Selective Electron Transfer in an Unsymmetric Acceptor--{Zn(II)}porphyrin--Acceptor Triad: Toward a Controlled Directionality of Electron Transfer from the Porphyrin {S$_2$} and {S$_1$} States as a Basis for a Molecular Switch},
  Author                   = {Wallin, S. and Monnereau, C. and Blart, E. and Gankou, J. R. and Odobel, F. and Hammarstr\"{o}m, L.},
  Journal                  = {J. Phys. Chem. A},
  Year                     = {2010},
  Number                   = {4},
  Pages                    = {1709--1721},
  Volume                   = {114},
  Doi                      = {10.1021/jp907824d}
}

@Article{feskov_jpca_11,
  Title                    = {Kinetics of nonequilibrium electron transfer in photoexcited ruthenium(II)-cobalt(III) complexes},
  Author                   = {S. V. Feskov and A. O. Kichigina and A. I. Ivanov},
  Journal                  = {J. Phys. Chem. A},
  Year                     = {2011},
  Number                   = {9},
  Pages                    = {1462--1471},
  Volume                   = {115},
  Doi                      = {10.1021/jp108607t}
}

@Article{feskov_jpca_13,
  Title                    = {The Efficiency of Intramolecular Charge Separation from the Second Excited State: Suppression of the Hot Charge Recombination by Electron Transfer to the Secondary Acceptor},
  Author                   = {S. V. Feskov and A. I. Ivanov},
  Journal                  = {J. Phys. Chem. A},
  Year                     = {2013},
  Number                   = {45},
  Pages                    = {11479--11489},
  Volume                   = {117},
  Doi                      = {10.1021/jp408516q}
}

@Article{barykov_jpcc_15,
  Title                    = {Effect of the excitation pulse carrier frequency on ultrafast photoinduced charge transfer kinetics: Effect of intramolecular high frequency vibrational mode excitation},
  Author                   = {V. Y. Barykov and V. N. Ionkin and A. I. Ivanov},
  Journal                  = {J. Phys. Chem. C},
  Year                     = {2015},
  Number                   = {6},
  Pages                    = {2989--2995},
  Volume                   = {119},
  Doi                      = {10.1021/acs.jpcc.5b00005}
}

@Article{newton_jpcb_15,
  Title                    = {Extension of {Hopfield's} Electron Transfer Model To Accommodate Site-Site Correlation},
  Author                   = {Marshall D. Newton},
  Journal                  = {J. Phys. Chem. B},
  Year                     = {2015},
  Number                   = {46},
  Pages                    = {14728-14737},
  Volume                   = {119},
  Doi                      = {10.1021/acs.jpcb.5b07456}
}

@Article{douhal_jppc_16,
  Title                    = {Ultrafast and Fast Charge Separation Processes in Real Dye-Sensitized Solar Cells},
  Author                   = {Mart\'{\i}n, C. and Zi\'{o}{\l}ek, M. and Douhal, A.},
  Journal                  = {J. Photochem. Photobiol. C},
  Year                     = {2016},
  Pages                    = {1--30},
  Volume                   = {26},
  Doi                      = {10.1016/j.jphotochemrev.2015.12.001}
}

@Article{feskov_cp_16,
  Title                    = {Effect of geometrical parameters of dyad {D-A} and triad {D-A1-A2} on the efficiency of ultrafast intramolecular charge separation from the second excited state},
  Author                   = { S. V. Feskov and A. I. Ivanov},
  Journal                  = {Chem. Phys.},
  Year                     = {2016},
  Number                   = {},
  Pages                    = {164-172},
  Volume                   = {478},
  Doi                      = {10.1016/j.chemphys.2016.03.013}
}

@Article{feskov_jppc_16,
  Title                    = {Non-equilibrium effects in ultrafast photoinduced charge transfer kinetics},
  Author                   = {S. V. Feskov and V. A. Mikhailova and A. I. Ivanov},
  Journal                  = {J. Photochem. Photobiol. C},
  Year                     = {2016},
  Pages                    = {48-72},
  Volume                   = {29},
  Doi                      = {10.1016/j.jphotochemrev.2016.11.001}
}

@Article{feskov_rjpca_16,
  Title                    = {Efficiency of Intramolecular Electron Transfer from the Second Excited State of the Donor in Molecular Triads {D}-{A1}-{A2}},
  Author                   = {S. V. Feskov and A. I. Ivanov},
  Journal                  = {Russ. J. Phys. Chem.},
  Year                     = {2016},
  Number                   = {1},
  Pages                    = {144--151},
  Volume                   = {90},
  Doi                      = {10.1134/S0036024416010106}
}

@Article{nazarov_jpcb_16,
  Title                    = {Effect of Intramolecular High Frequency Vibrational Mode Excitation on Ultrafast Photoinduced Charge Transfer and Charge Recombination Kinetics},
  Author                   = {Nazarov, A. E. and Barykov, V. Yu. and Ivanov, A. I.},
  Journal                  = {J. Phys. Chem. B},
  Year                     = {2016},
  Number                   = {12},
  Pages                    = {3196--3205},
  Volume                   = {120},
  Doi                      = {10.1021/acs.jpcb.6b00539}
}

@Article{ostroverkhova_cr_16,
  Title                    = {Organic optoelectronic materials: mechanisms and applications},
  Author                   = {O. Ostroverkhova},
  Journal                  = {Chem. Rev.},
  Year                     = {2016},
  Pages                    = {13279--13412},
  Volume                   = {116},
  Doi                      = {10.1021/acs.chemrev.6b00127}
}

@Article{bazlov_rjpcb_17,
  Title                    = {Effectiveness of Charge Separation from the Long-Lived Second Excited State of Donors},
  Author                   = {S. V. Bazlov and S. V. Feskov and A. I. Ivanov},
  Journal                  = {Russ. J. Phys. Chem. B},
  Year                     = {2017},
  Pages                    = {242--248},
  Volume                   = {11},
  Number                   = {2},
  Doi                      = {10.1134/S1990793117020026},
 }

@Article{kumpulainen_cr_17,
  Title                    = {Ultrafast Elementary Photochemical Processes of Organic Molecules in Liquid Solution},
  Author                   = {Kumpulainen, T. and Lang, B. and Rosspeintner, A. and Vauthey, E.},
  Journal                  = {Chem. Rev.},
  Year                     = {2017},
  Pages                    = {10826--10939},
  Volume                   = {117},
  Doi                      = {10.1021/acs.chemrev.6b00491}
}

@Article{feskov_jcp_18,
  Title                    = {Solvent-assisted multistage nonequilibrium electron transfer in rigid supramolecular systems: Diabatic free energy surfaces and algorithms for numerical simulations},
  Author                   = {S. V. Feskov and A. I. Ivanov},
  Journal                  = {J. Chem. Phys.},
  Year                     = {2018},
  Pages                    = {104107},
  Volume                   = {148},
  Number                   = {10},
  DOI                      = {10.1063/1.5016438}
}

@article{feskov_ctc_18,
  title                    = {The Green's function technique for numerical simulations of multichannel electron transfer reactions in electron-donor-acceptor complexes},
  author                   = {S. V. Feskov},
  journal                  = {Comput. Theor. Chem.},
  volume                   = {1145},
  pages                    = {15 - 21},
  year                     = {2018},
  issn                     = {2210-271X},
  doi                      = {10.1016/j.comptc.2018.10.007}
}

@Article{feskov_jcp_19,
  author                   = {Feskov, Serguei V. and Rogozina, Marina V. and Ivanov, Anatoly I. and Aster, Alexander and Koch, Marius and Vauthey, Eric},
  title                    = {Magnetic field effect on ion pair dynamics upon bimolecular photoinduced electron transfer in solution},
  journal                  = {J. Chem. Phys.},
  volume                   = {150},
  number                   = {2},
  pages                    = {024501},
  year                     = {2019},
  doi                      = {10.1063/1.5064802}
}

@article{feskov_jpcb_20,
  author                   = {Feskov, Serguei V. and Malykhin, Roman E. and Ivanov, Anatoly I.},
  title                    = {The Efficiency of Photoinduced Intramolecular Charge Separation from the Second Excited State: What Factors Can Control It?},
  journal                  = {J. Phys. Chem. B},
  volume                   = {124},
  number                   = {46},
  pages                    = {10442--10455},
  year                     = {2020},
  doi                      = {10.1021/acs.jpcb.0c07978}
}

@article{siplivy_jcp_20,
  author                   = {Siplivy, Nickolay B. and Feskov, Serguei V. and Ivanov, Anatoly I.},
  title                    = {Quantum yield and energy efficiency of photoinduced intramolecular charge separation},
  journal                  = {J. Chem. Phys.},
  volume                   = {153},
  number                   = {4},
  pages                    = {044301},
  year                     = {2020},
  doi                      = {10.1063/5.0013708}
}

@article{nazarov_jml_22,
  author                   = {Alexey E. Nazarov and Anatoly I. Ivanov and Arnulf Rosspeintner and Gonzalo Angulo},
  title                    = {Full relaxation dynamics recovery from ultrafast fluorescence experiments by means of the stochastic model: Does the solvent response dynamics depend on the fluorophore nature?},
  journal                  = {J. Mol. Liq.},
  volume                   = {360},
  pages                    = {119387},
  year                     = {2022},
  doi                      = {10.1016/j.molliq.2022.119387}
}

@Article{feskov_ijms_22,
  author                   = {Feskov, Serguei V.},
  title                    = {Semiclassical Theory of Multistage Nonequilibrium Electron Transfer in Macromolecular Compounds in Polar Media with Several Relaxation Timescales},
  journal                  = {Int. J. Mol. Sci.},
  volume                   = {23},
  year                     = {2022},
  number                   = {24},
  article-number           = {15793},
  ignored_url              = {https://www.mdpi.com/1422-0067/23/24/15793},
  issn                     = {1422-0067},
  doi                      = {10.3390/ijms232415793}
}

@Article{feskov_rjpcb_24,
  author                   = {Feskov, Serguei V.},
  title                    = {Method of Splitting Polarization Coordinates for the Description of Ultrafast Multistage Electron Transfer in a Non-Debye Medium},
  journal                  = {Russ. J. Phys. Chem. B},
  volume                   = {18},
  year                     = {2024},
  issue                    = {1},
  pages                    = {1--8},
  ignored_url              = {https://doi.org/10.1134/S1990793124010081},
  issn                     = {1990-7923},
  doi                      = {10.1134/S1990793124010081}
}

@article{wrobel_ccr_11,
  title         = {Photoinduced electron transfer processes in fullerene-organic chromophore systems},
  author        = {Danuta Wróbel and Andrzej Graja},
  journal       = {Coord. Chem. Rev.},
  volume        = {255},
  number        = {21},
  pages         = {2555-2577},
  year          = {2011},
  issn          = {0010-8545},
  doi           = {10.1016/j.ccr.2010.12.026}
}

@Article{guldi_csr_02,
  author        = "Guldi, Dirk M.",
  title         = "Fullerene–porphyrin architectures; photosynthetic antenna and reaction center models",
  journal       = "Chem. Soc. Rev.",
  year          = "2002",
  volume        = "31",
  issue         = "1",
  pages         = "22-36",
  publisher     = "The Royal Society of Chemistry",
  doi           = "10.1039/B106962B"
}

@Article{souza_cc_09,
  author        = "D’Souza, Francis and Ito, Osamu",
  title         = "Supramolecular donor–acceptor hybrids of porphyrins/phthalocyanines with fullerenes/carbon nanotubes: electron transfer, sensing, switching, and catalytic applications",
  journal       = "Chem. Commun.",
  year          = "2009",
  issue         = "33",
  pages         = "4913-4928",
  doi           = "10.1039/B905753F"
}

@article{machin2023,
  title         = {Artificial photosynthesis: current advancements and future prospects},
  author        = {Mach{\'\i}n, Abniel and Cotto, Mar{\'\i}a and Ducong{\'e}, Jos{\'e} and M{\'a}rquez, Francisco},
  journal       = {Biomimetics},
  volume        = {8},
  number        = {3},
  pages         = {298},
  year          = {2023},
  publisher     = {MDPI},
  doi           = {10.3390/biomimetics8030298}
}

@article{cherepanov2022,
  title         = {Current state of the primary charge separation mechanism in photosystem {I} of cyanobacteria},
  author        = {Cherepanov, Dmitry A. and Semenov, Alexey Yu. and Mamedov, Mahir D. and Aybush, Arseniy V. and Gostev, Fedor E. and Shelaev, Ivan V. and Shuvalov, Vladimir A. and Nadtochenko, Victor A.},
  journal       = {Biophys. Rev.},
  volume        = {14},
  number        = {4},
  pages         = {805--820},
  year          = {2022},
  publisher     = {Springer},
  doi           = {10.1007/s12551-022-00983-1}
}

@article{nguyen_sci_23,
   author       = {Hoang H. Nguyen  and Yin Song  and Elizabeth L. Maret  and Yogita Silori  and Rhiannon Willow  and Charles F. Yocum  and Jennifer P. Ogilvie },
   title        = {Charge separation in the photosystem II reaction center resolved by multispectral two-dimensional electronic spectroscopy},
   journal      = {Sci. Adv.},
   volume       = {9},
   number       = {18},
   pages        = {eade7190},
   year         = {2023},
   doi          = {10.1126/sciadv.ade7190}
}

@article{bottari_ccr_21,
  title         = {An exciting twenty-year journey exploring porphyrinoid-based photo- and electro-active systems},
  author        = {Giovanni Bottari and Gema {de la Torre} and Dirk M. Guldi and Tomás Torres},
  journal       = {Coord. Chem. Rev.},
  volume        = {428},
  pages         = {213605},
  year          = {2021},
  issn          = {0010-8545},
  doi           = {10.1016/j.ccr.2020.213605}
}

@article{wang_as_24,
  author        = {Wang, Lingsong and Zhu, Weigang},
  title         = {Organic Donor-Acceptor Systems for Photocatalysis},
  journal       = {Adv. Sci.},
  volume        = {11},
  number        = {10},
  pages         = {2307227},
  doi           = {10.1002/advs.202307227},
  year          = {2024}
}

@Article{scattergood_dt_14,
  author        = "Scattergood, Paul A. and Delor, Milan and Sazanovich, Igor V. and Bouganov, Oleg V. and Tikhomirov, Sergei A. and Stasheuski, Alexander S. and Parker, Anthony W. and Greetham, Gregory M. and Towrie, Michael and Davies, E. Stephen and Meijer, Anthony J. H. M. and Weinstein, Julia A.",
  title         = "Electron transfer dynamics and excited state branching in a charge-transfer platinum(ii) donor–bridge-acceptor assembly",
  journal       = "Dalton Trans.",
  year          = "2014",
  volume        = "43",
  issue         = "47",
  pages         = "17677-17693",
  doi           = "10.1039/C4DT01682C"
}

@article{canton_nc_15,
  title         = {Visualizing the non-equilibrium dynamics of photoinduced intramolecular electron transfer with femtosecond X-ray pulses},
  journal       = {Nat. Commun.},
  author        = {Canton, Sophie E. and Kjær, Kasper S. and Vankó, György and van Driel, Tim B. and Adachi, Shin-ichi and Bordage, Amélie and Bressler, Christian and Chabera, Pavel and Christensen, Morten and Dohn, Asmus O. and Galler, Andreas and Gawelda, Wojciech and Gosztola, David and Haldrup, Kristoffer and Harlang, Tobias and Liu, Yizhu and Møller, Klaus B. and Németh, Zoltán and Nozawa, Shunsuke and Pápai, Mátyás and Sato, Tokushi and Sato, Takahiro and Suarez-Alcantara, Karina and Togashi, Tadashi and Tono, Kensuke and Uhlig, Jens and Vithanage, Dimali A. and Wärnmark, Kenneth and Yabashi, Makina and Zhang, Jianxin and Sundström, Villy and Nielsen, Martin M.},
  volume        = {6},
  number        = {1},
  pages         = {6359},
  year          = {2015},
  issn          = {2041-1723},
  doi           = {10.1038/ncomms7359}
}

@article{rizzi_jpca_08,
  author        = {Rizzi, Alberto C. and van Gastel, Maurice and Liddell, Paul A. and Palacios, Rodrigo E. and Moore, Gary F. and Kodis, Gerdenis and Moore, Ana L. and Moore, Tom A. and Gust, Devens and Braslavsky, Silvia E.},
  title         = {Entropic Changes Control the Charge Separation Process in Triads Mimicking Photosynthetic Charge Separation},
  journal       = {J. Phys. Chem. A},
  volume        = {112},
  number        = {18},
  pages         = {4215-4223},
  year          = {2008},
  doi           = {10.1021/jp712008b}
}

@article{rego_jpcc_14,
  author        = {Rego, Luis G. C. and Hames, Bruno C. and Mazon, Kahio T. and Joswig, Jan-Ole},
  title         = {Intramolecular Polarization Induces Electron–Hole Charge Separation in Light-Harvesting Molecular Triads},
  journal       = {J. Phys. Chem. C},
  volume        = {118},
  number        = {1},
  pages         = {126-134},
  year          = {2014},
  doi           = {10.1021/jp408955e}
}

@article{manna_jpcl_15,
  author        = {Manna, Arun K. and Balamurugan, D. and Cheung, Margaret S. and Dunietz, Barry D.},
  title         = {Unraveling the Mechanism of Photoinduced Charge Transfer in Carotenoid–Porphyrin–C60 Molecular Triad},
  journal       = {J. Phys. Chem. Lett.},
  volume        = {6},
  number        = {7},
  pages         = {1231-1237},
  year          = {2015},
  doi           = {10.1021/acs.jpclett.5b00074}
}

@article{sun_jpcc_18,
  author        = {Sun, Xiang and Zhang, Pengzhi and Lai, Yifan and Williams, Kyle L. and Cheung, Margaret S. and Dunietz, Barry D. and Geva, Eitan},
  title         = {Computational Study of Charge-Transfer Dynamics in the Carotenoid–Porphyrin–C60 Molecular Triad Solvated in Explicit Tetrahydrofuran and Its Spectroscopic Signature},
  journal       = {J. Phys. Chem. C},
  volume        = {122},
  number        = {21},
  pages         = {11288-11299},
  year          = {2018},
  doi           = {10.1021/acs.jpcc.8b02697}
}

@article{hu_jpcb_20,
  author        = {Hu, Zhubin and Tong, Zhengqing and Cheung, Margaret S. and Dunietz, Barry D. and Geva, Eitan and Sun, Xiang},
  title         = {Photoinduced Charge Transfer Dynamics in the Carotenoid–Porphyrin–C60 Triad via the Linearized Semiclassical Nonequilibrium Fermi’s Golden Rule},
  journal       = {J. Phys. Chem. B},
  volume        = {124},
  number        = {43},
  pages         = {9579-9591},
  year          = {2020},
  doi           = {10.1021/acs.jpcb.0c06306}
}

@article{Ivanov_rusjpcb_08,
  author        = {Ivanov, Anatoly I. and Ionkin, Vladimir N. and Feskov, Serguei V.},
  title         = {Acceleration of the Recombination of Photoexcited Donor—Acceptor Complexes with a High-Frequency Vibrational Mode},
  journal       = {Russ. J. Phys. Chem.},
  volume        = {82},
  number        = {2},
  pages         = {303-309},
  year          = {2008},
  doi           = {10.1134/S0036024408020295}
}

@article{feskov_rjpca_17,
  author        = {Feskov, Serguei V. and Yudanov, Vladislav V.},
  title         = {Model of Multistep Electron Transfer in a Single-Mode Polar Medium},
  journal       = {Russ. J. Phys. Chem. A},
  volume        = {91},
  number        = {9},
  pages         = {1816-1823},
  year          = {2017},
  doi           = {10.1134/S0036024417090102}
}

@Book{Burdzinski_05,
   Title                    = {Organic Photochemistry and Photophysics},
   Author                   = {G. Burdzinski and J. Kubicki and A. Maciejewski and R.P. Steer and S. Velate and E.K.L. Yeow},
   Publisher                = {Taylor \& Francis},
   Year                     = {2005},
   pages                    = {312},
   Address                  = {New York},
   ISBN                     = {9780429126642},
   doi                      = {10.1201/9781420036992}
}

@Article{Vauthey_PCCP23,
    author ="Verma, Pragya and Tasior, Mariusz and Roy, Palas and Meech, Stephen R. and Gryko, Daniel T. and Vauthey, Eric",
    title  ="Excited-state symmetry breaking in quadrupolar pull-push-pull molecules: dicyanovinyl vs. cyanophenyl acceptors",
    journal  ="Phys. Chem. Chem. Phys.",
    year  ="2023",
    volume  ="25",
    issue  ="34",
    pages  ="22689-22699",
    publisher  ="The Royal Society of Chemistry",
    doi  ="10.1039/D3CP02810K"
}

@Article{Terenziani23,
    author ="Swathi, K. and Sujith, Meleppatt and Divya, P. S. and P, Merin Varghese and Delledonne, Andrea and Phan Huu, D. K. Andrea and Di Maiolo, Francesco and Terenziani, Francesca and Lapini, Andrea and Painelli, Anna and Sissa, Cristina and Thomas, K. George",
    title  ="From symmetry breaking to symmetry swapping: is Kasha{'}s rule violated in multibranched phenyleneethynylenes?",
    journal  ="Chem. Sci.",
    year  ="2023",
    volume  ="14",
    issue  ="8",
    pages  ="1986-1996",
    publisher  ="The Royal Society of Chemistry",
    doi  ="10.1039/D2SC05206G"
}

@article{IVANOVRev24,
    title = {Symmetry breaking charge transfer in excited multibranched molecules and dimers: A unified standpoint},
    author = {Anatoly I. Ivanov},
    journal = {J. Photochem. Photobiol. C},
    volume = {58},
    pages = {100651},
    year = {2024},
    issn = {1389-5567},
    doi = {10.1016/j.jphotochemrev.2024.100651}
}

@article{Siplivy24,
    author = {Siplivy, Nikolay B. and Ivanov, Anatoly I.},
    title = "{Symmetry breaking charge transfer and control of the transition dipole moment in excited octupolar molecules}",
    journal = {J. Chem. Phys.},
    volume = {160},
    number = {19},
    pages = {194302},
    year = {2024},
    month = {05},
    issn = {0021-9606},
    doi = {10.1063/5.0211030}
}

@article{Wasielewski2020,
    author = {Young, Ryan M. and Wasielewski, Michael R.},
    title = {Mixed Electronic States in Molecular Dimers: Connecting Singlet Fission, Excimer Formation, and Symmetry-Breaking Charge Transfer},
    journal = {Acc. Chem. Res.},
    volume = {53},
    number = {9},
    pages = {1957--1968},
    year = {2020},
    doi = {10.1021/acs.accounts.0c00397}
}

@article{les24,
    author = {Mikhailova, T. V. and Ivanov, A. I.},
    title = {Controlling the symmetry breaking charge transfer extent in excited quadrupolar molecules by tuning the locally excited state},
    journal = {J. Chem. Phys.},
    year = {2024},
    volume = {160},
    pages = {054302-1--054302-9},
    number = { },
    doi ={10.1063/5.0193532}
}

@article{MikhMikh24,
    author = {Mikhailova, Tatyana V. and Mikhailova, Valentina A. and Ivanov, Anatoly I.},
    journal = {J. Chem. Phys.},
    title = {Effect of locally excited state on fluorescence transition dipole moment in quadrupolar molecules subjected to symmetry breaking charge transfer},
    volume = {161},
    number = {15},
    pages = {154303},
    year = {2024},
    month = {10},
    issn = {0021-9606},
    doi = {10.1063/5.0237870}
}

@article{IvanovQubit,
    author = {Ivanov, Anatoly I.},
    title = {Electric dipole moment of excited octupolar molecules: Potential qubit implementation},
    journal = {J. Chem. Phys.},
    volume = {162},
    number = {2},
    pages = {024303},
    year = {2025},
    month = {01},
    issn = {0021-9606},
    doi = {10.1063/5.0243375}
}

@ARTICLE{Ivanov18,
    author = {Anatoly I. Ivanov},
    title = {Theory of Vibrational Spectra of Excited Quadrupolar Molecules with Broken Symmetry},
    journal = {J. Phys. Chem. C},
    year = {2018},
    volume = {122},
    pages = {29165--29172},
    number = {},
    doi = {10.1021/acs.jpcc.8b10985}
}

@article{Antipov22,
   author = {Antipov,Ivan F.  and Ivanov,Anatoly I. },
   title = {Minimal model of excited-state symmetry breaking in symmetric dimers and covalently linked dyads},
   journal = {J. Chem. Phys.},
   volume = {157},
   number = {22},
   pages = {224104},
   year = {2022},
   doi = {10.1063/5.0129697}
}

@book{blankenship_book_21,
   title             = {Molecular Mechanisms of Photosynthesis},
   author            = {Blankenship, R.E.},
   isbn              = {9781119800019},
   lccn              = {2021028963},
   year              = {2021},
   publisher         = {Wiley}
}

@article{ponseca_cr_17,
   author            = {Ponseca, Carlito S. Jr. and Chábera, Pavel and Uhlig, Jens and Persson, Petter and Sundstr{\"o}m, Villy},
   title             = {Ultrafast Electron Dynamics in Solar Energy Conversion},
   journal           = {Chem. Rev.},
   volume            = {117},
   number            = {16},
   pages             = {10940-11024},
   year              = {2017},
   doi               = {10.1021/acs.chemrev.6b00807}
}

@article{santos_cr_22,
   author            = {Santos, Elizabeth and Schmickler, Wolfgang},
   title             = {Models of Electron Transfer at Different Electrode Materials},
   journal           = {Chem. Rev.},
   volume            = {122},
   number            = {12},
   pages             = {10581-10598},
   year              = {2022},
   doi               = {10.1021/acs.chemrev.1c00583}
}

@article{lawrence_nrb_23,
   author            = {Lawrence, J.M. and Egan, R.M. and Hoefer, T. and Scarampi, Alberto and Shang, Linying and Howe, Christopher J. and Zhang, Jenny Z.},
   title             = {Rewiring photosynthetic electron transport chains for solar energy conversion},
   journal           = {Nat. Rev. Bioeng.},
   volume            = {1},
   pages             = {887–905},
   year              = {2023},
   doi               = {10.1038/s44222-023-00093-x}
}

@article{ballabio_acs_22,
   author            = {Ballabio, Marco and Cánovas, Enrique},
   title             = {Electron Transfer at Quantum Dot–Metal Oxide Interfaces for Solar Energy Conversion},
   journal           = {ACS Nanosci. Au.},
   volume            = {2},
   number            = {5},
   pages             = {367-395},
   year              = {2022},
   doi               = {10.1021/acsnanoscienceau.2c00015}
}

@article{xin_nrp_19,
   author            = {Xin, Na and Guan, Jianxin and Zhou, Chenguang and Chen, Xinjiani and Gu, Chunhui and Li, Yu and Ratner, Mark A. and Nitzan, Abraham and Stoddart, J. Fraser and Guo, Xuefeng},
   title             = {Concepts in the design and engineering of single-molecule electronic devices},
   journal           = {Nat. Rev. Phys.},
   volume            = {1},
   number            = {3},
   pages             = {211--230},
   year              = {2019},
   doi               = {10.1038/s42254-019-0022-x}
}

@article{kundu_jpcl_18,
   author            = {Kundu, Mainak and He, Ting-Fang and Lu, Yangyi and Wang, Lijuan and Zhong, Dongping},
   title             = {Short-Range Electron Transfer in Reduced Flavodoxin: Ultrafast Nonequilibrium Dynamics Coupled with Protein Fluctuations},
   journal           = {J. Phys. Chem. Lett.},
   volume            = {9},
   number            = {11},
   pages             = {2782-2790},
   year              = {2018},
   doi               = {10.1021/acs.jpclett.8b00882}
}

@article{lu_nc_20,
   author            = {Lu, Yangyi and Kundu, Mainak and Zhong, Dongping},
   title             = {Effects of nonequilibrium fluctuations on ultrafast short-range electron transfer dynamics},
   journal           = {Nat. Comm.},
   volume            = {11},
   number            = {1},
   pages             = {2822},
   year              = {2020},
   doi               = {10.1038/s41467-020-15535-y}
}

@article{matyushov_jml_18,
   author            = {Dmitry V. Matyushov},
   title             = {Fluctuation relations, effective temperature, and ageing of enzymes: The case of protein electron transfer},
   journal           = {J. Mol. Liq.},
   volume            = {266},
   pages             = {361-372},
   year              = {2018},
   issn              = {0167-7322},
   doi               = {10.1016/j.molliq.2018.06.087}
}

@article{torieda_jpca_04,
   author            = {Torieda, Hiroaki and Nozaki, Koichi and Yoshimura, Akio and Ohno, Takeshi},
   title             = {Low Quantum Yields of Relaxed Electron Transfer Products of Moderately Coupled Ruthenium(II)−Cobalt(III) Compounds on the Subpicosecond Laser Excitation},
   journal           = {J. Phys. Chem. A},
   volume            = {108},
   number            = {22},
   pages             = {4819-4829},
   year              = {2004},
   doi               = {10.1021/jp037259z}
}

@article{Hariharan25,
   author            = {Mazumder, Aniruddha and Vinod, Kavya and Thomas, Aivin Chemmarappallil and Hariharan, Mahesh},
   title             = {Accelerating Symmetry-Breaking Charge Separation in an Angular versus Linear Perylenediimide Dimer through the Modulation of Coulombic Coupling},
   journal           = {J. Phys. Chem. Lett.},
   volume            = {16},
   number            = {19},
   pages             = {4819-4827},
   year              = {2025},
   doi               = {10.1021/acs.jpclett.5c00372}
}

@article{dereka_jpcl_24,
   author            = {Dereka, Bogdan and Balanikas, Evangelos and Rosspeintner, Arnulf and Li, Zhiquan and Liska, Robert and Vauthey, Eric},
   title             = {Excited-State Symmetry Breaking and Localization in a Noncentrosymmetric Electron Donor–Acceptor–Donor Molecule},
   journal           = {J. Phys. Chem. Lett.},
   volume            = {15},
   number            = {32},
   pages             = {8280-8286},
   year              = {2024},
   doi               = {10.1021/acs.jpclett.4c01694}
}

@article{Nazarov_SB20,
   author            = {Nazarov, Alexey E. and Ivanov, Anatoly I.},
   title             = {Nonstationary Theory of Excited State Charge Transfer Symmetry Breaking Driven by Polar Solvent},
   journal           = {J. Phys. Chem. B},
   volume            = {124},
   number            = {47},
   pages             = {10787--10801},
   year              = {2020},
   doi               = {10.1021/acs.jpcb.0c07612}
}

@ARTICLE{Vauthey17,
    author           = {Bogdan Dereka and  Eric Vauthey},
    title            = {Solute-Solvent Interactions and Excited-State Symmetry Breaking: Beyond the Dipole-Dipole and the Hydrogen-Bond Interactions},
    journal          = {J. Phys. Chem. Lett.},
    year             = {2017},
    volume           = {8},
    pages            = {3927--3932},
    number           = {16},
    doi              = {10.1021/acs.jpclett.7b01821}
}

@article{phelan_jacs_19,
    author           = {Brian T. Phelan and Jinyuan Zhang and Guan-Jhih Huang and Yi-Lin Wu and Mehdi Zarea and Ryan M. Young and Michael R. Wasielewski},
    title            = {Quantum Coherence Enhances Electron Transfer Rates to Two Equivalent Electron Acceptors},
    journal          = {J. Am. Chem. Soc.},
    volume           = {141},
    number           = {31},
    pages            = {12236-12239},
    year             = {2019},
    doi              = {10.1021/jacs.9b06166}
}

@article{fukuzumi_acr_14,
    author           = {Fukuzumi, Shunichi and Ohkubo, Kei and Suenobu, Tomoyoshi},
    title            = {Long-Lived Charge Separation and Applications in Artificial Photosynthesis},
    journal          = {Acc. Chem. Res.},
    volume           = {47},
    number           = {5},
    pages            = {1455-1464},
    year             = {2014},
    doi              = {10.1021/ar400200u}
}

@article{loong_jpcb_22,
    author           = {Loong, Hao and Zhou, Jiadong and Jiang, Nianqiang and Feng, Yi and Xie, Guojing and Liu, Linlin and Xie, Zengqi},
    title            = {Photoinduced Cascading Charge Transfer in Perylene Bisimide-Based Triads},
    journal          = {J. Phys. Chem. B},
    volume           = {126},
    number           = {12},
    pages            = {2441-2448},
    year             = {2022},
    doi              = {10.1021/acs.jpcb.2c00965}
}

@article{brown_jpca_24,
    author           = {Brown, Paige J. and Qiu, Yunfan and Latawiec, Elisabeth I. and Phelan, Brian T. and Tcyrulnikov, Nikolai A. and Palmer, Jonathan R. and Krzyaniak, Matthew D. and Kopp, Sebastian M. and Huang, Yuheng and Young, Ryan M. and Wasielewski, Michael R.},
    title            = {Enhancing Photogenerated Radical Pair Properties in Donor-Chromophore-Acceptor Systems for Quantum Information Applications},
    journal          = {J. Phys. Chem. A},
    volume           = {128},
    number           = {43},
    pages            = {9371-9382},
    year             = {2024},
    doi              = {10.1021/acs.jpca.4c05015}
}

@article{fisher_jacs_24,
    author           = {Fisher, Jeremy M. and Williams, Malik L. and Palmer, Jonathan R. and Powers-Riggs, Natalia E. and Young, Ryan M. and Wasielewski, Michael R.},
    title            = {Long-Lived Charge Separation in Single Crystals of an Electron Donor Covalently Linked to Four Acceptor Molecules},
    journal          = {J. Am. Chem. Soc.},
    volume           = {146},
    number           = {14},
    pages            = {9911-9919},
    year             = {2024},
    doi              = {10.1021/jacs.4c00201}
}

@article{wang_jpcc_23,
    author           = {Wang, Xinyue and Wang, Hongxiang and Zhang, Meixia and Pullerits, Tõnu and Song, Peng},
    title            = {Charge-Separated States Determined Photoinduced Electron Transfer Efficiency in a D-D-A System in an External Electric Field},
    journal          = {J. Phys. Chem. C},
    volume           = {127},
    number           = {6},
    pages            = {2805-2817},
    year             = {2023},
    doi              = {10.1021/acs.jpcc.2c07364}
}

@article{lee_cr_23,
    author           = {Lee, Yong-Min and Nam, Wonwoo and Fukuzumi, Shunichi},
    title            = {Redox catalysis via photoinduced electron transfer},
    journal          = {Chem. Sci.},
    year             = {2023},
    volume           = {14},
    issue            = {16},
    pages            = {4205-4218},
    doi              = {10.1039/D2SC07101K}
}

@Article{santoro_mol_22,
    AUTHOR           = {Santoro, Antonio and Bella, Giovanni and Cancelliere, Ambra M. and Serroni, Scolastica and Lazzaro, Giuliana and Campagna, Sebastiano},
    TITLE            = {Photoinduced Electron Transfer in Organized Assemblies—Case Studies},
    JOURNAL          = {Molecules},
    VOLUME           = {27},
    YEAR             = {2022},
    NUMBER           = {9},
    ARTICLE-NUMBER   = {2713},
    DOI              = {10.3390/molecules27092713}
}

@article{hou_jmcc_19,
    author            = "Hou, Yuqi and Zhang, Xue and Chen, Kepeng and Liu, Dongyi and Wang, Zhijia and Liu, Qingyun and Zhao, Jianzhang and Barbon, Antonio",
    title             = "Charge separation{,} charge recombination{,} long-lived charge transfer state formation and intersystem crossing in organic electron donor/acceptor dyads",
    journal           = "J. Mater. Chem. C",
    year              = "2019",
    volume            = "7",
    issue             = "39",
    pages             = "12048-12074",
    doi               = "10.1039/C9TC04285G"
}

@article{zusman_jcp_99,
    author            = {Zusman, Leonid D. and Beratan, David N.},
    title             = {Electron transfer in three-center chemical systems},
    journal           = {J. Chem. Phys.},
    volume            = {110},
    number            = {21},
    pages             = {10468-10481},
    year              = {1999},
    month             = {06},
    doi               = {10.1063/1.478976}
}

@article{kirner_cs_15,
    author            = "Kirner, Sabrina V. and Henkel, Christian and Guldi, Dirk M. and Megiatto Jr, Jackson D. and Schuster, David I.",
    title             = "Multistep energy and electron transfer processes in novel rotaxane donor–acceptor hybrids generating microsecond-lived charge separated states",
    journal           = "Chem. Sci.",
    year              = "2015",
    volume            = "6",
    issue             = "12",
    pages             = "7293-7304",
    doi               = "10.1039/C5SC02895G",
}

\section*{Authors Biographies}

\textbf{S. V. Feskov} is a Professor and Leading Researcher at Volgograd State University (VolSU), where he received his Ph.D. (2000) and D.Sc. (2012) in Chemical Physics. He began his scientific career in 1995 as a member of Prof. A. I. Ivanov’s group, studying magnetic field effects in spin-selective radical reactions. From 2006 to 2009, he worked as a visiting researcher in the group of Prof. A. I. Burshtein at the Weizmann Institute of Science (Israel). His research is focused on theoretical and numerical modeling of ultrafast charge transfer processes in complex molecular systems, with particular interest in the role of environmental relaxation, magnetic field effects, and molecular diffusion in liquids.

\bigskip
\textbf{A. I. Ivanov} is a Professor of Physics at Volgograd State University. He graduated from the Department of Physics at Bashkir State University (Ufa, Russia) in 1973. Before joining VolSU in 1982, he was a researcher at the Institute of Chemistry, Russian Academy of Sciences (Ufa), where he also earned his Ph.D. in Physics. He later obtained his D.Sc. (habilitation) in Chemical Physics from the Physical-Technical Institute of the Russian Academy of Sciences. His early research was devoted to the theory of elementary chemical reactions, nonradiative transitions in polyatomic molecules, and electron capture in the gas phase. Since the mid-1990s, his work has centered on the theory of ultrafast charge transfer in solutions. Since 2016, his primary research interest has been symmetry breaking charge transfer in electronically excited molecular aggregates.

\end{document}